\pdfoutput=1

\documentclass
[10pt,preprintnumbers,twocolumn,nofootinbib,floatfix,aps,prd]{revtex4-1}
\usepackage{hyperref}
\usepackage{amssymb}
\usepackage{amsmath}
\usepackage{multirow}
\usepackage{mathrsfs}
\usepackage{subfig}
\usepackage{graphicx}
\usepackage{booktabs}
\usepackage{epstopdf}

\def\slashchar#1{\setbox0=\hbox{$#1$}           
   \dimen0=\wd0                                 
   \setbox1=\hbox{/} \dimen1=\wd1               
   \ifdim\dimen0>\dimen1                        
      \rlap{\hbox to \dimen0{\hfil/\hfil}}      
      #1                                        
   \else                                        
      \rlap{\hbox to \dimen1{\hfil$#1$\hfil}}   
      /                                         
   \fi}                                        %

\def\eq#1{eq.~(\ref{#1})}
\def\Eq#1{Eq.~(\ref{#1})}
\def\Eqs#1#2{Eqs.~(\ref{#1}) and (\ref{#2})}
\def\eqs#1#2{eqs.~(\ref{#1}) and (\ref{#2})}
\def\eqss#1#2#3{eqs.~(\ref{#1}), (\ref{#2}) and (\ref{#3})}

\def\Ref#1{ref.~\cite{#1}}

\def\Refs#1{refs.~\cite{#1}}

\newcommand{\ie}{{\it i.e.}, }

\newcommand{\gy}{g_{Y}}

\newcommand{\gbl}{g_{BL}}
\newcommand{\gz}{g_{Z}}
\newcommand{\gzp}{g_{Z'}}
\renewcommand{\bar}{\overline}
\renewcommand{\hat}{\widehat}

\begin{document}

\title{Heavy Fermion Non-Decoupling Effects in Triple Gauge Boson Vertices}
\author{Athanasios  Dedes$^{*}$ and Kristaq Suxho}
\email{adedes@cc.uoi.gr, csoutzio@cc.uoi.gr}
\affiliation{Division of Theoretical Physics, Physics Department, University of Ioannina,  GR 45110, Greece}

\begin{abstract}
Within a spontaneously broken 
gauge group we carefully analyse and calculate triple gauge boson
vertices dominated  by  triangle one-loop Feynman diagrams involving
heavy fermions compared to external momenta and gauge boson masses. 
We perform our calculation strictly in four dimensions 
and derive a general formula for 
the off-shell, one-particle  irreducible (1PI) effective
vertex which satisfies the relevant Ward Identities
and the Goldstone boson equivalence theorem. 
Our goal is to search for non-decoupling heavy fermion effects 
highlighting their synergy with gauge chiral anomalies.
Particularly in the Standard Model, we find that when the arbitrary anomaly parameters are fixed by
gauge invariance and/or Bose symmetry, the heavy fermion contribution
cancels its anomaly contribution leaving behind anomaly and mass independent
contributions from the light fermions.
We  apply these results in calculating the corresponding CP-invariant
one-loop induced corrections to triple gauge boson vertices in the SM,
minimal $Z'$models as well as their extensions with a fourth fermion generation,
and compare with experimental data. 
%

\end{abstract}

\maketitle


\section{Introduction}
\label{sec:intro}

In general, the Appelquist-Carazzone~\cite{Appelquist:Carazzone} theorem states
that the effect from a heavy fermion mass $m$
at low energy observables is suppressed by powers of $m$. However, this theorem does not hold for
theories with chiral gauge couplings or large mass splitting within gauge multiplets, 
a situation known to take place in the minimal Standard Model (SM) of particle physics~\cite{Glashow,Weinberg,Salam}.
Failure of the decoupling of a heavy fermion from radiative corrections
requires breaking of a local gauge symmetry and, in addition, 
breaking of a global symmetry by these
corrections~\cite{Peskin:delayed,Kennedy:1991wa}.     

Another aspect of theories with chiral gauge couplings is the Adler-Bell-Jackiw or chiral 
anomaly~\cite{Adler,Bell,Bardeen,Fujikawa}. This is the situation where certain classical Ward Identities (WIs) are violated
by quantum corrections (for reviews see~\cite{Harvey,Hill,Bilal}).  
For a model that is non-anomaly free, anomalous Ward Identities render it non-renormalizable
and non-unitary. This problem shows up in every symmetry breaking stage of the model. In order to
cancel chiral anomalies associated with axial (AAA) 
or vector-axial (VVA) currents in gauge theories,  we either need 
to stick to only by-construction anomaly-free gauge groups, or,
 to introduce additional chiral femionic fields~\cite{Bouchiat,Gross}. 

An energy region of  experimental interest corresponds to the case where a fermion mass
$m$ is very heavy, $m_Z^2 < s \ll m^2$, so that it cannot be pair-produced  
at Tevatron, LHC or a future lepton-collider. If this fermion 
is chiral \ie it receives its mass from the Higgs mechanism 
which is also responsible for the gauge boson mass, 
then the question of the decoupling of this particle 
would cause a  problem in anomaly cancellation and therefore to gauge invariance. 
This question
has been tackled in many papers in the literature most 
notably by D'Hoker and Farhi in \Ref{D'Hoker:SM,D'Hoker:GEN} :
decoupling of a fermion whose mass is generated by a Yukawa coupling induces an action functional of the Higgs field and 
gauge boson fields term, analogous to 
Wess-Zumino-Witten (WZW) term~\cite{Wess,Witten:Anom} in chiral Lagrangian. 
Then D'Hoker and Farhi showed that the theory 
without the decoupled fermion but with the WZW term is gauge invariant.   
Applications of this non-decoupling effect has been utilised in many physics projects 
from hadronic up to electroweak physics of the SM and beyond, 
see for example \Refs{Kaymakcalan:1983qq,HHH1,HHH2,DiazCruz,D'Hoker:1992,Lin,Feruglio:1992fp}. 
However,  to our knowledge, the above conclusion  has not been 
drawn in the broken phase of theories with spontaneous gauge symmetry
breaking like the SM. It is after all meaningful to discuss  non-decoupling
effects \emph{only} in theories where the physical masses appear explicitly.

The problem when discussing decoupling effects or in general physics 
associated with the fermionic triangle graph
is related to the question : {\it what is the correct result for such a graph?}
The answer depends on the physical set-up in which it arises~\cite{Jackiw:1999qq}. 
For example,
as we shall show below in the case of SM, gauge invariance and  
Bose symmetry 
are enough to set the triple neutral gauge boson vertices finite and well defined. 
Only then  can we reach the conclusions for the theory at the heavy fermion mass limit.

If the SM gauge group is extended by extra $U(1)$'s then 
anomaly cancellation conditions become more involved. 
Recently,  the authors of \Refs{Antoniadis:2009,Dudas:2009}  noted 
that such cancellations may occur inside a ``cluster'' of anomaly-free
heavy fermion sector which is not
accessible by the current colliders, leaving behind non-decoupling effects 
in trilinear gauge boson vertices of the extra massive gauge boson $Z'$ and those of the SM $Z'ZZ, Z'WW, Z'Z\gamma$ that may be observable at low energies.  
These effects are visible in the energy region where 
$M_{Z'} \sim g v < \sqrt{s} \ll m \sim \lambda v$. 
For these non-decoupling effects to occur it is necessary 
for fermions and gauge bosons to receive mass
from the same Higgs boson and there must be a hierarchy 
between Yukawa and gauge coupling, $\lambda\sim O(1) \gg g$.
In this paper we also elaborate on this issue categorising 
conditions among couplings where such a situation occurs.
We then present a few toy-model examples with two or three different external
gauge bosons.

We note in passing that, within field theory, 
mixed anomaly cancellations via 4d Green-Shwartz mechanism
have been discussed and analysed  
phenomenologically in many 
papers e.g. \cite{Preskill,Anastasopoulos,Wells,
Coriano1,Coriano2,Armillis}.

Our goal here is to construct a perturbative,  gauge invariant
one-loop proper effective vertex for three external
gauge bosons that incorporates both chiral anomaly 
ambiguities together with non-decoupling effects induced 
by heavy fermions in an explicit manner.  We would like to apply 
this effective vertex in order to:
\begin{itemize}
\item investigate the interplay between  chiral anomaly effects  and
non-decoupling effects of individual particles
 in  trilinear gauge boson vertices in the SM 
and  its extensions, 
\item categorise all possible models of mixed anomaly cancellations
and  non-decoupling effects of very heavy fermions that are directly 
unreachable at the LHC,
\item search for  
phenomenological implications at  colliders.
\end{itemize}

General Lorentz-invariant expressions for three gauge boson
vertices  have been analysed 
in detail in \Refs{Hagiwara,Gaemers}. One-loop corrections in 
the SM for the $VWW, V=Z,\gamma$ using dimensional regularisation
were  considered in~\cite{Peskin:delayed} with special emphasis 
on  the non-decoupling effects due to large doublet mass splittings. 
The first  correct calculation for the $Z\gamma\gamma$ vertex was
performed in \Ref{Rosenberg} while for $ZZ\gamma$ in
\Ref{Barroso:1984re}. Phenomenological studies including expectations 
for those interactions at hadron and lepton colliders were studied in detail in
\Refs{Baur,Gounaris:1999kf,Gounaris:2000tb,Gounaris:2000dn}.
Finally, a complete 1PI vertex for three off-shell gauge bosons is a 
useful tool in analysing low energy inelastic scattering processes
with a photon in the final state. Dark matter scattering off atomic electrons and nuclei 
mediated by light gauge boson particles is one application
among many  (see \Refs{Dedes:2009bk,Kopp:2009et,Quevedo:2011}).

The outline of our article is as follows: in section~\ref{sec:anal},  we first present the 
1PI effective action for the triple gauge boson vertex
and then in section~\ref{sec:ndeffects} we discuss all possible and general
non-decoupling effects from heavy fermions. 
These two sections are supplemented by three Appendices~\ref{sec:toy}, \ref{sec:app} and 
\ref{app:cate},
which contain all relevant details of our calculation. 
The generality of 1PI vertex, $\Gamma^{\mu\nu\rho}$, presented
in section~\ref{sec:anal}, is to some extend a new result.
In addition, the discussion of anomaly driven
non-decoupling effects given in section~\ref{sec:ndeffects}, is also, to the best of
our knowledge, a new material.
Section~\ref{TGBV} contains 
applications of the general vertex in the SM,
in minimal $Z'$ models and their extensions with a fourth sequential generation.
Special care has been given to the synergy between the chiral anomaly
and the non-decoupling contributions in order to clarify relevant issues in the literature.
Appendices~\ref{app:W} and \ref{app:int} deal with the evaluation of charged external
gauge boson triple vertices and with analytical 
expressions of various integrals, respectively.
Section~\ref{conc} concludes with a brief discussion of our findings.

%

\section{The Trilinear Gauge Boson Vertex}
\label{sec:anal}

In this section we briefly present the main results for
the three gauge boson 1PI vertex, $\Gamma^{\mu\nu\rho}$. 
The details of this calculation are given 
in Appendices~\ref{sec:toy} and \ref{sec:app}. Furthermore,
the behaviour  of $\Gamma^{\mu\nu\rho}(s)$ at high energies $s$,  
and issues on 
gauge invariance and  Goldstone boson  equivalence theorem
are discussed in the subsequent subsections.
\subsection{The construction of  $\Gamma^{\mu\nu\rho}$}
%
\begin{figure*}[t]
   \centering
   \includegraphics[height=1.6in]{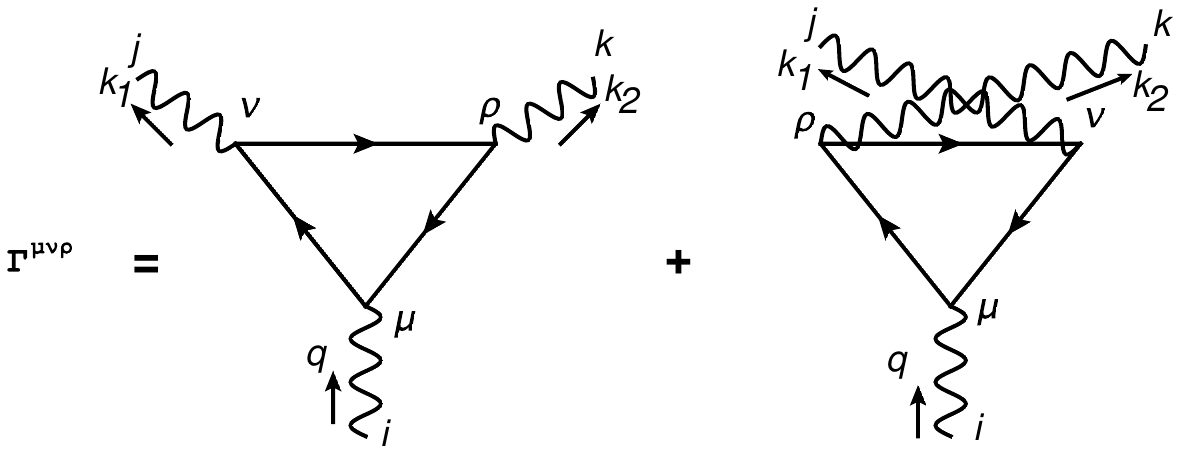} 
   \caption{The one-loop effective trilinear gauge boson vertex, $\Gamma^{\mu\nu\rho}$. The crossed diagram is obtained with the replacement
   $\{\nu,\rho \} \leftrightarrow \{\rho,\nu \}$ and $k_{1}\leftrightarrow k_{2}$. Indices $\{i,j,k\}$ denote
distinct external gauge bosons in general.}
   \label{fig:graph}
\end{figure*}
%

The relevant diagrams are depicted in Fig.~\ref{fig:graph}
and their evaluation is developed in Appendix~\ref{sec:app}.
What we basically need in order 
to calculate the diagrams in  Fig.~\ref{fig:graph}
is the interaction part of the Lagrangian 
\begin{equation}
\mathscr{L}_{int} \ \supset \ e\, \bar{\Psi}\: \gamma^{\mu}\,
 (\alpha + \beta \, \gamma_{5} ) \: \Psi \, A_{\mu} \;,
 \label{lint}
 \end{equation}
where $\Psi(x)$ is a 4-component 
spinor consisting of a pair of two Dirac fermions coupled chirally 
to a vector field $A_{\mu}(x)$.
Flavour or spinor indices  are silently  implied.
We shall assume a model interaction for \eq{lint} that
arises from a spontaneously broken Abelian gauge theory. A toy 
model as such is described in Appendix~\ref{sec:toy}. 
Then $\alpha$ and $\beta$ in \eq{lint} are real numbers (in units of $e$)
 related
to linear combinations of hypercharges [see for instance \eq{ab}].

The integral representation for this diagram is given in \eq{gmnr}.
By naive power counting this integral is linearly divergent. This means
that when we make a shift of integration variable, e.g.,  $p\to p+ a$, the 
result depends upon the choice of the arbitrary
 vector $a^{\mu}$. This change is only
reflected in the form factors proportional to $k_{1}$ and $k_{2}$ in Lorentz
invariant expansion of $\Gamma^{\mu\nu\rho}$ [see \eq{rose2} below]. 
As a result, the naive Ward Identities (WIs), \eqss{eq:A15}{WI2}{WI3} 
are violated by terms  that contain the arbitrary four vector $a^{\mu}$.
It is useful to write this four vector as a linear combination of the
two independent external momenta : 
$a^{\mu} =  z \, k_{1}^{\mu} + w \, k_{2}^{\mu}$, 
with $z,w$  arbitrary real parameters.  

In order to write out an explicit form for the trilinear gauge boson
vertex, say for three identical massive gauge bosons, we make
use of  an explicit expression for the triangle graphs first calculated 
by Rosenberg~\cite{Rosenberg}. The most general form of the 
axial tensor  $\Gamma^{\mu\nu\rho}$, 
consistent with Lorentz and parity symmetry, is, 
\begin{align}
&\Gamma^{\mu\nu\rho}(k_{1},k_{2};w,z) =  \biggl
[A_{1}(k_{1},k_{2};w)\, \varepsilon^{\mu\nu\rho\sigma}\, k_{2\sigma} 
\nonumber \\[3mm]  &+ 
A_{2}(k_{1},k_{2};z)\,\varepsilon^{\mu\nu\rho\sigma}\, k_{1\sigma}
 +
A_{3}(k_{1},k_{2})\,\varepsilon^{\mu\rho\beta\delta}\, k_{2}^{\nu}\, k_{1\beta}\, k_{2\delta}
\nonumber \\[3mm] &+
 A_{4}(k_{1},k_{2})\,
\varepsilon^{\mu\rho\beta\delta}\, k_{1}^{\nu}\, k_{1\beta}\, k_{2\delta}
 + 
A_{5}(k_{1},k_{2})\, \varepsilon^{\mu\nu\beta\delta}\, k_{2}^{\rho}\, k_{1\beta}\, k_{2\delta} \nonumber \\[3mm]  &+  
A_{6}(k_{1},k_{2})\, \varepsilon^{\mu\nu\beta\delta}\, k_{1}^{\rho}\,k_{1\beta}\, k_{2\delta}\biggr ] \;.
\label{rose2}
\end{align}
By naive power counting the dimensionless form factors $A_{1,2}$ are infinite. They can be
rendered finite by forcing them to obey the relevant, albeit anomalous, Ward Identities. 
However,  $A_{1,2}$ 
are in general \emph{undetermined}
 since they depend on arbitrary parameters $w$ and $z$.
This arbitrariness can be fixed by physical requirements 
like for example conservation of charge.
On the other hand, the form factors (or integrals)  $A_{3,..6}$ 
are finite having dimension of inverse mass square. 
The latter can be found independently 
by direct diagrammatic methods. The whole procedure is 
described in detail in Appendix~\ref{sec:app}.

Therefore, non-decoupling effects should originate solely from the 
 $A_{1}$ and $A_{2}$ parts of $\Gamma^{\mu\nu\rho}$ but without any further 
 physical input they are undetermined. A direct calculation of $A_{1,2}$ 
 with dimensional regularisation~\cite{'tHooft:1972fi} or 
 with Pauli-Villars regularisation is not a good choice
 when shifting integration variables within linearly (and above) 
 divergent Feynman integrals in four dimensions~\cite{Rohrlich,Pugh:1969kn,Elias:1982sq}.
 The outcome for a single 
external gauge boson ($i=j=k$ in Fig.~\ref{fig:graph}) triangle graph
is appended in \eqss{FWI1}{FWI2}{FWI3}. From these expressions and from 
\eq{rose2} we obtain $A_{1}(k_{1},k_{2};w)$ and 
$A_{2}(k_{1},k_{2};z)$ in terms of the finite integrals $A_{3..6}$.
The corresponding results,
in the case of three external identical gauge bosons,
are given by \eqs{A1}{eq:A33}  
while the finite integrals $A_{3..6}$ by \eqss{eq:A28}{A4}{A5}.

Furthermore, although Bose symmetry could constrain the arbitrary 
numbers $w$ and $z$, it is not enough to eliminate them
altogether: a physical condition is needed, e.g.,  conservation
of electric charge for fermions coupled to external  photons 
or vanishing triangle graph for
on-shell momenta of massive gauge bosons or, even, a pure theoretical
reason, like the  decoupling property.  

It is straightforward, albeit tedious, to generalize $\Gamma^{\mu\nu\rho}$
in \eq{rose2}  to the case of three distinct
external, massive or massless,
 gauge bosons ($i\ne j \ne k$ in Fig.~\ref{fig:graph}). 
\begin{widetext}
%
With the assignments depicted in Fig.\ref{fig:graph}, the generalised
Ward Identities 
for vertices $\mu, \nu, \rho$ are written respectively as\footnote{In order not to clutter the
notation we suppress indices $i,j,k$ in the following expressions for 
$\Gamma$'s.}
\begin{subequations}
\label{WIs}
\begin{align}
 q_{\mu}\: \Gamma^{\mu\nu\rho}(k_{1},k_{2},w,z) &=  i \, m_{A_{i}}\: \Gamma^{\nu\rho}(k_{1},k_{2})  +
\frac{e^{3}[(\alpha_{i}\alpha_{j}+\beta_{i}\beta_{j})\beta_{k}+(\alpha_{i}\beta_{j}+\alpha_{j}\beta_{i})\alpha_{k}]}{4\pi^{2}}\varepsilon^{\lambda\nu\rho\sigma}k_{1\lambda} k_{2\sigma} (w-z),  \label{GWI1} \\[3mm]
-k_{1\nu}\: \tilde{\Gamma}^{\nu\rho\mu}(k_{1},k_{2},w,z) &= i\, m_{A_{j}}\: \tilde{\Gamma}^{\rho\mu}(k_{1},k_{2})+\frac{e^{3}[(\alpha_{j}\alpha_{k}+\beta_{j}\beta_{k})\beta_{i}+(\alpha_{j}\beta_{k}+\alpha_{k}\beta_{j})\alpha_{i}]}{4\pi^{2}}
\varepsilon^{\lambda\mu\rho\sigma}
k_{1\lambda} k_{2\sigma} (w-1), \label{GWI2} \\[3mm]
 -k_{2\rho}\: \hat{\Gamma}^{\rho\mu\nu}(k_{1},k_{2},w,z) &= i \, 
 m_{A_{k}}\: \hat{\Gamma}^{\mu\nu}(k_{1},k_{2})
+\frac{e^{3}[(\alpha_{k}\alpha_{i}+\beta_{k}\beta_{i})\beta_{j}
+(\alpha_{k}\beta_{i} +\alpha_{i}\beta_{k})\alpha_{j}]}{4\pi^{2}}
\varepsilon^{\lambda\mu\nu\sigma}k_{1\lambda} k_{2\sigma} (z+1),
\label{GWI3}
\end{align}
\end{subequations}
where the corresponding $\Gamma, \tilde{\Gamma}$, and 
$\hat{\Gamma}$ are appended in \eqs{GG3}{GG2}. 
It is remarkable here to note the $i$'th gauge boson 
mass, $m_{A_{i}}=-2\beta_{i} e v$, in front of the
pseudoscalar 1PI function $\Gamma^{\nu\rho}$.
This term and  
the analogous in \eqs{GWI2}{GWI3} are the 
source of heavy fermion mass non-decoupling effects
since in the formal limit of $m\to \infty$ there is a remaining piece of order
$e^{3} \varepsilon^{\lambda\nu\rho\sigma}k_{1\lambda} k_{2\sigma}/4\pi^{2}$
in $\Gamma^{\mu\nu\rho}$ for example. 
On the other hand, it shows that currents which are associated to 
unbroken symmetry generators \ie to massless gauge bosons,
do not provide any non-decoupling effect in $\Gamma^{\mu\nu\rho}$.
Moreover, $\Gamma^{\nu\rho}, \tilde{\Gamma}^{\rho\mu}, \hat{\Gamma}^{\mu\nu}$ 
depend linearly upon the Yukawa coupling $\lambda$,
that is responsible for the fermion mass through the Higgs mechanism
and  vanishes in the limit of $\lambda\to 0$\footnote{Throughout, we  assume chiral 
fermions that receive mass through Yukawa interactions with the Higgs field.}.

Using the WI's for the vertices ${\nu}$ and ${\rho}$, \ie
\eqs{GWI2}{GWI3} as well as \eq{rose2},
 we obtain the following expressions for the
integrals $A_{1}$ and $A_{2}$:
\begin{subequations}
\label{A12}
\begin{align}
A_{1}(k_{1},k_{2};w) &= 
(k_{1}\cdot k_{2}) A_{3}+k_{1}^{2} A_{4} -\frac{e^3
m^{2}\beta_{j}}{\pi^{2}}I_{1}(k_{1},k_{2},m)
+\frac{e^3[(\alpha_{j}\alpha_{k}+\beta_{j}\beta_{k})\beta_{i}+(\alpha_{j}\beta_{k}+\alpha_{k}\beta_{j})\alpha_{i}]}{4\pi^{2}}(w-1) \;, 
\label{GENA1} \\[3mm]
A_{2}(k_{1},k_{2};z) &= (k_{1} \cdot k_{2}) A_{6}+ k_{2}^{2}
A_{5}-\frac{e^3
m^{2}\beta_{k}}{\pi^{2}}I_{2}(k_{1},k_{2},m)+\frac{e^3[(\alpha_{i}\alpha_{k}+\beta_{i}\beta_{k})\beta_{j}+(\alpha_{i}\beta_{k}+\alpha_{k}\beta_{i})\alpha_{j}]}{4\pi^{2}}(z+1)
\;,
\label{GENA2}
\end{align}
\end{subequations}
 where the ``non-decoupled'' integrals are given by
\begin{subequations}
\label{I12}
\begin{align}
I_{1}(k_{1},k_{2},m) &=\int_{0}^{1}dx\int_{0}^{1-x}dy\frac{-(\alpha_{i}\alpha_{k}+\beta_{k}\beta_{i})+2
x \beta_{i}\beta_{k}}{\Delta}\;, \label{GI1} \\[3mm]
I_{2}(k_{1},k_{2},m) &=\int_{0}^{1}dx\int_{0}^{1-x}dy\frac{(\alpha_{i}\alpha_{j}+\beta_{i}\beta_{j})-2
y \beta_{i}\beta_{j}}{\Delta}\;, \label{GI2}
\end{align}
\end{subequations}
with
\begin{eqnarray}
\Delta \equiv \Delta(k_{1},k_{2})= x(x-1)k_{2}^2+y(y-1)k_{1}^2-2xyk_{1}\cdot k_{2}+m^{2}.
\end{eqnarray}
The following limits,
\begin{subequations}
\label{API12}
\begin{align}
\lim_{m\rightarrow \infty} m^{2} I_{1}(k_{1},k_{2},m) &= -\frac{1}{6}\: (3 \alpha_{i}
\alpha_{k} + \beta_{i}\beta_{k}) \;, \\[3mm]
\lim_{m\rightarrow \infty} m^{2} I_{2}(k_{1},k_{2},m) &= \frac{1}{6}\:(3 \alpha_{i}
\alpha_{j} + \beta_{i}\beta_{j}) \;,
\end{align}
\end{subequations}
are also useful in simplifying formulae when discussing 
synergies of anomalous  and non-decoupling terms.

We are now ready to complete $\Gamma^{\mu\nu\rho}$ in \eq{rose2} 
by reading directly from \eq{GG3} the finite (in four dimensions) 
terms $A_{3..6}$. We find:
\begin{subequations}
\label{A3456}
\begin{align}
A_{3}(k_{1},k_{2}) &=
-\frac{e^3[(\alpha_{i}\alpha_{j}+\beta_{i}\beta_{j})\beta_{k}
+(\alpha_{i}\beta_{j} +\beta_{i}\alpha_{j})
\alpha_{k}]}{\pi^{2}}\int_{0}^{1}dx\int_{0}^{1-x}dy\, \frac{xy}{\Delta}
\;, \label{GENA3}
\\[3mm]
A_{4}(k_{1},k_{2}) &= \frac{e^3[(\alpha_{i}\alpha_{j}+\beta_{i}\beta_{j})\beta_{k}+(\alpha_{i}\beta_{j}+\beta_{i}\alpha_{j})\alpha_{k}]}{\pi^{2}}\int_{0}^{1}dx\int_{0}^{1-x}dy\, \frac{y(y-1)}{\Delta}\;, \label{GENA4}\\[3mm]
A_{5}(k_{1},k_{2})&= -\frac{e^3[(\alpha_{i}\alpha_{j}+\beta_{i}\beta_{j})\beta_{k}+(\alpha_{i}\beta_{j}+\beta_{i}\alpha_{j})\alpha_{k}]}{\pi^{2}}\int_{0}^{1}dx\int_{0}^{1-x}dy\, \frac{x
(x-1)}{\Delta} \;, \label{GENA5}
\\[3mm]
A_{6}(k_{1},k_{2}) &= -A_{3}(k_{1},k_{2}) \label{GENA6} \;.
\end{align}
\end{subequations}
One could guess the expressions above with $i\neq j \neq k$ 
 from the ones with a single 
identical gauge boson $i=j=k$
 by exploiting simple combinatoric 
algebra 
in \eqss{eq:A28}{A4}{A5} and \eqs{A1}{eq:A33}. One can check that
all the above form factors obey the Bose symmetry specified in 
 \eqss{Bose1}{Bose2}{Bose3}.

In summary, our main result is the trilinear gauge boson 
vertex $\Gamma^{\mu\nu\rho}$ of  \eq{rose2}, supplemented by
form factor components $A_{i=1..6}$ 
read from  \eqs{A12}{A3456}. \Eq{rose2}
satisfies the relevant  Ward Identities stated in \eq{WIs}
which originate from the partial conservation of vector and axial vector 
symmetries in (\ref{syms}).

\subsection{Unitarity}
\label{app:unit}

We can make full use of the effective vertex $\Gamma^{\mu\nu\rho}$ in order 
to calculate, as an example,  the matrix element for the process 
$ZZ\longrightarrow ZZ $ with an intermediate massive vector boson $Z'$. 
We perform the calculation in the 
center of mass frame with the following kinematics:
\begin{align*}
 p_{1} &=(E,0,0,p)\;,  \quad p_{2}=(E,0,0,-p) \;, \quad 
 k_{1} =(E,p \sin\theta,0,p \cos\theta)\;, \quad
 k_{2}=(E,-p\sin\theta,0,-p\cos\theta)\;, 
 \\
\varepsilon_{}(p_{1})&=\frac{1}{m_{Z}}(p,0,0,E)\;, \quad
\varepsilon_{}(p_{2})=\frac{1}{m_{Z}}(p,0,0,-E)\;, \quad \\
\varepsilon_{}(k_{1})&=\frac{1}{m_{Z}}(p,E\sin\theta,0,E\cos\theta)\;,
\quad
\varepsilon_{}(k_{2})=\frac{1}{m_{Z}}(p,-E\sin\theta,0,-E\cos\theta)\;,
\end{align*}
where
$p_{1}$ and $p_{2}$ are the four-momenta of 
incoming particles, $k_{1}$ and $k_{2}$ the four-momenta of outgoing particles,
$\varepsilon(p_{1})$, $\varepsilon(p_{2})$, 
$\varepsilon(k_{1})$, $\varepsilon(k_{2})$ are the polarisation vectors 
of the incoming and outgoing particles respectively 
and $\theta$ is the scattering angle of the outgoing $Z$ boson in the 
center of mass frame.
Non-zero contributions arise only from  $t$ and $u$-channels
since the s-channel amplitude vanishes in this frame.
Working in the unitary gauge, 
we find a contribution to $ZZ\to ZZ$ due to loop-induced
$\Gamma^{\mu\nu\rho}_{Z'ZZ}$ of \eq{rose2} as,
\begin{eqnarray}
 \mathscr{M}\ = \ \mathscr{M}_{t}+ \mathscr{M}_{u} \ &=& \
 \bigg(\frac{E^{2} \sin^{2}\theta}{t-m_{Z^{'}}^{2}} \bigg) 
 \bigg[(A_{1}-A_{2})  + 
p^{2}\, (1-\cos\theta) \,(A_{3}-A_{6})\bigg]^{2} 
\nonumber \\[3mm]
&+&
\bigg(\frac{E^{2} \sin^{2}\theta}{u-m_{Z^{'}}^{2}} \bigg) 
 \bigg[(A_{1}-A_{2})  + 
p^{2}\, (1+\cos\theta)\, (A_{3}-A_{6})\bigg]^{2}  \;, \label{Matuni}
\end{eqnarray}
where $t=(p_{1}-k_{1})^{2}=-2 p^{2}(1-\cos\theta)$ 
and $u=(k_{1}-p_{2})^{2}=-2 p^{2}(1+\cos\theta)$.
The factors $A_{1}$ and $A_{2}$ in \eq{A12} 
are dimensionless and, in the limit of
$E^{2}\to \infty$ vary at worse as 
constants while from \eq{A3456} we have
 $A_{3}=-A_{6}$ which asymptotically    
goes like  $E^{-2}$. 
Therefore at high energies $E^{2}\to \infty$, 
terms inside the square brackets in \eq{Matuni} behave like
constants and so the amplitude does at high energies. This means
that unitarity is satisfied as is of course expected for a renormalised theory.
It is worthwhile noting that in
the limit  $E^{2}\to \infty$ we obtain $(A_{1}-A_{2}) \propto c (w-z)$, 
where $c$ is the anomaly pre-factor  present in the second term in the r.h.s
of \eq{GWI1}. 
There is still however a finite and non-vanishing 
constant contribution from the $A_{3,6}$ form factors
in \eq{Matuni}  which for every particle contribution reads,
\begin{equation}
\lim_{E^{2}\to \infty} \mathscr{M} =  -\left ( \frac{c}{4\pi^{2}} \right )^{2} 
\sin^{2}\theta \left [ 1 + 2 (w-z) + \frac{(w-z)^{2}}{2\sin^{2}\theta} \right ]\;. \label{eq:asym}
\end{equation}
We observe that the unknown parameters $w$ and $z$ still remain
in the amplitude.
Only  the relation $w=z$ removes them from the asymptotic limit.
We shall come back at this point when discussing the $Z'^{*}ZZ$-vertex in 
section~\ref{sec:zprime}.

\subsection{Goldstone boson Equivalence Theorem 
and $R_{\xi}$ - independence}
\label{GBET}

There are literally $N$-ways to derive the Ward Identities of \eq{WIs}.
 A classical method is to  demand invariance of the path integral
 under the combined local vector and axial-vector 
 gauge transformations (\ref{syms}). We can then represent  these 
 WI's diagrammatically to prove the Goldstone Boson equivalence 
 theorem~\cite{Cornwall,Vayonakis,Lee}.  
 This is most clearly explained 
 in Lorentz gauge ($\xi=0$) where the gauge fixing term (\ref{gaf})
 does not involve the Goldstone boson field $\varphi$.
  \begin{figure*}[htbp]
     \centering
     \includegraphics{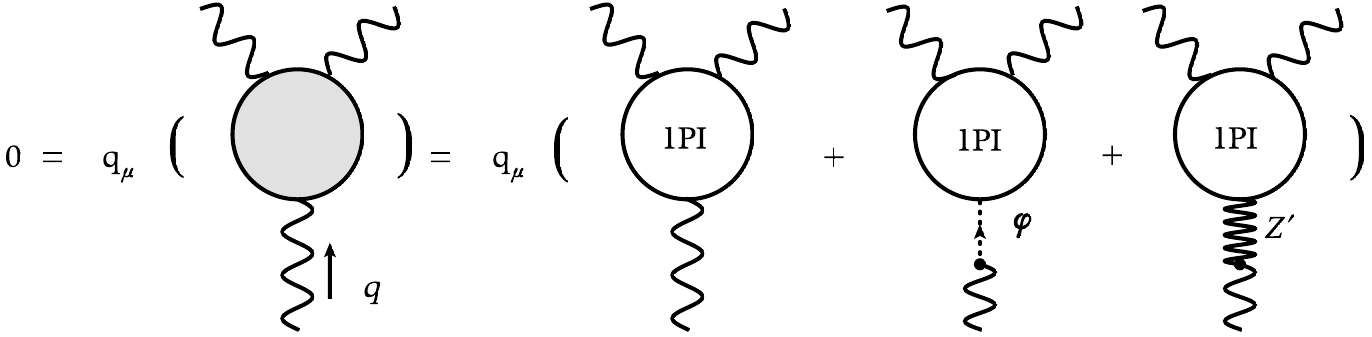} 
     \caption{Graphical representation of the WI in \eq{GWI1}.}
     \label{fig2}
  \end{figure*}
Then conservation of the gauge current implies that $q^{\mu}$ can be
contracted directly with $\Gamma^{\mu\nu\rho}$  and also 
with the derivatively coupled Goldstone boson to $\Gamma^{\nu\rho}$.
In principle there is a third contribution from possible mixings with
other gauge bosons, say $Z'$,  that couple to the same fermions in the vertex.
This last mixing must necessarily be proportional to 
$(g_{\mu\lambda}-q_{\mu}q_{\lambda}/q^{2})$ 
and when contracted with $q^{\mu}$, vanishes. 
Therefore, by using rules from the toy model in Appendix~\ref{sec:toy} 
it is straightforward to see that we recover the classical  WI
(\ref{GWI1}), without the anomalous term. 
While a possible 
gauge boson mixing contributes to $\Gamma^{\mu\nu\rho}$, 
it does not contribute to WIs  in (\ref{WIs}).
At very high energy, the longitudinal polarization vector 
is $\varepsilon_{{L\mu}}(q) \simeq q_{\mu}/m_{A}$, where $m_{A}$ is 
the gauge boson mass. In other words for an anomaly-free model, \eq{GWI1}
or the sum of the diagrams in Fig.~\ref{fig2}, can be written as,
\begin{equation}
\epsilon_{L\mu}(q)\: \Gamma^{\mu\nu\rho} \ = \ i\, \Gamma^{\nu\rho}\;.
\label{eq:GBET}
\end{equation}
This equation tells us that at the high energy limit, the physical amplitude
with the gauge boson in vertex $\mu$ is replaced by the vertex with a 
 Goldstone boson
that `has been eaten'. However, as is evident from \eq{GWI1}, 
the relation (\ref{eq:GBET}) is broken by possible gauge anomalies.
This is another reason why the latter 
should be absent.

One can easily check by studying
for example the fermion-antifermion 
annihilation process to two gauge bosons with the toy 
model of Appendix~\ref{sec:toy}, that \Eq{eq:GBET} is 
the required condition for the amplitude to be 
gauge $\xi$-independent. 
Again the anomalous term \emph{must} be absent.

\end{widetext}

\section{Non-Decoupling Effects}
\label{sec:ndeffects}

Heavy fermion non-decoupling effects can be cast in two classes : 
\begin{itemize}
\item[A)] effects that arise from a large mass splitting between
particles within an anomaly-free multiplet. 
\item[B)] {\it anomaly driven} effects that originate from decoupling a whole
anomaly-free multiplet.  
\end{itemize}
In case (A), formal decoupling 
of the heavy particle that participates in the anomaly cancellation mechanism  
will leave at low energies an effective Lagrangian $\Delta \Gamma^{\mu\nu\rho}$
that accounts for the anomaly cancellation missing 
piece~\cite{D'Hoker:SM,D'Hoker:GEN,HHH1}. 
In case (B)  the Higgs coupling to fermions will be much larger
than the gauge coupling with the latter being approximately zero when the 
fermion mass is going to  infinity~\cite{Antoniadis:2009,Dudas:2009}.

\vspace{3mm}

\subsection{Non-Decoupling due to large mass splitting}
We are going to focus first on the simplest case with three external identical 
gauge bosons. This means we set $i=j=k$ in the
Ward Identities of \eq{WIs} or else we look directly at expressions, 
 (\ref{FWI1}) - (\ref{FWI3}).
In order to carry out a systematic study of non-decoupling effects
and their interplay with chiral anomalies it is essential to keep track of 
the anomalous terms
that depend on the arbitrary parameters $w$ and $z$.
By exploiting Bose symmetries for on-shell external gauge bosons,
and specifically, (\ref{Bose}) among legs $j$ and $k$ we find $w=-z$, while
with (\ref{Bosee}) among legs $i$ and $j$ we find (after some tedious algebra)
$2 w -z -1=0$. The solution of this system, 
  \begin{equation}
  w \ = \ -z \ = \ \frac{1}{3} \;,
  \label{concon}
  \end{equation}
finally fixes the arbitrary parameters $w$ and $z$.
Our observation is that these fixed values for the arbitrary parameters 
correspond to the case of a particle decoupling from the effective
action, \ie
\begin{eqnarray} 
\lim_{m\to \infty} \Gamma^{\mu\nu\rho}(k_{1},k_{2};w,z) = 0 \ 
\Rightarrow w = -z =  \frac{1}{3} \;.
\label{cond1}
\end{eqnarray}
We elaborate this point in what follows.
The WIs now take the form: 
\begin{widetext}
  \begin{subequations}
  \label{DWI}
  \begin{eqnarray}
q_{\mu}\Gamma^{\mu\nu\rho}(k_{1},k_{2}; w=1/3)  &=&
-\frac{e^{3} \beta m^{2}}{\pi^{2}}\:
\varepsilon^{\lambda\nu\rho\sigma}\, k_{1\lambda}\, k_{2\sigma}
\, I_{0}(k_{1},k_{2};m)
+ \frac{e^{3}(\beta^{3}+3\alpha^{2}\beta)}{6
\pi^{2}}\: \varepsilon^{\lambda\nu\rho\sigma}\:
\: k_{1\lambda}\,k_{2\sigma}\;.\label{DWI1}
 \\[3mm]
-k_{1\nu}\widetilde{\Gamma}^{\nu\rho\mu}(k_{1},k_{2};w=1/3)
 &=& - \frac{ e^{3} \beta m^{2}}{\pi^{2}}\,
 \varepsilon^{\lambda\mu\rho\sigma}\, k_{1\lambda}\, k_{2\sigma}\,
 I_{1}(k_{1},k_{2};m)
-\frac{e^{3}(\beta^{3}+3\alpha^{2}\beta)}{6\pi^{2}}\varepsilon
^{\lambda\mu\rho\sigma}\:   \, k_{1\lambda} k_{2\sigma}\;,
\label{DWI2}
 \\[3mm]
-k_{2\rho}\widehat{\Gamma}^{\rho\mu\nu}(k_{1},k_{2};w=1/3) &=&
-\frac{e^{3} \beta m^{2}}{\pi^{2}}\,
\varepsilon^{\lambda\mu\nu\sigma} \, k_{1\lambda}\, k_{2\sigma}\, I_{2}(k_{1},k_{2};m)
+\frac{e^{3}(\beta^{3}+3
\alpha^{2}\beta)}{6
\pi^{2}}\: \varepsilon^{\lambda\mu\nu\sigma}\: k_{1\lambda} k_{2\sigma}\;,\label{DWI3}
\end{eqnarray}
\end{subequations}
\end{widetext}
where the integrals $I_{0,1,2}$ are defined in \eqss{I0}{I1}{I2}
respectively.
The anomalous  terms in (\ref{DWI}) are then allocated ``democratically'' in the three
legs of $\Gamma^{\mu\nu\rho}$ as one would have naively expected.
 Note also that since 
$\lim_{m\to \infty}m^{2} I_{0} = - \lim_{m\to \infty}m^{2} I_{1} =
\lim_{m\to \infty} m^{2 }I_{2} = \frac{1}{6}(\beta^{2}+3 \alpha^{2})$ the
r.h.s of \eqss{DWI1}{DWI2}{DWI3} cancels identically, verifying our statement in 
\eq{cond1}. Therefore,
for a Dirac  fermion pair 
circulating the loop as shown in Fig.~\ref{fig:graph} and for three identical
external gauge bosons, 
\emph{at the formal decoupling limit, the finite contributions 
are equal and opposite
to the anomaly contributions in the vertex}.
 In a Lorentz gauge, terms in $\Gamma^{\mu\nu\rho}$
  proportional to $I_{0,1,2}$ arise from the mixing between 
  the Goldstone boson $\varphi$ and the gauge boson
 as it is shown in Fig.~\ref{fig2}. We should notice however, that 
 our calculation of WIs in (\ref{DWI}) given in Appendix~\ref{sec:app}
  contains no reference to 
 a particular gauge choice.

For a Lorentz-invariant and renormalizable 
chiral gauge theory 
the anomalous terms {\it i.e.,} the last terms on the 
r.h.s of eqs.~(\ref{DWI}), have to be absent. The only 
way\footnote{Of course there is the trivial case of  vector
multiplets \ie $\beta=0$.}, 
consistent with renormalizability\footnote{We are not going to consider 
here the situation~\cite{Preskill}
of incorporating non-renormalizable counterterms to cancel 
the anomalies at the expense of introducing a cut-off scale $\Lambda \sim
4\pi v$.}~\cite{Gross,Bouchiat}, to remove the anomaly terms,
is to add a new Dirac fermion pair with 
opposite $\beta$ \ie opposite hypercharges $Y_{L}$ and $Y_{R}$. 
A consistent way to  describe heavy fermion decoupling
effects is to perform the calculation directly in the broken
phase of the theory where physical masses appear explicitly.
Assuming that the mass of the second (heavy) pair and the 
energy, $s=(k_{1}+k_{2})^{2}$,  is much bigger
than the first (light) fermion pair, say, $m^{2}_{2}\gg s \gg m^{2}_{1}\approx 0$, 
there is a non-decoupled term  in the 1PI effective
action which can be read off from
 \eqss{rose}{A1}{eq:A33} [or \eqs{rose2}{A12} for $i=j=k$] to be,  
\begin{eqnarray}
\Delta \Gamma^{\mu\nu\rho}(k_{1},k_{2}) \approx \frac{e^{3} \: (\beta^{3} + 3 \alpha^{2}\beta)}{6\pi^{2}} \: \varepsilon^{\mu\nu\rho\sigma}\, (k_{1}-k_{2})_{\sigma} \;.\nonumber \\
\label{diff}
\end{eqnarray}
This term remains in the 1PI effective 
function for the light particle. In the  heavy mass limit ($m_{2}\to \infty$), the
form factors $A_{i=3,...6}(k_{1},k_{2})$ vanish as $1/m^{2}$ leaving only
the term (\ref{diff}) in the low energy effective action which
 has no `memory'  anymore from the heavy mass $m_{2}$. 
 Although, the exact non-kinematic 
 prefactor in \eq{diff}, depends upon  model 
 details, its magnitude (in $e$-units) is approximately, 
$\alpha/\pi$ and could be observable.
 Furthermore, the non-decoupling term (\ref{diff})
does not depend on the regularization scheme, \ie on the parameters
$w$ and $z$ in \eqs{A1}{eq:A33}, 
since the model is by construction  anomaly-free. 

\subsection{Anomaly Driven non-decoupling effects}
\label{sub-ano}

This is a category of possible non-decoupling effects 
for  models possessing an anomaly-free 
cluster of heavy  particles just above those  known 
from the Standard Model.
We systematically then check anomaly cancellation conditions
in Ward Identities (\ref{WIs}) by demanding
the pre-factors of $I_{1,2}$ integrals in \eqs{GENA1}{GENA2}  to be
non-zero. We are seeking for minimal models with 
 up-to three different gauge bosons and
up to the least $n$-Dirac fermions.

\begin{table*}[t]
\begin{tabular}{|c|c|c|c|}
\hline
\multicolumn{1}{|c|}{} 
& $\psi_{1}$ & $\psi_{2}$ & $\psi_{3}$ 
\\ \hline
$U(1)_{X}$ & $\alpha=e,\beta=-e $ & $\alpha=e,\beta=e $ & $\alpha=0,\beta=0 $ 
\\ \hline
$U(1)_{Y}$ & $\alpha=-e,\beta=-e $ & $\alpha=0,\beta=0 $ & $\alpha=e,\beta=e $ 
\\ \hline
\end{tabular}
\caption{Charges of an anomaly-free model with non-decoupling remnants 
in three gauge boson vertices $XXY$ and $YYX$.
}
\label{tab:2}
\end{table*}

A model that contains one gauge boson $X$, 
with V-A couplings as in \eq{lint},
coupled to only one fermion is 
impossible to exist because it  is anomalous (except the trivial 
case of a vector-like particle where $\beta=0$).
Adding  an extra fermion with the same mass 
 but with opposite axial-vector coupling ($\beta$) renders the model anomaly-free.
 Such a simple particle content does not lead to non-decoupling
 effects because all these effects are
proportional to an odd power  of the axial-vector coupling ($\sim\beta^{2k+1}$) 
and therefore the sum over the two 
fermions vanishes. Similar situation arises when
more fermions are circulating in the loop.

More interesting is the case where
one has 
two, distinct, external gauge bosons, X and Y,  either massive or massless. 
The cancelation of trilinear anomalies requires 
the existence of at least two fermions 
with opposite axial-vector couplings but again
it is impossible to satisfy instantaneously the 
mixed anomaly and non-decoupling conditions [see below].
We first  obtain the  general  
conditions for an anomaly-free model with two gauge 
bosons $X$ and $Y$.  In notation of \eq{lint} these conditions read,
\begin{subequations}
\label{aXY}
\begin{eqnarray}
 \sum_{i=1}^{n}(\beta_{X}^{3}+3 \alpha_{X}^{2}\beta_{X})_{i} &=& 0\;, 
 \label{an1}\\
 \sum_{i=1}^{n}(\beta_{Y}^{3}+3 \alpha_{Y}^{2}\beta_{Y})_{i}&=&0\;, 
 \label{an2} \\
 \sum_{i=1}^{n}(\beta_{X}^{2}\beta_{Y}+2 \alpha_{X}\alpha_{Y}\beta_{X}+\alpha_{X}^{2}\beta_{Y})_{i} &=& 0\;,  \label{an3} \\
\sum_{i=1}^{n}(\beta_{Y}^{2}\beta_{X}+2 \alpha_{X}\alpha_{Y}\beta_{Y}+\alpha_{Y}^{2}\beta_{X})_{i}&=&0\;,  \label{an4}
\end{eqnarray}
\end{subequations}
where $n$ is the total number of fermions. 
Starting from trilinear anomalies (\ref{an1}) or (\ref{an2}) we see that 
the case $n=1$
requires only vectorial couplings, $\beta_{X} =\beta_{Y}=0$. 
Therefore for $n=1$ there is no non-trivial solution. For $n=2$
the non-zero couplings must satisfy the following conditions: 
\begin{align}
\beta_{X2} = -\beta_{X1}, &\qquad \alpha_{X2} = \pm \alpha_{X1} \nonumber \\
\beta_{Y2} = -\beta_{Y1}, &\qquad \alpha_{Y2} = \pm \alpha_{Y1}\;.
\label{e15}
\end{align}
Turning to mixed anomalies (\ref{an3}) and (\ref{an4}),
it is amusing first to note that they are satisfied even with one internal
fermion ($n=1$),   iff 
\begin{align}
\beta_{X} =   \alpha_{X}\;, &\qquad \beta_{Y} = - \alpha_{Y}\;, 
\label{e16}\\
&\mathrm{or} \nonumber \\
\beta_{X} = - \alpha_{X}\;, &\qquad \beta_{Y} =  \alpha_{Y} \;.
\label{e17}
\end{align}

Non-decoupling conditions are derived by the requirement that
the pre-factors of  $I_{1}$ and $I_{2}$ integrals  in 
\eqs{GENA1}{GENA2} 
are non-zero.
Hence, in the limit of
$k_{1}^{2},k_{2}^{2}\simeq {s}\ll m^{2}$ at 
least one of the following algebraic  expressions, 
\begin{align}
 \sum_{i=1}^{n}(\beta_{X}^{2}\beta_{Y}+3 \alpha_{X}\alpha_{Y}\beta_{X})_{i} \;, &\quad 
  \sum_{i=1}^{n}(\beta_{X}^{2}\beta_{Y}+3 \alpha_{X}^{2}\beta_{Y})_{i} \;, 
  \nonumber \\
   \sum_{i=1}^{n}(\beta_{Y}^{2}\beta_{X}+3 \alpha_{X}\alpha_{Y}\beta_{Y})_{i} \;, &\quad 
  \sum_{i=1}^{n}(\beta_{Y}^{2}\beta_{X}+3 \alpha_{Y}^{2}\beta_{X})_{i} \;,
  \nonumber \\ \label{ndXY}
  \end{align}
must be non-vanishing.
For $n=1$ the choice (\ref{e16}) [or (\eq{e17})] which eliminates
the mixed anomalies sets also eqs. (\ref{ndXY}) 
to a non-zero value. However,
to cancel the $XXX$ and $YYY$ anomalies one needs  at least $n=2$
fermions to satisfy the  conditions (\ref{e15}). These set
the non-decoupling expressions (\ref{ndXY}) back to zero.
The first non-trivial solution of the system \eqs{aXY}{ndXY} arises
with three pairs of chiral Dirac fermions ($n=3$) with an example of 
quantum numbers given in Table~\ref{tab:2}. Here, we use 
(\ref{e16}) and (\ref{e17}) to cancel mixed anomalies for $\psi_{1}$.
The other two particles $\psi_{2}$ and $\psi_{3}$ are singlets
under $U(1)_{Y}$ and $U(1)_{X}$, respectively.  
Plug these into   \eqss{rose}{A12}{API12}, 
we obtain the non-vanishing operators 
at the decoupling limit: 
\begin{subequations}
\begin{align}
\Gamma^{\mu\nu\rho}_{XYY} &= \Gamma^{\mu\nu\rho}_{YXX} = 
\frac{e^{3}}{3\pi^{2}} 
\varepsilon^{\mu\nu\rho\sigma} \: (k_{2}-k_{1})_{\sigma} \;, \\
\Gamma^{\mu\nu\rho}_{XYX} &= \Gamma^{\mu\nu\rho}_{YXY} = -  \frac{e^{3}}{3\pi^{2}} 
\varepsilon^{\mu\nu\rho\sigma} \: ( 2 k_{2} + k_{1})_{\sigma} \;, \\
\Gamma^{\mu\nu\rho}_{XXX} &= \Gamma^{\mu\nu\rho}_{YYY} =0 \;.
\end{align}
\end{subequations}

Next is a model example with $n=4$ Dirac fermions
charged under the product of gauge groups $U(1)_{X}\times U(1)_{Y}$.
This toy model has been examined in~\Ref{Antoniadis:2009}. 
Charge assignments are given in 
\begin{table*}[t]
\begin{tabular}{|c|c|c|c|c|}
\hline
\multicolumn{1}{|c|}{} 
& $\psi_{1}$ & $\psi_{2}$ & $\chi_{1}$ & $\chi_{2}$ 
\\ \hline
$U(1)_{X}$ & $\alpha=e_{1},\beta=0 $ & $\alpha=e_{2},\beta=0 $ & $\alpha=\frac{e_{3}+e_{4}}{2},\beta=\frac{e_{3}-e_{4}}{2} $ 
& $\alpha=\frac{e_{3}+e_{4}}{2},\beta=-\frac{e_{3}-e_{4}}{2} $ \\ \hline
$U(1)_{Y}$ & $\alpha=0,\beta=-q_{1} $ & $\alpha=0,\beta=q_{1} $ & $\alpha=q_{2},\beta=0 $ & $\alpha=-q_{2},\beta=0 $
\\ \hline
\end{tabular}
\caption{Charges of all fermions with respect to the gauge groups
$U(1)_{X}\times U(1)_{Y}$.}
\label{tab:anton}
\end{table*}
Table~\ref{tab:anton}. They are chosen in such a way 
that triangular anomalies $[U(1)_{X}]^{3}$ and $[U(1)_{Y}]^{3}$ are canceled separately. The cancelation
of mixed anomalies requires the extra condition
$q_{2}=q_{1}\frac{(e_{1}^{2}-e_{2}^{2})}{(e_{3}^2-e_{4}^2)}$. Charges
in Table~\ref{tab:anton} follow the general rules of eqs. (\ref{e15}).
If we assume that all  extra fermions have a common mass $m$ and 
are all very heavy, then in the low energy limit 
we find the following expressions for the effective vertices 
with different combinations 
of external gauge bosons:
\begin{subequations}
\begin{align}
 \Gamma_{XXX_{}}^{\mu \nu \rho} &= 
 \Gamma_{YYY_{}}^{\mu \nu \rho}=0 \;, \\ 
 \Gamma_{XXY_{}}^{\mu \nu \rho} &=
 \frac{q_{1}(e_{1}^{2}-e_{2}^{2})}{4 \pi^{2}} 
(2 k_{1}+k_{2})_{\sigma} \varepsilon^{\mu \nu \rho \sigma} \;, \\ 
 \Gamma_{YXX_{}}^{\mu \nu \rho} 
 &=\frac{q_{1}(e_{1}^{2}-e_{2}^{2})}{4 \pi^{2}} 
( k_{2}-k_{1})_{\sigma} \varepsilon^{\mu \nu \rho \sigma} \;, \\ 
 \Gamma_{XYY_{}}^{\mu \nu \rho} &=
 \Gamma_{YXY_{}}^{\mu \nu \rho} =0 \;.  
 \end{align}
\end{subequations}
These contributions arise from terms that are proportional
to $I_{1}$ and $I_{2}$-integrals when taking into account  that this 
model is anomaly-free.  
Such a situation should never occur
in the SM. The basic difference 
is that neither gauge bosons $X$ and $Y$ is purely vector-like 
for the entire fermionic sector {\it i.e.,} 
$X$ and $Y$ must be \emph{strictly  massive}.
This is a crucial difference that leads to the existence of remnants
in the low energy limit. 
On the contrary, the existence of the photon in the SM 
leads to a term related to $I_{1}$ or $I_{2}$ 
which always vanishes for an anomaly-free model.

We have also worked out the case with three different gauge bosons.
The corresponding 10 independent anomaly-free, and, 
18 independent non-decoupling conditions,
are quite involved and are 
presented separately in Appendix~\ref{app:cate}.
Again the non-decoupling effects arise for $n\ge 3$. 
The new feature that appear in this category is the fact that
one can exploit non-decoupling effects where one of the 
gauge bosons is massless.
Such a minimal ($n=3$) example  
comes into sight if we adopt the charge assignments shown
in Table~\ref{tab:3}.
Notice that all fermions have $\beta_{Y}=0$ \ie the $Y$ 
couples purely to a vector current.
\begin{table*}[t]
\begin{tabular}{|c|c|c|c|}
\hline
\multicolumn{1}{|c|}{} 
& $\psi_{1}$ & $\psi_{2}$ & $\psi_{3}$ 
\\ \hline
$U(1)_{X}$ & $\alpha=e,\beta=e $ & $\alpha=e,\beta=-e $ & $\alpha=0,\beta=0 $ 
\\ \hline
$U(1)_{Y}$ & $\alpha=e,\beta=0 $ & $\alpha=e,\beta=0 $ & $\alpha=e,\beta=0 $ 
\\ \hline
$U(1)_{Z}$ & $\alpha=e,\beta=-e $ & $\alpha=0,\beta=0 $ & $\alpha=e,\beta=e $ 
\\ \hline
\end{tabular}
\caption{Charges of an anomaly-free model with non-decoupling remnants
in three gauge boson vertex $XYZ$.}
\label{tab:3}
\end{table*}
 We can easily check 
that the conditions (\ref{eq:aXYZ})
 for an anomaly-free model   are satisfied 
 while at the same time some of the expressions in 
(\ref{eq:bXYZ}) are non zero.
The non-zero  effective vertices can be written in the form,
\begin{subequations}
\begin{align}
\Gamma_{XXZ}^{\mu\nu\rho} &=
-\Gamma_{ZZX}^{\mu\nu\rho} =
\frac{e^{3}}{3 \pi^{2}}(2 k_{1}+k_{2})_{\sigma} \varepsilon^{\mu\nu\rho\sigma}\;, \\  
\Gamma_{XZX}^{\mu\nu\rho}&=
-\Gamma_{ZXZ}^{\mu\nu\rho}= 
-\frac{e^{3}}{3 \pi^{2}}(2 k_{2}+k_{1})_{\sigma} \varepsilon^{\mu\nu\rho\sigma}\;, \\ 
\Gamma_{ZXX}^{\mu\nu\rho} &=
- \Gamma_{XZZ}^{\mu\nu\rho} =
\frac{e^{3}}{3 \pi^{2}}( k_{1}-k_{2})_{\sigma} \varepsilon^{\mu\nu\rho\sigma}\;, \\
\Gamma_{YXZ}^{\mu\nu\rho} &=
 \Gamma_{YZX}^{\mu\nu\rho} = 
 \frac{e^{3}}{2 \pi^{2}}( k_{1}+k_{2})_{\sigma} \varepsilon^{\mu\nu\rho\sigma}\;,  \\
\Gamma_{XYZ}^{\mu\nu\rho} &=
-\Gamma_{ZYX}^{\mu\nu\rho} =
\frac{e^{3}}{2 \pi^{2}} k_{{1}_{\sigma}} \varepsilon^{\mu\nu\rho\sigma}\;,
\label{ZgZp1} \\
\Gamma_{XZY}^{\mu\nu\rho} &=
\Gamma_{ZXY}^{\mu\nu\rho} =
-\frac{e^{3}}{2 \pi^{2}} k_{{2}_{\sigma}} \varepsilon^{\mu\nu\rho\sigma}\;.
\label{ZgZp2}
\end{align}
\end{subequations}
As an example, we observe that heavy fermion non-decoupling effects appear
in \Eqs{ZgZp1}{ZgZp2}. If a model like this 
with $X=Z', Y=\gamma, Z=Z$ can be embedded in
the SM, then it would in principle allow for decays  like $Z' \to Z\gamma$
that do not depend on the heavy fermion masses. 

We should  finally remark that in models considered in 
Tables~\ref{tab:2}-\ref{tab:3},
gravitational anomalies cancel out  
since it is always $\sum_{f} \beta^{X}_{f} =0$ for a given 
axial vector coupling between a vector boson $X$ and a fermion $f$.


\section{Applications}
\label{TGBV}

\subsection{Standard Model}
\label{sec:sm}

Focusing first in the Standard Model  with neutral, $Z$ or $\gamma$
 triple gauge boson
vertices we need only to consider the interaction Lagrangian 
with fermions. This  reads as
\begin{equation}
\mathscr{L}_{int} =   \sum_{f} \alpha^{\gamma}_{f} A_{\mu} 
\overline{\Psi}_{f}
\gamma^{\mu} \Psi_{f}  +   \sum_{f} Z_{\mu}
\overline{\Psi}_{f} \gamma^{\mu} 
 ( \alpha_{f}^{Z} +\beta_{f}^{Z} \gamma_{5} ) \Psi_{f} 
\;, \label{lint2}
\end{equation}
where the factors $\alpha_{f}^{V}, \beta_{f}^{V} $ 
with $V=\gamma, Z$ are
\begin{align}
\alpha^{\gamma}_{f} &= e \, Q_{f} \;, \quad   \beta_{f}^{\gamma} = 0\;,
\nonumber \\[2mm] 
\alpha^{Z}_{f} &=
\frac{\gz}{2}\, (T_{{f_{L}}}^{3} - 2\, s_{w}^{2}\, Q_{f})\;, \quad
\quad \beta_{f}^{Z} = - \frac{\gz}{2}\, T_{f_{L}}^{3} \;, 
\end{align}
and $T_{f_{L}}^{3}$ and $Q_{f}$ are the 
third component of weak isospin 
and charge of the SM Dirac fermions $f=\nu, e, u, d$, respectively. 
Explicitly in the SM,  the prefactors $\alpha_{f}^{V}$ and $\beta_{f}^{Z}$ 
take the form:
\begin{align}
\alpha_{u}^{\gamma} &= \frac{2}{3} e \;, \quad \alpha_{u}^{Z} = \frac{\gz}{2} (\frac{1}{2} - \frac{4}{3} s_{w}^{2} ) \;, \quad \beta_{u}^{Z} = -\frac{\gz}{4}\,, \nonumber \\
\alpha_{d}^{\gamma} &= -\frac{1}{3} e \;, \quad \alpha_{d}^{Z} = \frac{\gz}{2} (-\frac{1}{2} + \frac{2}{3} s_{w}^{2} ) \;, \quad \beta_{d}^{Z} = \frac{\gz}{4}\,, \nonumber \\
\alpha_{e}^{\gamma} &= -e \;, \quad \alpha_{e}^{Z} = \frac{\gz}{2} 
(-\frac{1}{2}
+2 s_{w}^{2} ) \,, \quad \beta_{e}^{Z} = \frac{\gz}{4} \nonumber \\
\alpha_{\nu}^{\gamma} &= 0\;, \quad \alpha_{\nu}^{Z} = \frac{\gz}{4}
\;, \quad \beta_{\nu}^{Z} = - \frac{\gz}{4}\,, 
\label{SM:assi}
\end{align}
where $\gz=e/s_{w}$ is the weak boson gauge 
coupling and $s_{w}, c_{w}$ are the sinus and
cosinus of the weak mixing angle.

\subsubsection{$V^{*}ZZ$}
\label{sub:VZZ}
%
Our first application refers to
the vertex $V^{*}ZZ$ with $V=\gamma,Z$
being off-shell. This interaction 
has been searched for at LEP and Tevatron
 while is  currently under scrutiny at the LHC. 
At one-loop level the only CP-conserving contribution arises from
the triangle graph in Fig.~\ref{fig:graph}. Applying our general form 
of the 1PI vertex in \eq{rose2} and making use of the Bose 
symmetry 
${\nu \leftrightarrow \mu , k_{1} \leftrightarrow k_{2}}$ 
 as in \eq{Bose}, we find
\begin{align}
\Gamma_{V^{*}ZZ}^{\mu\nu\rho}(k_{1},k_{2};w) &= 
\left [ \epsilon^{\mu\nu\rho\sigma} (k_{1}-k_{2})_{\sigma} \left (
-A_{1} + \frac{s}{2} A_{3} \right ) \right. \nonumber \\[3mm] 
&+ \left. A_{3}\: q^{\mu} \epsilon^{\rho\beta\nu\delta} 
k_{1\beta} k_{2\delta} \right ]\;, \label{VZZ}
\end{align}
where the polarization
vectors $\epsilon_{\nu}^{*}(k_{1})\epsilon_{\rho}^{*}(k_{2})$
outside the square brackets have been omitted, and also,  
we set $A_{1} \equiv A_{1}(k_{1},k_{2})...$ etc for simplicity.
More specifically, $A_{1}$  is ambiguous:
it depends on how the momentum is
 routing the loop \ie the parameter $w$.
This arbitrariness (or regularization scheme dependence if you wish) 
is further fixed by exploiting the fact that the $ZZZ$ on-shell 
boson vertex vanishes by Bose symmetry.
The latter  requires $w=1/3$. 
On the other hand for the
 vertex $\gamma ZZ$,  conservation of 
 the vector current and Bose symmetry implies that $w=z=0$. 

Having specified the arbitrary parameters $w$ and $z$
we apply our general expressions for $A_{1}$ and $A_{3}$ 
found in \eqs{GENA1}{GENA3}, specifically to  the vertices $Z^{*}ZZ$ and 
$\gamma^{*}ZZ$
and sum over all  SM fermions. 
By ignoring (see below however), 
the last term proportional to $q^{\mu}$ in \eq{VZZ},
we can easily find,
\begin{widetext}
\begin{align}
\Gamma^{\mu\nu\rho}_{Z^{*}ZZ}(k_{1},k_{2}) \ &= \ 
 \epsilon^{\mu\nu\rho\sigma} (k_{1}-k_{2})_{\sigma}
\sum_{f=u,d,e,\nu} \left [ m_{Z}^{2} (A_{3 f} - A_{4 f }) + \frac{m_{f}^{2} \beta_{f}^{Z}}{\pi^{2}} \, I_{1f} + \frac{1}{6\pi^{2}} \left ( \beta_{f}^{Z\, 3} + 3 \beta_{f}^{Z} \alpha_{f}^{Z\, 2} \right ) \right ]
\nonumber \\
&\equiv \epsilon^{\mu\nu\rho\sigma} (k_{1}-k_{2})_{\sigma} \: \Gamma_{Z^{*}ZZ}(s)  \;, \label{ZZZ}
\\[3mm]
\Gamma^{\mu\nu\rho}_{\gamma^{*}ZZ}(k_{1},k_{2}) \ &= \
 \epsilon^{\mu\nu\rho\sigma} (k_{1}-k_{2})_{\sigma}
\sum_{f=u,d,e,\nu} \left [ m_{Z}^{2} (A_{3 f} - A_{4 f })
 + \frac{m_{f}^{2} \beta_{f}^{Z}}{\pi^{2}} \, I_{1f} + \frac{1}{2\pi^{2}}    \alpha_{f}^{\gamma} \alpha_{f}^{Z} \beta_{f}^{Z}  \right ] 
 \nonumber \\
 &\equiv \epsilon^{\mu\nu\rho\sigma} (k_{1}-k_{2})_{\sigma}
 \Gamma_{\gamma^{*}ZZ}(s) 
  \label{gZZ} \;,
\end{align}
\end{widetext}
 where  $s=(k_{1}+k_{2})^{2}$ and $I_{1f}$ is given by \eq{GI1}.
The last term in \eqs{ZZZ}{gZZ} is the anomaly contribution,
while the second term is a non-decoupling one in the limit of
heavy fermion mass, $m_{f}\to \infty$. Again we should notice here
that in this limit and for one fermion contribution,
the last two terms mutually  cancel while the
first term vanishes as $m_{Z}^{2}/m_{f}^{2}$. 
Therefore, the decoupling of  heavy fermions  
in $V^{*}ZZ$ vertex is operative even if those fermions have
vastly different, but always much greater than the  EW scale,
masses among each other. 
In the SM for example, what is left behind after 
the decoupling of the top quark 
is a theory with an anomalous (sometimes called {\it Chern-Simons}) term
that is necessary to render the effective low energy 
theory gauge invariant. 

Especially for $\gamma^{*}ZZ$ one can go one step further and write
the whole effective vertex in terms of one integral only, namely
\begin{equation}
\Gamma_{\gamma^{*}ZZ}(s) \ = \ \frac{s}{2} \, \sum_{f=u,d,e,\nu} \, A_{3f}(s) \;.
\label{eq:gZZ}
\end{equation} 
Now bringing back the last term on the r.h.s of \eq{VZZ} we find
a perfectly fine and gauge invariant form for $\gamma^{*}ZZ$-vertex
\begin{eqnarray}
\Gamma^{\mu\nu\rho}_{\gamma^{*}ZZ}&&(s) = 
\sum_{f}\frac{s A_{3f}}{2} \times \nonumber \\ &&\left [
\epsilon^{\mu\nu\rho\sigma} (k_{1}-k_{2})_{\sigma} -
\frac{\epsilon^{\nu\rho\beta\sigma} q^{\mu} q_{\beta}}{s} (k_{1}-k_{2})_{\sigma}\right ].
\label{eq:gZZ2}
\end{eqnarray}
This vertex \emph{must} be proportional to $s$ in order to cancel the pole contribution
arising at $s=q^{2}=0$~\cite{Hagiwara}. This is a generic statement for all $\gamma^{*}VV$
vertices we address below.
One should recall  that this expression 
 has been derived only after fixing the anomaly coefficients, 
$w$ and $z$, by symmetry requirements. We could have done the reverse:
to fix $w,z$ from the requirement of no pole contribution in \eq{eq:gZZ2}.
In a way, the anomaly and the non-decoupled terms have been
absorbed in the finite integral $A_{3}$. It is now evident
from \eqs{eq:gZZ}{A3456} that $\Gamma_{\gamma^{*}ZZ}(s\to 0) =0$ 
for every fermion contribution, independently. Furthermore, as expected, for asymptotic values 
of $s$ we  also observe, $\Gamma_{\gamma^{*}ZZ}(s\to \infty) =0$, after summing over
all SM fermion contributions.

\begin{figure*}[ht!]
       \centering
        \subfloat[][]{
            \label{fig:a}
            \includegraphics[width=0.4\textwidth]{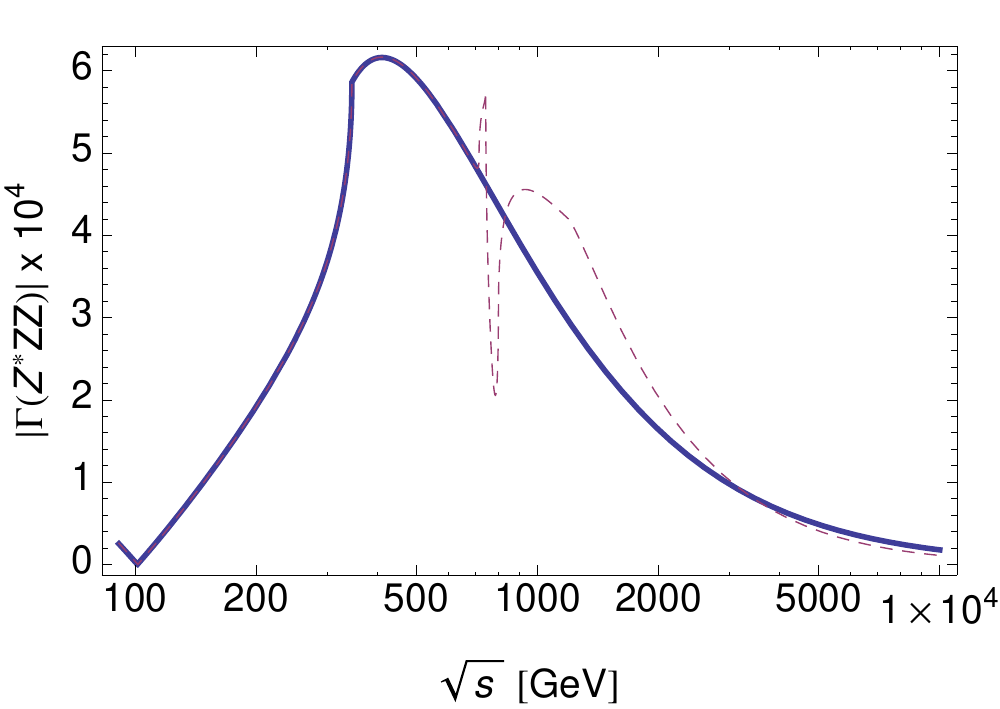}
        }
        \quad
        \subfloat[][]{
           \label{fig:b}
           \includegraphics[width=0.4\textwidth]{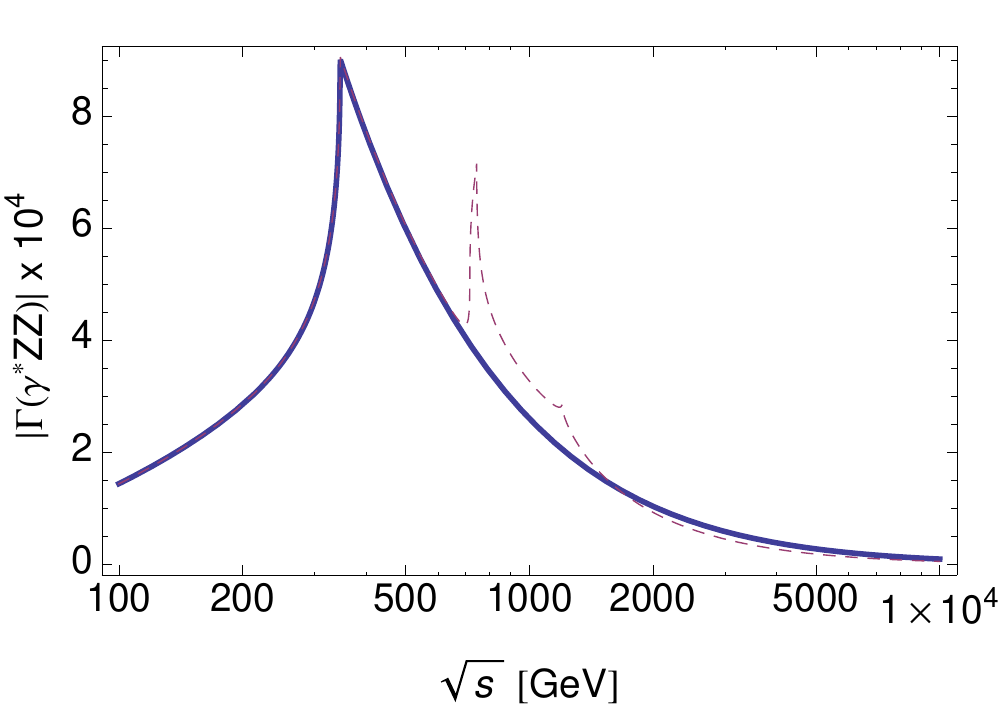}
        }
        \quad
        \subfloat[][]{
            \label{fig:c}
            \includegraphics[width=0.4\textwidth]{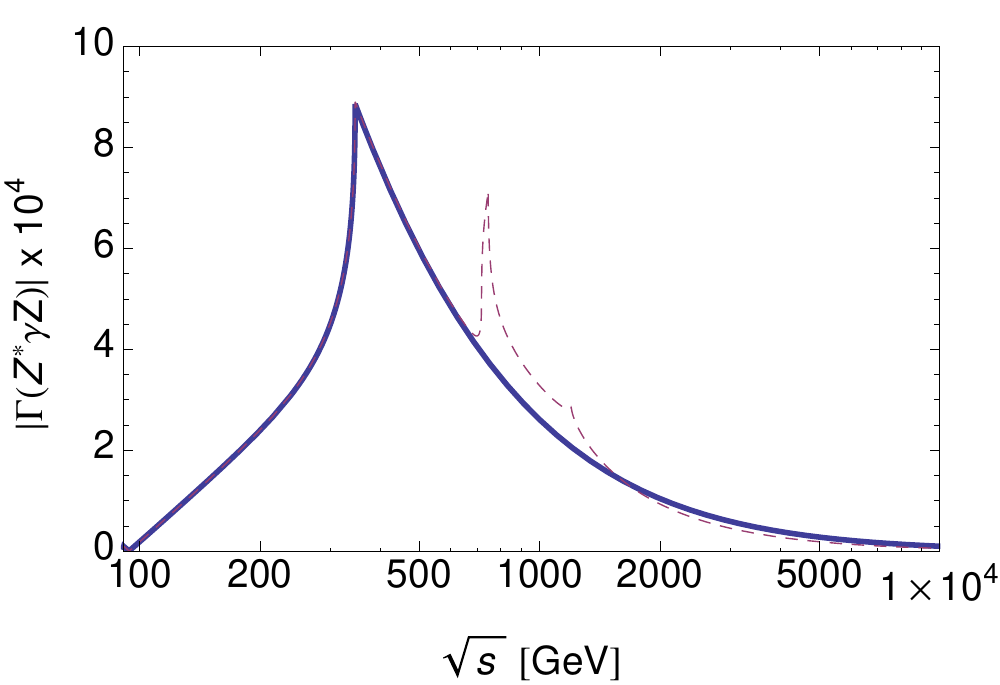}
        } 
        %
        \subfloat[][]{%
            \label{fig:d}
            \includegraphics[width=0.4\textwidth]{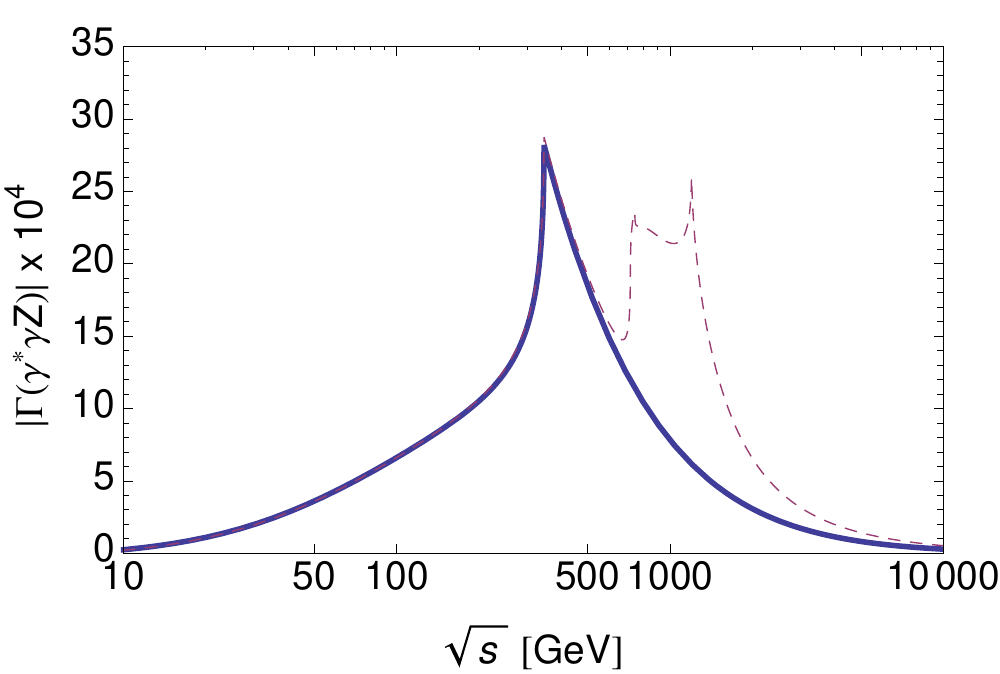}
        }%
\\
	\subfloat[][]{%
            \label{fig:e}
            \includegraphics[width=0.4\textwidth]{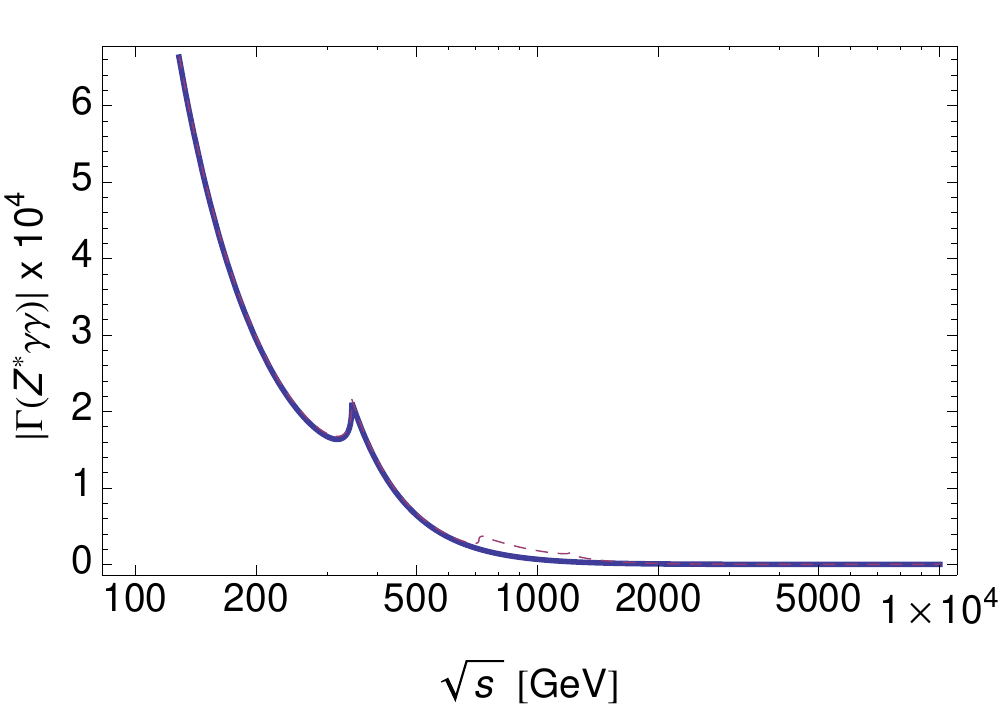}
        }
    \caption{
        The dependence of 
        $|\Gamma_{V^{*}VV}(s)|$ with $\sqrt{s}$ for different
 gauge bosons combinations, $V=\gamma, Z$ : (a) $Z^{*}ZZ$, 
 (b) $\gamma^{*}ZZ$, (c) $Z^{*}\gamma Z$, (d) $\gamma^{*}\gamma Z$,
 (e) $Z^{*}\gamma \gamma$. 
 The solid curve corresponds
to the SM, the dashed curve corresponds to the 
SM + 4th generation fermion model. Masses for light quarks and leptons
are neglected while $m_{t}=173~\rm{GeV}$. Fourth generation quarks and 
lepton masses are taken as in (\ref{4genmasses}).}
 \label{fig:gammas}
\end{figure*}

Within one generation of fermions, 
the SM is a chiral, gauge, and, anomaly-free
Quantum Field Theory (QFT). As 
a result, contributions to $\Gamma_{V^{*}ZZ}$ from (approximately)
massless generations, vanish identically (recall that form factors 
$A_{3,4}$  
are proportional to the anomaly factors, [see  \eqs{GENA3}{GENA4}]
and the second term vanishes in the massless case). 
Therefore to a good approximation, for $\sqrt{s}\gtrsim 2 M_{Z}$,
 the only non-negligible contribution to $\Gamma_{V^{*}ZZ}$ 
 arises from the third generation and is due to the large
  mass difference between
 the top quark and all other fermions.  
The top quark influences mainly the last two terms in the
square bracket of $\Gamma_{Z^{*}ZZ}$ and $\Gamma_{\gamma^{*}ZZ}$
in \eqs{ZZZ}{gZZ}.  If we make the (numerically crude) 
approximation of $m_{Z}^{2} \ll s < m_{t}^{2}$ and exploit \eq{ApI1}
from the Appendix~\ref{app:int} we find ($N_{c}=3$ is the color factor), 
\begin{align}
\frac{m_{t}^{2}\beta_{t}^{Z}}{\pi^{2}} I_{1t} &\approx 
- \frac{N_{c}}{6\pi^{2}} \left ( \beta_{t}^{Z\, 3} + 3 \beta_{t}^{Z} 
\alpha_{t}^{Z\, 2} \right ) \nonumber \\[2mm]
&- \frac{N_{c}}{120 \pi^{2}}\: (\beta_{t}^{Z\, 3} + 5 \beta_{t}^{Z} 
\alpha_{t}^{Z\, 2}) \frac{s}{m_{t}^{2}} \;.\label{ap29}
\end{align}
The first term is just the opposite of the top quark anomaly contribution
in $\Gamma_{Z^{*}ZZ}$ and they both cancel out 
in the limit of heavy top quark. One can prove easily this statement 
for all SM vertices, 
$\Gamma_{V^{*}VV}, V=Z,\gamma$ appearing 
below in this article and we
claim, following the arguments of  section~\ref{sec:ndeffects}, that
this is a general theorem: {\it a heavy particle cancels its own 
anomaly contribution in a triple gauge boson vertex and at the (non-perturbative) limit of $m\to \infty$ leaving no trace from itself behind}.
Of course in the top-less SM  the
last term in $\Gamma_{Z^{*}ZZ}$ does not vanish since the particle
content ($\tau, \nu_{\tau}, b$) is now anomalous.  
It is also evident from \eq{ap29}
that  the behaviour of $\Gamma_{Z^{*}ZZ}(s)$ at $s\approx m_{t}^{2}$
rises approximately linearly with $s$ as $s/m_{t}^{2}$. This is also verified
from our numerical results  shown in Fig.\ref{fig:gammas}a.
Similar conclusions one can derive for $\Gamma_{\gamma^{*}ZZ}$ and
Fig.\ref{fig:gammas}b but  this 
is rather obvious now because of \eq{eq:gZZ}.
  
Furthermore, it is also instructive to study  the behaviour of 
the vertices $\Gamma_{V^{*}ZZ}(s)$  in the asymptotic region,  
$s\gg m_{t}^{2} > m_{Z}^{2}$. By exploiting \eq{D14}  
and keeping only terms of order $m_{f}^{2}/s$ 
we arrive at the following expression,
\begin{align}
&\Gamma_{Z^{*}ZZ}(s\gg m_{t}^{2}) \approx N_{c}
\frac{m_{t}^{2}}{s} \left \{ \frac{2\beta_{t}^{Z\,3}}{\pi^{2}} 
\left (2- \ln\frac{s}{m_{t}^{2}} - i \pi \right ) \right.
\nonumber \\
&+ \left. \frac{\beta_{t}^{Z\,3}+\alpha_{t}^{Z\,2}\beta_{t}^{Z}}{\pi^{2}} 
\left (\frac{1}{2}\ln^{2}\frac{s}{m_{t}^{2}} - \frac{\pi^{2}}{2} +
 i \pi \ln\frac{s}{m_{t}^{2}} \right ) \right \} \;,\label{ASYMP}
\end{align}
in which both real and imaginary parts vanish
at asymptotic values of $s$ as they should  following unitarity 
arguments. The effect of a ``heavy'' particle (here the top quark)
is to just delay  the ``falling off'' of $|\Gamma_{Z^{*}ZZ}(s)|$ 
[see Fig.\ref{fig:gammas}a.] as $s\to \infty$.  Finally, it is also obvious that the 
real and imaginary part of $\Gamma_{Z^{*}ZZ}$ are of the same  
order of magnitude, a situation which remains true everywhere after 
the threshold energy, $s \gtrsim 4 m_{t}^{2}$. 

Translating our numerical 
results for the SM to the notation of \Ref{Hagiwara}\footnote{
We multiply $\Gamma_{V^{*}ZZ}(s)$ in \eqs{ZZZ}{gZZ}
with  $e\, m_{Z}^{2}/(s-m_{V}^{2})$.}
that is usually followed by the
 theoretical and experimental literature, 
we find for $m_{t}=173$ GeV and LEP energies, that
\begin{eqnarray}
f_{5}^{Z}(\sqrt{s}=200~{\rm GeV}) &=& 1.8 \times 10^{-4}\;,
\\
f_{5}^{\gamma}(\sqrt{s}=200~{\rm GeV}) &=& 2.1 \times 10^{-4} \;,
\end{eqnarray}
where we have neglected  
small imaginary part contributions from light quark
and lepton mass thresholds. 
These results agree with those quoted in \Ref{Gounaris:2000tb}.
Unfortunately, they are  too small to have been reached by 
LEP~\cite{DELPHI:2007pq}.

Just above the top quark threshold
 energies $s \ge 4 m_{t}^{2}$, the vertex develops a significant
absorptive part. This is  apparent from our analytical expressions 
in Appendix~\ref{app:int} for  integrals $A_{3..6}$ and $I_{1,2}$ and 
the discussion above. 
For $\sqrt{s}=500$ GeV we find :
\begin{eqnarray}
f_{5}^{Z}(\sqrt{s}=500~{\rm GeV}) &=& (0.4-0.5 i) \times 10^{-4}\;,
\\
f_{5}^{\gamma}(\sqrt{s}=500~{\rm GeV}) &=& (-0.3+0.3 i) \times 10^{-4} \;.
\end{eqnarray}
Note again that  the imaginary 
part of the amplitude is of the same order of magnitude as the 
real part.  


\subsubsection{$V^{*}\gamma Z$}
\label{subsec:VgZ}

Another non-trivial class among trilinear neutral gauge
boson vertices that have been and being searched for at 
colliders is the amplitude $V^{*}\gamma Z$.  
In the notation of Fig.~\ref{fig:graph}, we assign 
$V^{*}_{\mu}(q)$, $\gamma_{\nu}(k_{1})$ and $Z_{\rho}(k_{2})$
to the 1PI effective vertex $\Gamma^{\mu\nu\rho}_{V^{*}\gamma Z}$
of \eq{rose2} with  $V=Z, \gamma$.
When the photon and the Z-gauge boson are both on-shell
we find:
%
\begin{eqnarray}
&&\Gamma_{V^{*}\gamma Z}^{\mu\nu\rho}(k_{1},k_{2}) \ = \
\epsilon^{\mu\nu\rho\sigma} \, k_{1\sigma} \left (A_{2} 
+ \frac{s+m_{Z}^{2}}{2}\, A_{3} \right ) \nonumber \\[2mm]
&+& \epsilon^{\mu\rho\beta\delta} q^{\nu}\, q_{\beta}\, k_{2\delta}\,
\left (A_{3} + A_{6} \right ) 
+  \epsilon^{\nu\rho\beta \delta} q^{\mu} k_{1\beta}
 k_{2\delta}\, A_{3}\,.\nonumber \\ \label{VgZ}
\end{eqnarray} 
%
We have seen however in  \eq{GENA6} that
$A_{3} = -A_{6}$ and therefore, the second term in \eq{VgZ} vanishes
at one-loop. Furthermore, the last term when coupled to a  
light quark or lepton vector current,
is proportional to the mass of the incoming fermions and for current 
collider architectures this contribution is negligible\footnote{This term however
is important for gauge invariance to be preserved, 
as in \eq{eq:gZZ2} before.}. 
Hence, only the first term remains with potentially  visible effects.
When \emph{all} 
external particles are on-shell, Bose symmetry and gauge
invariance require the vertex $V\gamma Z$ to vanish. Bose symmetry
relations among form factors and gauge invariance  fix the
arbitrary parameters  $w$ and $z$ to be:
\begin{align}
Z\gamma Z ~:& \qquad  w=1\;, \quad z=0 \;,\label{26} \\[2mm]
\gamma \gamma Z ~: & \qquad  w=1 \;, \quad  z=1 \;. \label{27}
\end{align}

By substituting the form in $A_{2}$ from the general expression
of  (\ref{GENA2}) we obtain:
\begin{widetext}
\begin{align}
\Gamma^{\mu\nu\rho}_{Z^{*}\gamma Z}(k_{1},k_{2}) \ &= \
 \epsilon^{\mu\nu\rho\sigma} k_{1\sigma}
\sum_{f=u,d,e,\nu} \left [ m_{Z}^{2} (A_{3 f} + A_{5 f }) - \frac{m_{f}^{2} \beta_{f}^{Z}}{\pi^{2}} \, I_{2f} + \frac{1}{2\pi^{2}}\alpha_{f}^{Z}\alpha_{f}^{\gamma}\beta_{f}^{Z}  \right ] \equiv 
 \epsilon^{\mu\nu\rho\sigma} k_{1\sigma} \:
 \Gamma_{Z^{*}\gamma Z}(s) \;, \label{ZgZ}
\\[3mm]
\Gamma^{\mu\nu\rho}_{\gamma^{*}\gamma Z}(k_{1},k_{2}) \ &= \
 \epsilon^{\mu\nu\rho\sigma} k_{1\sigma}
\sum_{f=u,d,e,\nu} \left [ m_{Z}^{2} (A_{3 f} + A_{5 f })
 - \frac{m_{f}^{2} \beta_{f}^{Z}}{\pi^{2}} \, I_{2f} + \frac{1}{2\pi^{2}}    \alpha_{f}^{\gamma} \alpha_{f}^{\gamma}  \beta_{f}^{Z}  \right ] 
 \equiv  \epsilon^{\mu\nu\rho\sigma} k_{1\sigma}\:
\Gamma_{\gamma^{*}\gamma Z}(s) 
  \label{ggZ} \;.
\end{align}
\end{widetext}
One should notice that the square bracket of 
$\Gamma_{Z^{*}\gamma Z}$ is
approximately equal to $\Gamma_{\gamma^{*}ZZ}$ since in this
case $A_{5} \simeq - A_{4}$ and $I_{1} \simeq - I_{2}$.

It is amusing to see how greatly the $\gamma^{*}\gamma Z$-vertex is simplified. 
Placing back the last term of \eq{VgZ} in order to restore gauge invariance, we find,
\begin{align}
\Gamma^{\mu\nu\rho}_{\gamma^{*}\gamma Z}(s) = \sum_{f}\,   s\, A_{3f} \times
\left [\epsilon^{\mu\nu\rho\sigma} k_{1\sigma} - \frac{\epsilon^{\nu\rho\beta\sigma} 
q^{\mu} k_{2\beta} k_{1\sigma}}{s} \right ] \;.
\label{eq:VgamZ}
\end{align}
The  s-factor outside the vertex is expected because it must
 cancel the pole behaviour of the second term  in the square bracket. 
Once again, the ``physical'' choice of $w,z$ in the anomalous terms
played a crucial role in \eq{eq:VgamZ} like in the case of $\gamma^{*}ZZ$ vertex. 
Regarding  decoupling effects, \eq{eq:VgamZ} is self explained:
for every particle contribution,  a synergy between 
anomalous  and non-decoupling terms results in a well defined integral
$s A_{3f}$ that vanishes asymptotically due to the anomaly-free condition.
If however, the energy $\sqrt{s}$ is between two particle masses which combined render
the model anomaly-free then there should be non decoupling effects in this regime.
One the other hand, adding to the SM, 
anomaly-free and heavy
chiral fermions, there should be no-nondecoupling effects
remaining in the low energy regime where $\sqrt{s} \lesssim 2 m_{t}$. 

One can go one step further also in the case of $Z^{*}\gamma Z$ of
\eq{ZgZ}. In fact, we can  eliminate $I_{2f}$ and the anomaly factors from
\eq{ZgZ} leaving only the finite integrals $A_{3}$ and $A_{5}$, as
\begin{eqnarray}
\Gamma_{Z^{*}\gamma Z}(s) = \frac{1}{2}\sum_{f} \left [ (s+m_{Z}^{2}) A_{3f} +
m_{Z}^{2} A_{5f} \right ] \;.
\end{eqnarray}
After using few integral tricks, like for example the ones of \eq{inttricks},
it is easy to show that $\Gamma_{Z^{*}\gamma Z}(s)$
behaves like $(s-m_{Z}^{2}) A_{3f}$ near the $Z$-pole. 
In general,  $\Gamma_{V^{*}\gamma Z} \propto \sum_{f}(s-m_{V}^{2}) A_{3f}$
near the pole,  is clearly verified when performing
 the full numerical evaluation of the 
integrals as in Figs.~\ref{fig:gammas}c,\ref{fig:gammas}d.

One can easily see from further working out \eqs{ZgZ}{ggZ}
that due to the fact that the SM is an anomaly-free QFT, 
 the whole contribution arises to a very good approximation 
from particles of the third generation. Numerically,
in the conventions of \Ref{Hagiwara} [see also footnote 4],
 we find for LEP  energies 
\begin{eqnarray}
h_{3}^{Z}(\sqrt{s}=200~{\rm GeV}) &=& 2.1 \times 10^{-4}\;,\label{30}
\\
h_{3}^{\gamma}(\sqrt{s}=200~{\rm GeV}) &=& 7.2 \times 10^{-4} \;,
\label{31}
\end{eqnarray}
up to tiny small imaginary parts.
These results are in agreement with those presented in \Ref{Gounaris:2000tb}. As we have noticed above, 
it is also confirmed numerically that 
 $|f_{5}^{\gamma}| \simeq |h_{3}^{\gamma}|$.
 SM predictions of \eqs{30}{31} are in the best case
[for $h_{3}^{\gamma}$] two orders of magnitude below 
the published LEP bounds~\cite{DELPHI:2007pq}.

For comparison, at higher energies the SM predicts: 
\begin{eqnarray}
h_{3}^{Z}(\sqrt{s}=500~{\rm GeV}) &=& (0.3-0.6 \, i) \times 10^{-4}\;,\label{30}
\\
h_{3}^{\gamma}(\sqrt{s}=500~{\rm GeV}) &=& (0.9-1.8\, i) \times 10^{-4} \;.
\label{31}
\end{eqnarray}
%
Full numerical results for $|\Gamma_{V^{*}\gamma Z}(s)|$ are 
represented by solid lines in 
Figs.~\ref{fig:gammas}c,\ref{fig:gammas}d. We observe that 
in the neighborhood of the top threshold, 
$|\Gamma_{\gamma^{*}\gamma Z}(s)|$ is  one order of magnitude bigger
than $|\Gamma_{Z^{*}\gamma Z}(s)|$, but still in the region $10^{-3}$.
%
%
They are both however far below the 
current Tevatron and LHC 
sensitivity~\cite{Abazov:2008wg,Aaltonen:2011zc}.
For example, both ATLAS~\cite{Giraud:2012ja} and CMS~\cite{Martelli:2012ea} 
experiments at LHC currently report
bounds on trilinear, $V^{*}\gamma Z$, gauge boson 
vertices $|h_{3}^{Z,\gamma}|$ that in the best case 
are not less than 5\%. These experiments present bounds
w.r.t the scale $\Lambda$ in which the new physics enters.
Following the projecting sensitivity calculated in \Ref{Baur},
and setting  $\Lambda\sim m_{t}$ for the SM,
 LHC sensitivity  for $V^{*} Z\gamma$
 with $\sqrt{s}=14$ TeV will not be better
 than $\sim 10^{-2}$ and this makes its observation extremely 
difficult within SM, even for $\gamma \gamma Z$-vertex. 

\subsubsection{$V^{*}\gamma\gamma$}
\label{subsec:Vgg}

We now turn our discussion to the last SM 
neutral triple gauge boson vertex,
 the $V^{*}\gamma\gamma$.
Of course, thanks to Furry's theorem  only the case $V=Z$ 
is valid (for $V=\gamma$
all three currents are vector-like, \ie $\beta_{i}=0$). 
However, even in  $Z^{*}\gamma\gamma$ 
there are  no non-decoupling effects
since there is no would be Goldstone boson associated with
the unbroken $U(1)_{\rm em}$, \ie the final particles are massless.
Nevertheless one can write a simple $Z^{*}\gamma\gamma$ 1PI vertex.
We  obtain:  
\begin{align}
\Gamma_{Z^{*}\gamma\gamma}^{\mu\nu\rho}(k_{1},k_{2}) 
\ &= \ \epsilon^{\nu\rho\beta\delta} q^{\mu} k_{1\beta} k_{2\delta} \left
 [ A_{3} \right ] 
\nonumber \\[2mm]
&+ \frac{\beta_{f}^{Z} (\alpha_{f}^{\gamma})^{2}}{4\pi^{2}} 
\, \epsilon^{\mu\nu\rho\sigma} \left [ (w-1)\, k_{2} + (z+1)\, k_{1} 
\right ]_{\sigma} 
\;. \label{Zgg}
\end{align}
Landau~\cite{Landau} and Yang~\cite{Yang} say that the on-shell amblitute, $\epsilon_{\mu}(q)\Gamma^{\mu\nu\rho}_{Z^{*}\gamma\gamma}(k_{1},k_{2})$ 
must vanish due to selection rules on
space inversion and angular momentum conservation.
This fixes the arbitrary parameters
$w=-z=1$ for every fermion contribution $f$. One obtains the 
same values for $w$ and $z$  from $U(1)_{em}$ 
gauge invariance, \ie 
satisfaction of Ward Identities.
%
%
Although it is necessary to preserve gauge invariance, this remaining 
contribution is negligible for light s-channel incoming particles 
e.g., $e^{+}e^{-} \rightarrow \gamma \gamma$, but nevertheless it
may be important for heavy external particles like for example dark
matter particles or heavy neutrinos annihilating into photons (see related work in  \Refs{Aminneborg:1990xw,Rudaz:1989ij}).

Defining $\Gamma_{Z^{*}\gamma\gamma}(s) \equiv \sum_{f}m_{Z}^{2} \, A_{3f}(s)$ and summing over  the SM particles,
we find numerically,
\begin{eqnarray}
\Gamma_{Z^{*}\gamma\gamma} (\sqrt{s} &=& 200 \text{ GeV}) = 2.9 \times 10^{-4} 
\;,\\
\Gamma_{Z^{*}\gamma\gamma} (\sqrt{s} &=& 500 \text{ GeV}) = 
(3.2 - 5.6\, i) \times
10^{-5} \;. \nonumber \\
\end{eqnarray}
For various values of $s$, the function
 $|\Gamma_{Z^{*}\gamma\gamma}(s)|$  is plotted in
Fig.~\ref{fig:gammas}e. Notably, at very small $s$ this quantity 
behaves like $1/s$  and in contrary to the previous $Z^{*}VV$ vertices 
does not vanish at $s=m_{Z}^{2}$. 
For general values of $s$,  
and  $k_{1}^{2} = k_{2}^{2}=0$,
$\Gamma_{Z^{*}\gamma\gamma}(s)$ is easily written as
\begin{equation}
\Gamma_{Z^{*}\gamma\gamma}(s) = \sum_{f} \frac{\beta_{f}^{Z}(\alpha_{f}^{\gamma})^{2}}{2\pi^{2}} \, \frac{m_{Z}^{2}}{s} \, \xi_{f} \, J(\xi_{f})\;,
\label{exzgg}
\end{equation}
where $\xi_{f} \equiv 4 m_{f}^{2}/m_{Z}^{2}$ 
and the function $J({\xi_{f}})$ is 
appended in \eq{exjxi}.
%
For energies $(s)$ below the top quark 
threshold, $\Gamma_{Z^{*}\gamma\gamma}(s)$, approximately 
takes the form,  
\begin{align}
&\Gamma_{Z^{*}\gamma\gamma}(s) \equiv 
\sum_{f} m_{Z}^{2} A_{3f}(m_{Z}^{2} < s < m_{t}^{2}) \approx
\nonumber \\[2mm]
&- N_{c}\frac{ \beta_{t}^{Z} \alpha_{t}^{\gamma\, 2}}{\pi^{2}}
 \:  \left [\frac{m_{Z}^{2}}{2s}+ 
\left (\frac{m_{Z}^{2}}{m_{t}^{2}}
\right )
\left ( \frac{1}{24} + \frac{1}{180} \frac{s}{m_{t}^{2}}\right ) \right ]\;,
\end{align}
a behaviour which shows  decoupling of a heavy top-quark mass. 
This follows our general
statement just below \eq{ap29}: since the anomalous term in \eq{Zgg}
vanishes due to the physical choice of $w$ and $z$, there is no 
non-decoupled remnant to cancel it.  In the asymptotic region 
we find
\begin{align}
&\Gamma_{Z^{*}\gamma\gamma}(s\gg m_{Z}^{2},m_{t}^{2} ) \approx
\nonumber \\[2mm]
&N_{c}\frac{\beta_{t}^{Z}\alpha_{t}^{\gamma\, 2}}{2\pi^{2}}\: \left (
\frac{m_{Z}^{2}m_{t}^{2}}{s^{2}} \right ) \: \left [
\ln^{2}\frac{s}{m_{t}^{2}} - \pi^{2} + 2 i \pi \ln\frac{s}{m_{t}^{2}} \right ]\;.
\end{align}
Therefore, $\Gamma_{Z^{*}\gamma\gamma}(s)$ behaves asymptotically
as $1/s^{2}$, while all other neutral vertices behave like $1/s$.
This fast drop with  $s$ is also verified by comparing the 
 solid lines between Figs.~\ref{fig:gammas}a,b,c,d and 
Fig.~\ref{fig:gammas}e.

\subsubsection{$V^{*}W^{-}W^{+}$}
\label{sec:VWW}

Just for completeness, 
we study the chiral CP-invariant part of the $(\gamma,Z)^{*}WW$ vertex. 
For on-shell $W$'s and in momentum space
this corresponds to  operators of the form,  
\begin{eqnarray}
f_{5}^{V} \epsilon^{\mu\nu\rho\sigma} \, (k_{1}-k_{2})_{\sigma} \;.
\label{eq:f5V}
\end{eqnarray}
There are of course CP-invariant, non-chiral operators
generated from our fermion triangle graph 
that have the form~\cite{Gaemers,Hagiwara},
\begin{eqnarray}
f_{1}^{V} (k_{1}-k_{2})^{\mu} g^{\nu\rho} - \frac{f_{2}^{V}}{m_{W}^{2}} 
(k_{1}-k_{2})^{\mu} q^{\nu} q^{\rho} \nonumber 
\\ + f_{3}^{V} (q^{\nu} g^{\mu\rho} - q^{\rho} g^{\mu\nu})\;.
\label{eq:f123V}
\end{eqnarray}
In the SM, note  that both $f_{1}$  and $f_{3}$, exist at tree level.
We are interested here only on chiral, one-loop (triangle) induced
operators (\ref{eq:f5V}).

\begin{figure}[htbp]
   \centering
   \includegraphics[scale=0.75]{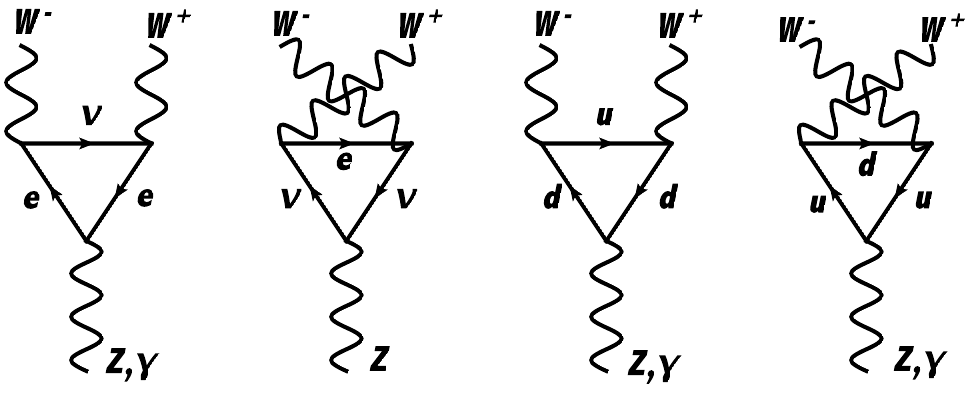} 
   \caption{Standard Model fermion contributions to $(Z,\gamma)WW$ one-loop vertex.}
   \label{fig:VWW}
\end{figure}

The numerical calculation of the $(\gamma^{*},Z^{*})W^{-}W^{+}$ effective 
vertices are somehow more complicated  than the neutral ones. 
There are two masses  and two different neutral vertices involved, 
making the triangle diagram looking differently than  its crossed
counterpart (see Fig.~\ref{fig:VWW}). We follow the same steps as we did for the neutral
vertices  and  present our results (and technical details)
in Appendix~\ref{app:W}. The chiral CP-invariant part of the effective 
vertex, $\Gamma^{\mu\nu\rho}$, is the same as in \eq{rose2}. 
The finite form factors $A_{3..6}$ need to be slightly  modified by the mass difference
of the two fermions involved; analogously for $A_{1,2}$. 
Our main conclusion for a general vertex that contains external charged  gauge bosons
is given by  \eqs{A12W}{I12W}.

The relevant couplings $\alpha^{W}_{ff'}$, and  $\beta^{W}_{ff'}$ can be read
from the charged current part of the SM Lagrangian,
\begin{align}
\mathscr{L} \supset \gz (W_{\mu}^{+}\, J_{W}^{\mu\, +} +
W_{\mu}^{-}\, J_{W}^{\mu\, -} ) \;,
\end{align}
with the $J_{W}^{\pm}$-currents being
\begin{equation}
J_{W}^{\mu\, +} = (J_{W}^{\mu\, -})^{\dagger} = \frac{1}{2\sqrt{2}}\:
\left [ \bar{\nu} \gamma^{\mu} (1-\gamma_{5}) e + 
\bar{u} \gamma^{\mu} (1-\gamma_{5}) d \right  ]\;.
\end{equation}
Hence $\alpha^{W}_{ff'} = - \beta^{W}_{ff'} = 
\frac{\gz}{2\sqrt{2}}$ for the pairs $(ff') = (\nu, e), (u,d)$, respectively.
For simplicity, we ignore quark and lepton mixing effects,
but these can  easily be included.

We therefore set $\alpha_{j,k} = -\beta_{j,k} =\frac{\gz}{2\sqrt{2}}$ in \eqs{A12W}{I12W}.
The neutral gauge boson-fermion couplings, $\alpha_{f}^{V}, \beta_{f}^{V}$, are taken 
from  \eq{SM:assi}.
Assuming CP-conservation,  the 1PI effective action 
$\Gamma^{\mu\nu\rho}_{V^{*}WW}$ with $V=\gamma, Z$ looks exactly 
the same as in \eq{VZZ} with the only difference being  the form factors 
$A_{1,3}$ must be replaced by those given in \eq{A12W} [and the 
paragraph below ({\ref{A12W})]. 
Therefore we write\footnote{Our notation for $\Gamma_{V^{*}W^{-}W^{+}}(s)$ is related
to the standard form factor of \Ref{Hagiwara}, as $\Gamma_{V^{*}W^{-}W^{+}}(s) = 
- g_{VWW} f_{5}^{V}(s)$.} , 
\begin{equation}
\Gamma^{\mu\nu\rho}_{V^{*}W^{-}W^{+}}(k_{1},k_{2}) \equiv  \epsilon^{\mu\nu\rho\sigma}\, (k_{1}-k_{2})_{\sigma} \, \Gamma_{V^{*}W^{-}W^{+}}(s)\;,
\end{equation}
where
\begin{widetext}
\begin{align}
\Gamma_{V^{*}W^{-}W^{+}}(s)  \ &= \  
 \sum_{\mathrm{doublets}}\left [ m_{W}^{2}(A_{3} -A_{4}) \ + \ 
\frac{\gz^{2} \alpha_{f_{d}}^{V}}{16\pi^{2}} \, \mathcal{I}_{1} \ + \
\frac{\gz^{2}\beta_{f_{d}}^{V}}{16\pi^{2}} \, \mathcal{I}_{2} \ + \
\frac{\gz^{2}}{32\pi^{2}} \, (\alpha_{f_{d}}^{V} - \beta_{f_{d}}^{V})\, (w-1) 
+ (f_{u}\leftrightarrow f_{d}) \right ] \;.
\end{align}
In this formula we abbreviate 
$A_{3,4} \equiv A_{3,4}(m_{f_{u}}^{2},m_{f_{d}}^{2})$ and 
$\mathcal{I}_{1,2} \equiv \mathcal{I}_{1,2}(m_{f_{u}}^{2},m_{f_{d}}^{2})$, 
with 
\begin{subequations}
\begin{align}
\mathcal{I}_{1} &= \int_{0}^{1} dx \int_{0}^{1-x} dy \, 
\frac{-(x+y) \Delta m^{2}  + m_{f_{u}}^{2}}{x(x-1)m_{W}^{2} + y(y-1)m_{W}^{2}
-xy(s-2m_{W}^{2}) -(x+y) \Delta m^{2} + m_{f_{u}}^{2}}\;,  \\[3mm]
\mathcal{I}_{2} &= \int_{0}^{1} dx \int_{0}^{1-x} dy \, 
\frac{2x \,m_{f_{d}}^{2} + (x+y) \Delta m^{2}  - m_{f_{u}}^{2}}{x(x-1)m_{W}^{2} + y(y-1)m_{W}^{2}
-xy(s-2m_{W}^{2}) -(x+y) \Delta m^{2} + m_{f_{u}}^{2}}\;,
\end{align}
\end{subequations}
\end{widetext}
where $\Delta m^{2} \equiv m_{f_{u}}^{2} -m_{f_{d}}^{2}$.
In the limit of heavy masses, $m^{2}= m_{f_{u}}^{2} = m_{f_{d}}^{2} \gg s, m_{W}^{2}$, 
we obtain, 
\begin{eqnarray}
\lim_{m^{2}\to \infty} \mathcal{I}_{1} = \frac{1}{2} \;,\quad 
\lim_{m^{2}\to \infty} \mathcal{I}_{2} = -\frac{1}{6} \;.
\end{eqnarray}
%
%
Lets examine the $\gamma^{*}W^{-}W^{+}$ case first. We must set 
$\beta^{\gamma}_{f_{u,d}} = 0$.  In this case gauge invariance [see \eq{WIsW}] 
implies $w=z$ and CP-invariance $w=-z$, and therefore $w=z=0$. Having
fixed the anomalous term the result for this vertex turns out to be simply, 
\begin{equation}
\Gamma_{\gamma^{*}W^{-}W^{+}}(s)  =   
\frac{1}{2}\, s\sum_{\mathrm{doublets}}
\biggl [ A_{3} (m_{f_{u}}^{2},m_{f_{d}}^{2}) + (f_{u}\leftrightarrow f_{d}) \biggr ]\;,
\label{eq:gWW}
\end{equation}
where $A_{3}$ is a form factor  defined in the Appendix~\ref{app:W}.
We should note here that $\Gamma_{\gamma^{*}W^{-}W^{+}}(s=0) =0$
as it should be~\cite{Hagiwara,Gaemers}, \ie there is no pole at $q^{2}=0$.
This is a  special case where the anomaly term conspires with 
$\mathcal{I}_{1}$-term such that the final result contains no non-decoupling
terms. In order for gauge invariance to be non-anomalous, the last terms in 
the WIs system (\ref{WIsW}), must vanish. This implies a relation among 
fermion charges, 
\begin{equation}
\sum_{f=e,\nu, d,u} \alpha_{f}^{\gamma} = Q_{e} + Q_{\nu} + 3\, Q_{d} + 3\, Q_{u} = 0 \;,
\label{cqc}
\end{equation}
which is exactly the charge conservation condition. 
%
Then, in the asymptotic limit, $s\gg m_{W}^{2}, m_{f_{u,d}}^{2}$, the amplitude 
for $\Gamma_{\gamma^{*}W^{-}W^{+}}(s\to \infty)$
vanishes, thanks to \eq{cqc}. This is obvious from the numerical outcome in 
Fig.~\ref{fig:gWW}. It also shows an enhanced threshold behaviour around 
$\sqrt{s}\approx 2 m_{t}$ (solid line).  
%
\begin{figure}[htbp]
   \centering
   \includegraphics[width=3.1in]{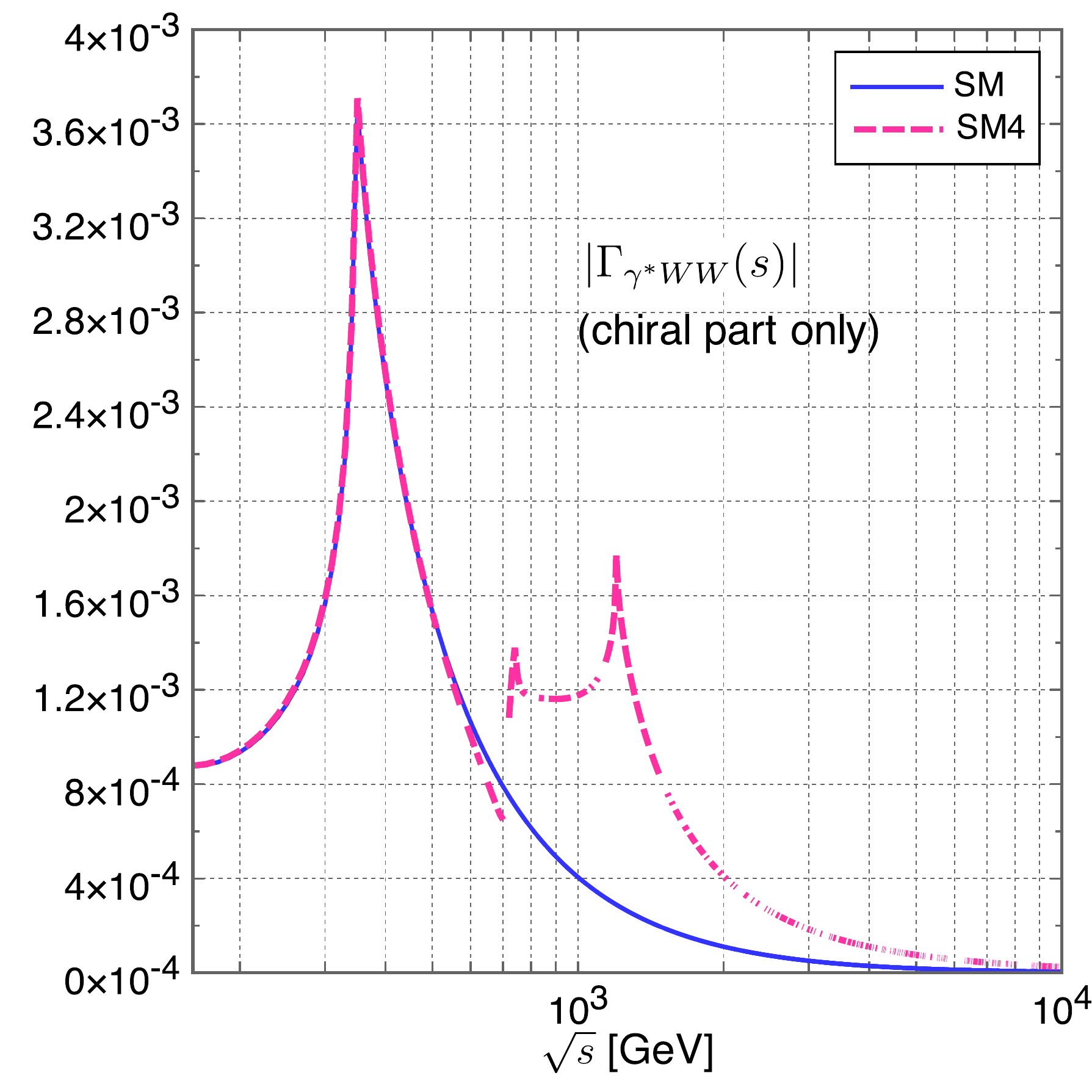} 
   \caption{The effective vertex $|\Gamma_{\gamma^{*}WW}(s)|$ in the minimal SM (solid line) 
  and in
   SM  with an extra fourth fermion  generation (SM4),  (dashed line).} 
   \label{fig:gWW}
\end{figure}
Quantitatively, this can be seen  from \eq{eq:gWW}
by expanding $A_{3}$ around the threshold. 
%
%
Compared to $\Gamma_{\gamma^{*}ZZ}(s)$, there is an additional 
contribution due to the large mass difference $\Delta m^{2} =m_{t}^{2}-m_{b}^{2}
\approx m_{t}^{2}$, in the numerical factor that multiplies $s/m_{t}^{2}$.
Our evaluation of integrals contains one numerical integration and follows the
procedure of  Appendix B in \Ref{Peskin:delayed}. Our analytic
formulae in Appendix~\ref{app:int}, 
at the limit of $m_{W}=0$, are in full agreement with these results.   
Few representative values are,
\begin{eqnarray*}
\Gamma_{\gamma^{*}WW}(\sqrt{s}=200~\mathrm{GeV}) &=& 
(6.8- 6.4\, i)\times 10^{-4} \;, \\
\Gamma_{\gamma^{*}WW}(\sqrt{s}=500~\mathrm{GeV}) 
&=& (-1.5 +  15\, i)\times 10^{-4} \;.
\end{eqnarray*} 
Comparing with $\gamma^{*}ZZ$ vertex we see here that the mass
splitting generates a sizeable absorptive part that dominates the 
vertex after $\sqrt{s}\gtrsim 2 m_{W}$.  

\begin{figure}[htbp]
   \centering
   \includegraphics[width=3.1in]{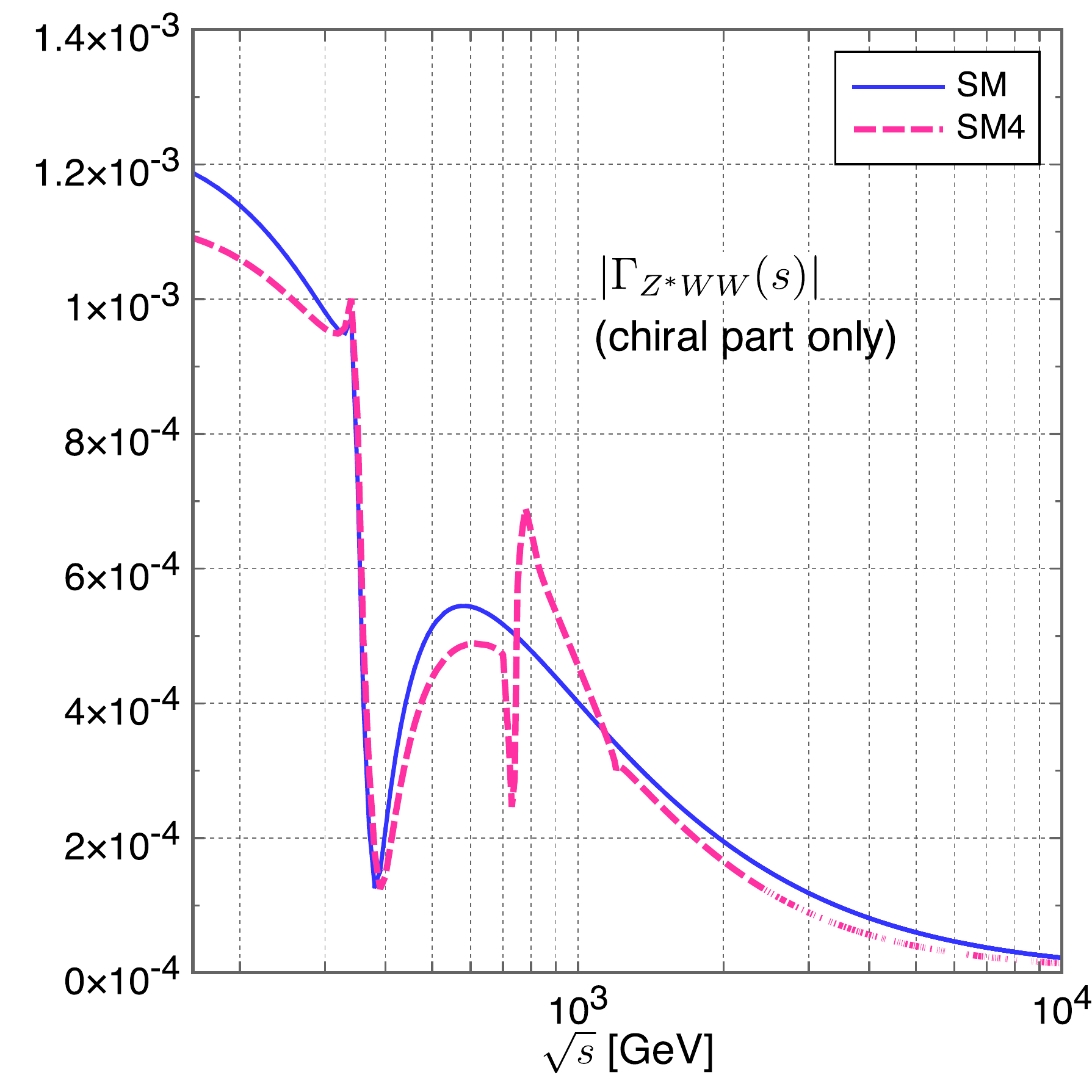} 
   \caption{The effective vertex $|\Gamma_{Z^{*}WW}(s)|$ in the minimal SM (solid line) 
  and in
   SM  with an extra fourth fermion  generation (SM4), (dashed line).} 
   \label{fig:ZWW}
\end{figure}

We now turn to the $Z^{*}W^{-}W^{+}$ vertex. 
This time we have only CP-symmetry at our disposal which 
sets  only the constraint $w=-z$. 
At the broken limit there is no other symmetry remaining  
in order to fix  the parameter $w$ alone.
However, in the exact $SU(2)$-limit, where 
$[g', s_{w}\to 0, \alpha_{f}=-\beta_{f}]$,  this vertex
should be exactly the same as the $Z^{*}ZZ$-vertex. There, the arbitrary 
parameters are fixed by Bose symmetry to be $w=-z=1/3$.   
For this choice of $w$ and at the heavy mass limit,  
$m^{2}= m_{f_{u}}^{2} = m_{f_{d}}^{2} \gg s, m_{W}^{2}$, 
the vertex is proportional to $\alpha_{f}+\beta_{f} \propto s_{w}^{2}$,
 for every 
fermion contribution, which in turn is proportional to $SU(2)$-breaking effects.
Another, equally good, choice would be $w=0$, for example. 
The physical  requirement here is the decoupling of a particle from the 
$\Gamma_{Z^{*}WW}$-vertex.

In conclusion, the $Z^{*}WW$ vertex is \emph{undetermined}: there is only
CP-symmetry, that is not enough to fix
two arbitrary parameters. However, for the anomaly-free SM this
arbitrariness is irrelevant since it is cancelled when the whole 
fermion contribution is taken into account.  We shall meet this 
situation again in the $Z'VV$-vertex below.

Our numerical evaluation of the  SM $|\Gamma_{Z^{*}WW}(s)|$ is shown 
in Fig.~\ref{fig:ZWW}. This time, the top quark threshold destructively adds
to the vertex.  As in previous cases, we present few representative values,  
\begin{eqnarray}
\Gamma_{Z^{*}WW}(\sqrt{s}=200~\mathrm{GeV}) &=& 
-(8.5+7.6\, i)\times 10^{-4} \;,\nonumber \\
\Gamma_{Z^{*}WW}(\sqrt{s}=500~\mathrm{GeV}) 
&=& -(3.8 +  3.5\, i)\times 10^{-4} \;, \nonumber
\end{eqnarray} 
that show similar order of magnitude  values for the real part 
as in the  $Z^{*}ZZ$ vertex but an enhanced absorptive part. The latter
is due to custodial symmetry breaking effects \ie the large mass
difference between the top and the bottom quarks.   
Although there is an intense experimental ongoing analyses at 
LEP~\cite{Abdallah:2010zj}, Tevatron~\cite{Abazov:2011rk} and 
LHC~\cite{Giraud:2012ja,Martelli:2012ea} for the first 
three CP-invariant  non-chiral
operators, $f_{i=1..3}^{V}$ of \eq{eq:f123V}, we are not aware of a similar
experimental  search on  the chiral $f_{5}^{V}$ of \eq{eq:f5V}.  

 \subsection{Models with a sequential fourth fermion generation}

In our first departure from the SM we assume a fourth generation matter
of quarks and leptons.
Apart from the fact that the 4th generation neutrino has to weight more
than 45 GeV, a certain tuning to avoid EW constraints is needed.
More specifically, one extra doublet of degenerate 
leptons contributes a piece 
of approximately  $1/6\pi \approx 0.05$ into the 
S-parameter~\cite{PeskinTakeuchi} while the current 
fit~\cite{Baak:2011ze} to the EW data gives,
\begin{equation}
S \ = \ 0.04 \pm 0.10 \;.
\end{equation}  
Therefore, a 4th, mass degenerate, fermion 
generation will contribute a $4/6\pi \approx 0.2$ piece 
to $S$-parameter which is incompatible with the fit. A certain 
mass difference or else a certain weak isospin violation is needed
which is parameterized by the $T$ parameter~\cite{PeskinTakeuchi}.  
A consistent parameter space with EW precision data and published 
direct searches is 
\begin{eqnarray} 
 m_{\nu4}&=& 400~\rm{GeV}\;, \quad m_{e4}= 660~\rm{GeV}\;, 
 \nonumber \\ 
 m_{t4} &=& 358~\rm{GeV}\;, \quad  m_{b4}=372~\rm{GeV}\;.
 \label{4genmasses}
 \end{eqnarray}
This mass
 spectrum corresponds to Tevatron experiments allowed region, where the 
analyses from CDF~\cite{Aaltonen:2011vr} have excluded 
$t_{4}$ and $b_{4}$ quarks to 
have masses smaller than the values quoted above\footnote{Currently,
the sequential 4th generation is under siege from LHC~\cite{Peskin:2011zz}. 
If there exist
new heavy SM type quarks, they will contribute a factor of up to $N_{c}^{2}=9$ into the 
Higgs production cross section for the (triangle) process $gg \to H$. 
The current cross section sensitivity  at the LHC is within a few of the 
SM prediction and therefore it sets an indirect
bound over the whole exclusion Higgs area,
up to 550-600 GeV. Other direct bounds from the LHC 
on 4th generation top and bottom quarks involve assumptions
about their mass difference to be smaller than the W-mass. These caveats  
are discussed in some detail with complete references in \Ref{Wingerter:2011dk}.}.
The 
leptons mass spectrum is chosen such that it does not 
contribute significantly to the oblique parameters, e.g., for these 
values of lepton masses one has $\Delta S_{l}\simeq 0$~\cite{Baak:2011ze}. 

Due to the fact that
the charges are the same as in the SM,  the anomalies are canceled
in each generation. It is important to notice here 
that if all  the extra fermions were 
very heavy and had the same mass, 
no effect would be left back and the decoupling would work perfectly.
The reason is, first of all, that 
the sum over all extra fermions of expressions 
that contain the finite integrals 
$A_{3}$,$A_{4}$ or $A_{5}$ vanishes because the 
integrand factors out  a term $\sum_{f} c_{f}$,
where $c_{f}$ is the pre-anomaly factor 
of each fermion. But this sum is equal to zero for
 an anomaly-free generation.
On the other hand, terms proportional 
to $I_{1}$ or $I_{2}$ in \eq{A12},
in the limit of large fermion mass, are canceled exactly by
the anomalous term for special values of $w$ and $z$ parameters 
that are fixed by the Bose symmetry in each case. 
But this constraint is not necessary, e.g., if an
anomaly-free generation of very heavy mass degenerate  
chiral fermions 
is added to the SM,
it has no effects at low energies, no 
matter what the values of $w$ and $z$ are.
This is guaranteed by the fact that the extra
generation is anomaly-free. 
 
 The numerical 
analysis for the three gauge bosons vertices
is the same as previously. Using the approximate integral 
expressions from Appendix~\ref{app:int}, we  
draw plots for the amplitudes $|\Gamma_{V^{*}VV}(s)|$
and $|\Gamma_{V^{*}WW}(s)|$
versus $\sqrt{s}$ in  different 
combinations of the external gauge
bosons $V=\gamma, Z$. These plots are collected in Fig.~\ref{fig:gammas}, and
Figs.~\ref{fig:gWW},\ref{fig:ZWW}, respectively [dashed line].

The extra generation has a significant contribution to $\Gamma$'s, 
in the region near  twice the threshold of each extra fermion, where
the amplitude rises until those values 
(shown as peaks in every combination of external gauge bosons) and  drops fast as 
$1/s$ (apart from $V^{*}\gamma\gamma$ which drops as $1/s^{2}$). 
We see that for small values of energy the two curves (the
curve that corresponds to the SM case and the curve that
corresponds both to the SM and the 4th generation) have the same
form. In this energetic region ($\sqrt{s} \lesssim\,600\,{\rm GeV}$) the
dominant feature is the first peak that corresponds to the
threshold energy for the creation of the top quark
$(\sqrt{s}\approx 350\,\mathrm{GeV}\approx 2\,m_{t})$. 
In addition,
the contribution from the extra fermionic generation is negligible,
because all the extra fermions are heavy compared to the
energy, \ie $(2\,m_{f}>\sqrt{s})$. 
%
 These extra fermions have more or less similar masses.
As before with the top-quark mass, there is a cancellation
between the anomaly contributions and the $I_{1,2}$ parts
of the amplitude for each fermion separately.
 As a result, the total contribution from
the fourth generation is negligible as we can see from
Fig.~\ref{fig:gammas}.

 The situation is
different when $\sqrt{s}$ runs over the mass spectrum of the extra
fermionic generation. Firstly for $(\sqrt{s} \gtrsim \,600\,\mathrm{GeV})$ we see
different peaks that correspond to the threshold energy for the
creation of the extra fermions $(\sqrt{s}\approx 2\,m_{i})$. When
$(2\,m_{i}<\sqrt{s}<2\,m_{j})$, there is a non-zero contribution
to the total amplitude. In this case,  fermions whose masses are very
heavy compared to $\sqrt{s}$, exhibit the same behaviour as
previously \ie the anomalous term cancels out against the finite contribution.

Reading our results from Fig.~\ref{fig:gammas}, 
the best case  for
observing triple gauge boson vertex is $\gamma^{*}\gamma Z$
where $h_{3}^{\gamma}(\sqrt{s} = 500~\mathrm{GeV}) \approx 10^{-4}$.   
This is by two orders of magnitude below  the expected
 LHC sensitivity (with $\Lambda \sim 1$ TeV)~\cite{Baur}.  




\subsection{Minimal $Z'$ models}
\label{sec:zprime}

Grand Unified Theories (GUTs) with 
rank larger than four could break to the SM gauge group times additional
$U(1)'$s : $SU(3)\times SU(2) \times U(1)\times U(1)^{\prime n}$. 
This symmetry is broken down to $U(1)_{em}$ and 
therefore there is a possibility of additional forces mediated by the 
$Z'$ gauge bosons associated with the broken $U(1)'$ symmetries
(for a review see \Ref{Langacker:2008yv}).

We shall concentrate here on minimal  models with one additional
neutral gauge boson, the $Z'$. Minimal here means models that 
contain no-additional \ie no exotic, matter particles apart from
the SM ones and right handed 
neutrinos. The latter play a crucial role in cancelling anomalies due
to the  additional $U(1)'$ and in producing viably small neutrino masses.  
These models were devised first in 
\Ref{Appelquist:2002mw} and later elaborated in
\Refs{Salvioni:2009mt,Salvioni:2009jp}. Following the notation 
of~\cite{Salvioni:2009mt} we can describe these models with three additional parameters: 
the mass of the new gauge boson, $M_{Z'}$,
and the couplings $\gy$ and $\gbl$. The latter enter into the current
which couples to the unmixed $Z_{0}'$ gauge boson as
\begin{equation}
J_{Z_{0}'}^{\mu}  = 
 \sum_{f=f_{L},f_{R}} \left [ \gy Y_{f} + \gbl \, (B-L)_{f} \right ] \bar{f} \gamma^{\mu} f \;.
\end{equation}
From this, it is easy to construct $\mathscr{L}_{int}$ in \eq{lint2}  
with 
\begin{subequations}
\label{zpc}
\begin{align}
\alpha_{f}^{Z} &= \cos\theta' \, \alpha_{f}^{Z_{0}} - \sin\theta'
\, \alpha_{f}^{Z'_{0}} \;, \\
\alpha_{f}^{Z'} &= \sin\theta' \, \alpha_{f}^{Z_{0}} + \cos\theta'
\, \alpha_{f}^{Z'_{0}} \;, \\
\beta_{f}^{Z} &= \cos\theta' \, \beta_{f}^{Z_{0}} - \sin\theta'
\, \beta_{f}^{Z'_{0}} \;, \\
\beta_{f}^{Z'} &= \sin\theta' \, \beta_{f}^{Z_{0}} + \cos\theta'
\, \beta_{f}^{Z'_{0}} \;,
\end{align}
\end{subequations}
where $\theta'$ is the mixing angle between $Z$ and $Z'$ gauge 
bosons given by,
\begin{equation}
\tan\theta' \ = \ -\frac{\gy}{\gz} \:
 \frac{M_{Z_{0}}^{2}}{M_{Z'}^{2}-M_{Z_{0}}^{2}} \;,
\end{equation}
with $M_{Z_{0}}^{2} = g_{Z}^{2} v^{2}/4$  the `SM' $Z$-boson mass. 
Also in \eq{zpc} we obtain for $\alpha_{f}^{Z_{0}'}, \beta_{f}^{Z_{0}'}$,
\begin{align}
\alpha_{u}^{Z_{0}'} &= \frac{1}{2}\, \left (\frac{5}{6}\gy + \frac{2}{3}
\gbl \right )\;, \quad \beta_{u}^{Z_{0}'} = \frac{ \gy}{4}\;, 
\nonumber \\
\alpha_{d}^{Z_{0}'} &= \frac{1}{2}\, \left (-\frac{1}{6}\gy + \frac{2}{3}
\gbl \right )\;, \quad \beta_{d}^{Z_{0}'} = - \frac{ \gy}{4}\;, 
\nonumber \\
\alpha_{e}^{Z_{0}'} &= \frac{1}{2}\, \left (-\frac{3}{2}\gy - 2
\gbl \right )\;, \quad \beta_{e}^{Z_{0}'} = -\frac{\gy}{4}\;, 
\nonumber \\
\alpha_{\nu}^{Z_{0}'} &= \frac{1}{2}\, \left (-\frac{1}{2}\gy - 
2 \gbl \right )\;, \quad \beta_{\nu}^{Z_{0}'} =  \frac{\gy}{4}\;,
\label{as}
\end{align}
while the corresponding expressions for $\alpha_{f}^{Z_{0}}, \beta_{f}^{Z_{0}}$
are given by \eq{SM:assi}.
This parameterisation through $\gy$ and $\gbl$
helps us to very easily  incorporate several models that
have been studied in the literature: $Z_{B-L}$ when the $U(1)_{B-L}$
charges of the SM fermions are proportional to $(B-L)$ quantum numbers,
$Z_{\chi}$ a GUT inspired $SO(10) \to SU(5)\times U(1)_{\chi}$ model
and finally, $Z_{3R}$ where the corresponding $U(1)_{3R}$ charges 
are proportional to $T_{3R}$ generator 
of the global $SU(2)_{R}$ symmetry.  We summarise the couplings
of these models in the following table:

\begin{table}[h]
\begin{tabular}{|c|c|c|c|}
\hline
              &  $Z_{B-L}$  &  $Z_{\chi}$ & $Z_{3R}$ \\ \hline
 $\gy$   &  0     & $-\frac{2}{\sqrt{10}} \gzp$  
 & $- \gzp $ \\ \hline
 $\gbl$ & $\sqrt{\frac{3}{8}} \gzp$ & $\frac{5}{2\sqrt{10}} \gzp$ & $\frac{1}{2} \gzp$ \\ \hline
 \end{tabular}
 \end{table}
 %
 
Here, we wish to calculate the effective vertices
 $\Gamma_{Z^{'*}\gamma Z}$ and 
$\Gamma_{Z^{'*} Z Z}$ for those models. Recalling \eqs{VgZ}{VZZ}
with $i=Z',\, j=\gamma \mathrm{~or~} Z$ and $k=Z$ respectively, 
we obtain
\begin{widetext}
\begin{align}
\Gamma^{\mu\nu\rho}_{Z^{'*}\gamma Z}(s) &\approx
 \epsilon^{\mu\nu\rho\sigma} k_{1\sigma}\, 
 \sum_{f=u,d,e,\nu} \left [ m_{Z}^{2} (A_{3 f} 
+ A_{5 f }) 
- \frac{m_{f}^{2} \beta_{f}^{Z}}{\pi^{2}} \, I_{2f} 
 +   \frac{(z+1)}{4\pi^{2}}\,(\alpha_{f}^{Z'}\beta_{f}^{Z}+
 \alpha_{f}^{Z}\beta_{f}^{Z'})\,\alpha_{f}^{\gamma}  \right ] \nonumber \\[2mm]
&\equiv   \epsilon^{\mu\nu\rho\sigma} k_{1\sigma}\, \Gamma_{Z^{'*}\gamma Z}(s)
 \;, 
  \label{ZpgZ}
  \\[3mm]
\Gamma^{\mu\nu\rho}_{Z^{'*}ZZ}(s)  &=
 \epsilon^{\mu\nu\rho\sigma} (k_{1}-k_{2})_{\sigma}
 \sum_{f=u,d,e,\nu} \left [ m_{Z}^{2} (A_{3 f} 
- A_{4 f }) 
+ \frac{m_{f}^{2} \beta_{f}^{Z}}{\pi^{2}} \, I_{1f} 
 -  \frac{(w-1)}{4\pi^{2}} \,[
  (\alpha_{f}^{Z})^{2}\beta_{f}^{Z'}+
 (\beta_{f}^{Z})^{2}\beta_{f}^{Z'}+ 2\, \alpha_{f}^{Z'}\alpha_{f}^{Z}
 \beta_{f}^{Z}]  \right ] \nonumber \\[2mm]
 &\equiv \epsilon^{\mu\nu\rho\sigma} (k_{1}-k_{2})_{\sigma}  \,
 \Gamma_{Z^{'*}ZZ}(s) \;,
  \label{ZpZZ}
\end{align}
\end{widetext}
\begin{figure*}[ht!]
       \centering
        \subfloat[][]{
            \label{fig:Zpa}
            \includegraphics[width=0.4\textwidth]{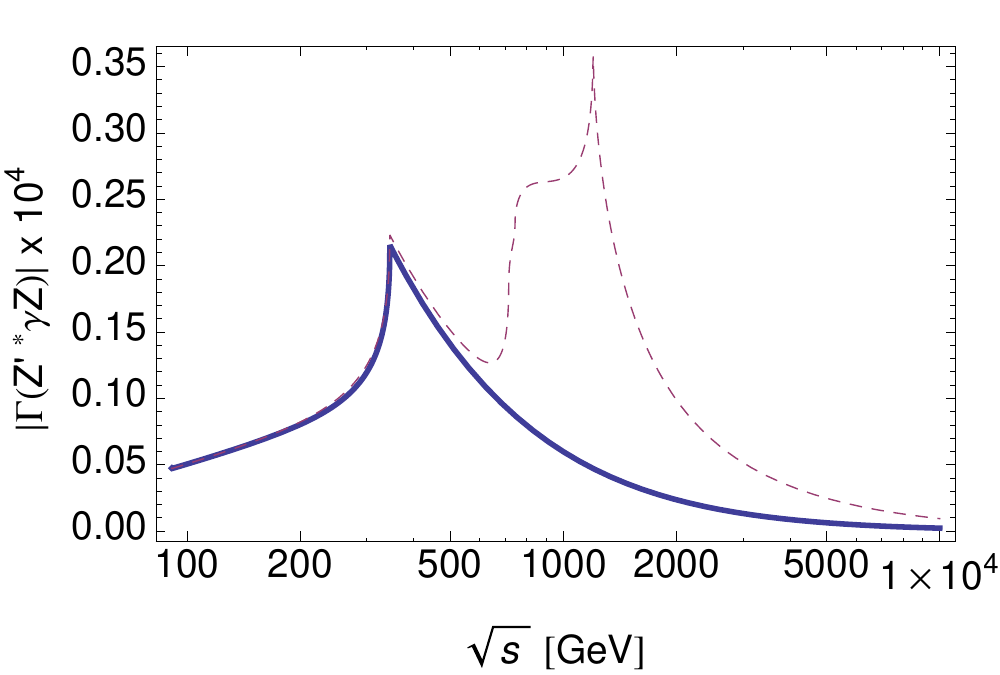}
        }
        \quad
        \subfloat[][]{
           \label{fig:Zpb}
           \includegraphics[width=0.4\textwidth]{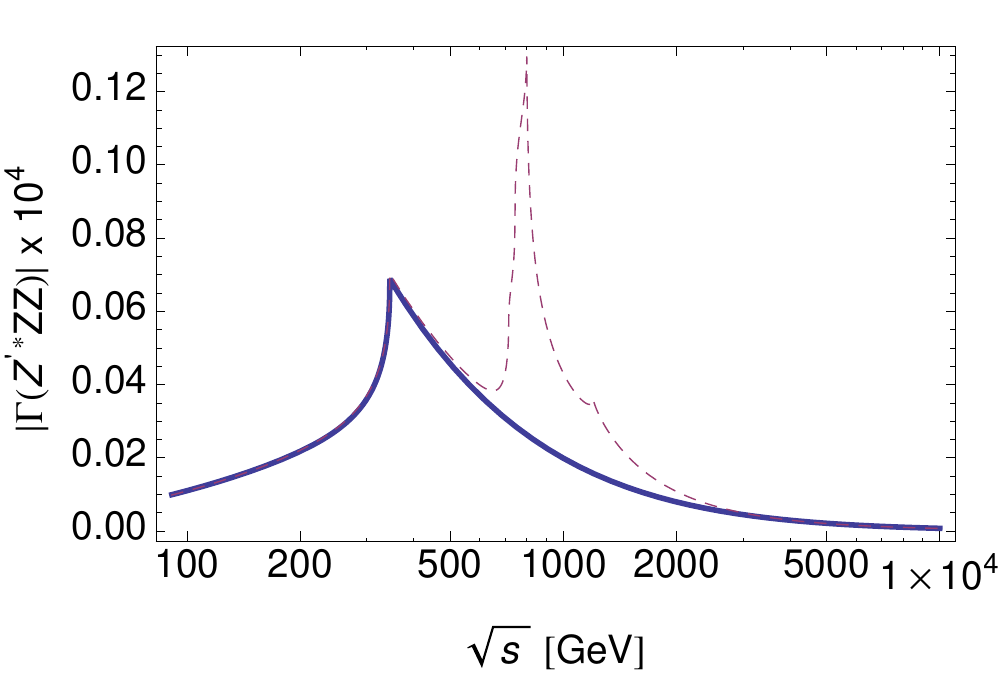}
        }
       \quad
        \subfloat[][]{
            \label{fig:Zpc}
            \includegraphics[width=0.4\textwidth]{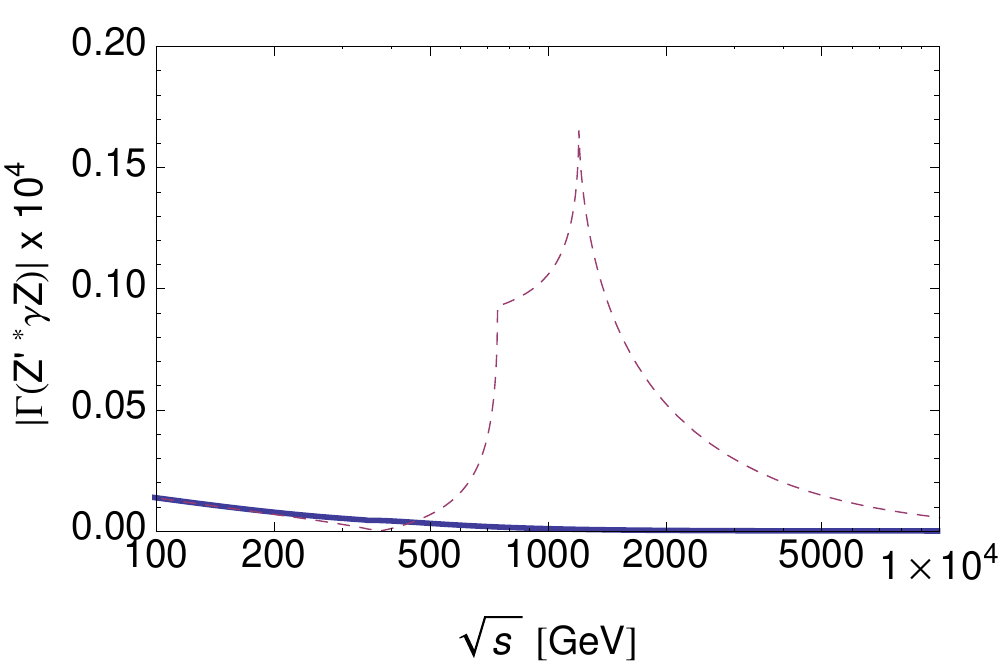}
        } 
        %
        \subfloat[][]{%
            \label{fig:Zpd}
            \includegraphics[width=0.4\textwidth]{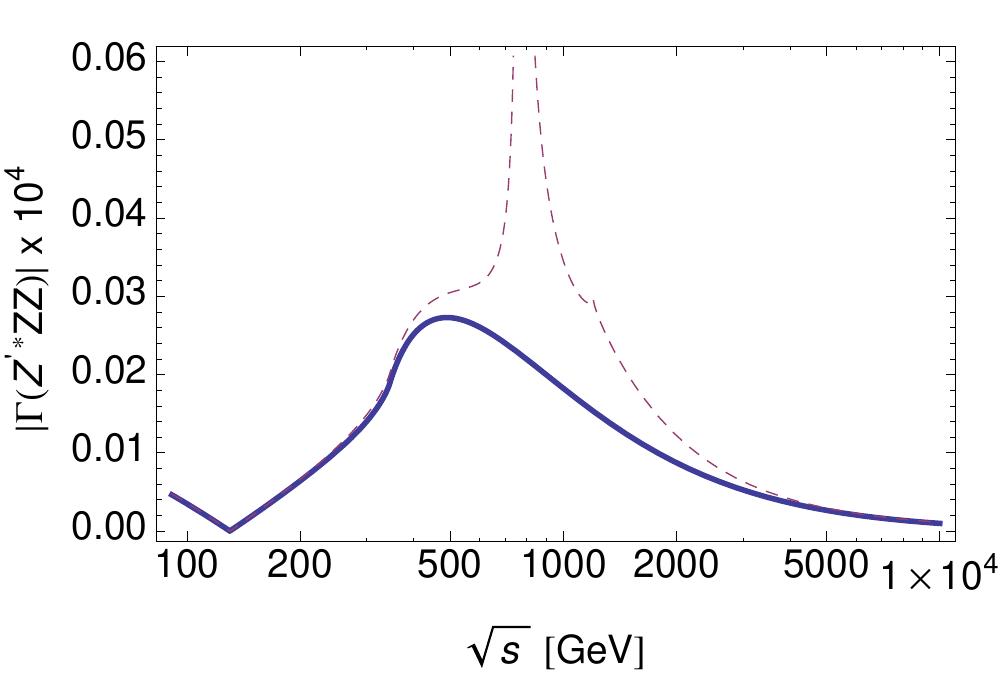}
        }%
         \quad
        \subfloat[][]{
            \label{fig:Zpe}
            \includegraphics[width=0.4\textwidth]{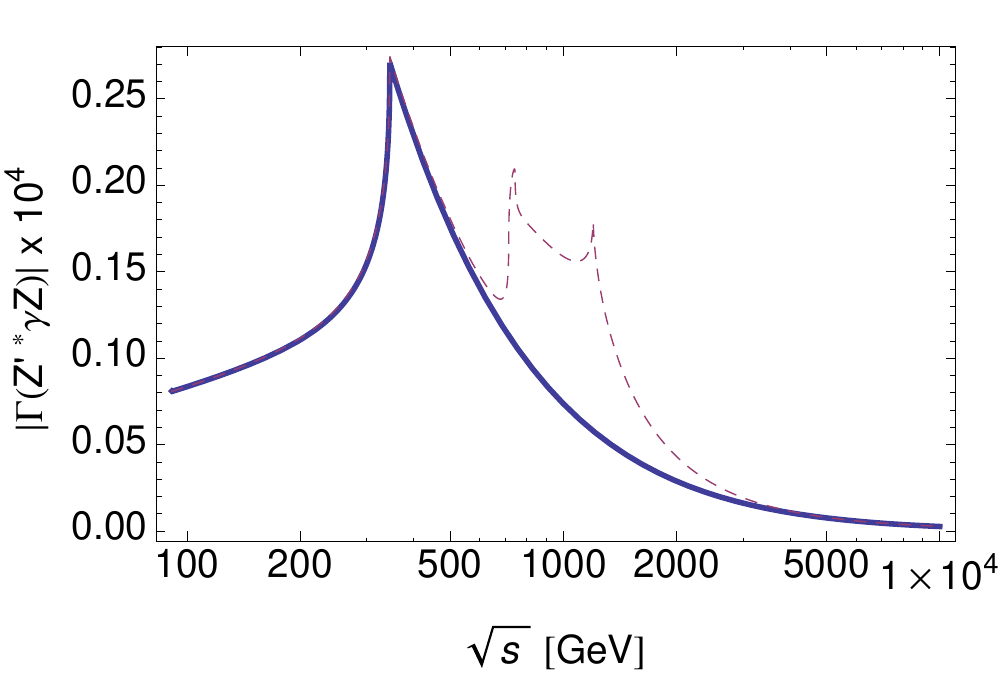}
        } 
        \subfloat[][]{%
            \label{fig:Zpf}
            \includegraphics[width=0.4\textwidth]{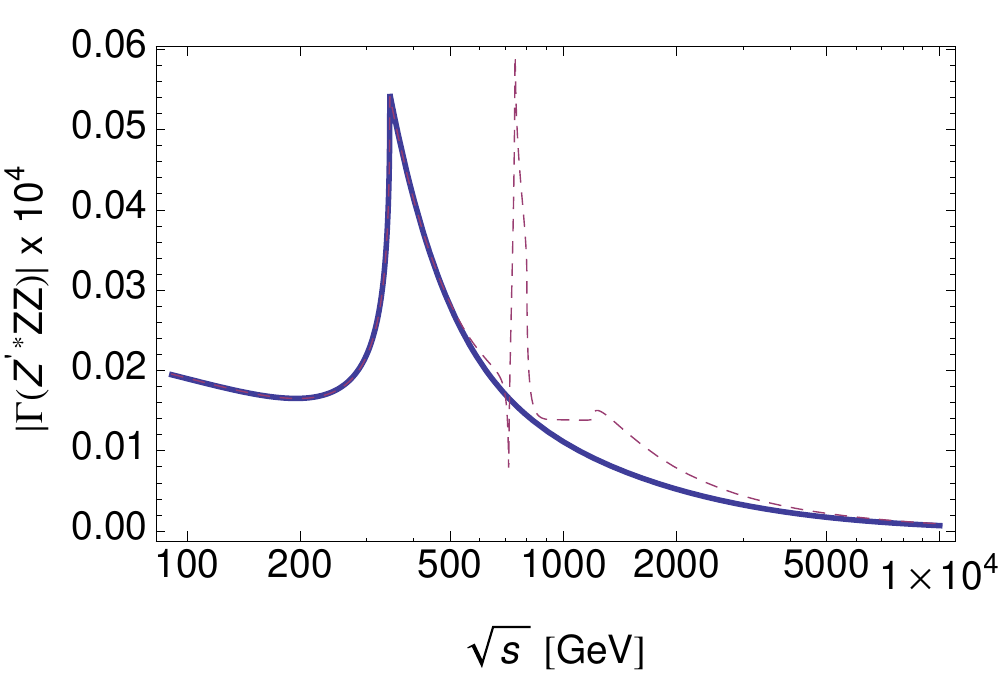}
        }%
    \caption{a,b) 
        $\mid\Gamma_{Z'VV}(s)\mid$ versus $\sqrt{s}$ for different
 gauge bosons combinations as they are given by 
 \eqs{ZpgZ}{ZpZZ}. The solid curve corresponds
to the SM spectrum with an extra $U(1)_{B-L}$,
  while the dashed curve corresponds to the same
   but with a  4th sequential fermion 
generation added as in Fig.~\ref{fig:gammas}. 
We take $M_{Z'}= 1~\mathrm{TeV}$ and $\gzp = \alpha_{\mathrm{em}}$.
c,d) The same as (a,b) but with $U(1)_{\chi}$. (e,f) The same as (a,b) but with 
$U(1)_{3R}$.}%
 \label{fig:Zpgammas}
\end{figure*}
with $\alpha_{f}$, and  $\beta_{f}$ given in \eqs{zpc}{as}.
Again the last terms on the r.h.s of \eqs{ZpgZ}{ZpZZ} 
arrive from the chiral anomaly of individual fermion contributions.
These anomalous terms cancel out when we sum over 
all SM fermions (here we also need the right handed neutrino).
This also removes the arbitrariness due to the unknown parameters
$w,z$. Contrary to the SM vertices, 
we cannot use here any physical arguments in order to remove completely 
both $w$ and $z$ parameters. We only have  $U(1)_{em}$
gauge invariance for $Z^{'*}\gamma Z$ and  Bose symmetry for $Z^{'*}ZZ$
while in the SM we 
have two neutral gauge bosons 
and two symmetries. 
But lets for the moment keep the anomalous terms. 
Obviously they are multiplied by  arbitrary parameters 
$(z+1)$ (for $Z^{'*}\gamma Z$) and $(w-1)$ (for $Z^{'*}ZZ$). 
Focusing on the $Z'_{B-L}$ model, where the mixing angle $\theta'$
vanishes, we observe that for any single heavy fermion contribution 
the 2nd and the 3rd term on the r.h.s of \eqs{ZpgZ}{ZpZZ}  
mutually cancel and what remains back is the effective theory
with the low mass fermion contributions but \emph{together with}
their anomalous terms included. The latter do not depend on particle masses. 
The choices for the arbitrary parameters are 
$w=z=1$ for $Z'\gamma Z$ and $w=z=0$ for $Z'ZZ$. 
The last condition can be interpreted as follows:
for the amplitude $ZZ\to ZZ$ to hold 
for asymptotic values of energies, 
  \eq{eq:asym} requires $w=z$  but  Bose symmetry requires $w=-z$. 
This conclusion does not stand firm in the case of mixing between
$Z$ and $Z'$ \ie in models $Z_{\chi}, Z_{3R}$ of the table above,
and the contribution of a heavy mass particle
is undetermined. Of course anomalies do cancel when
\emph{all} model fermions are added.  

In Fig.~\ref{fig:Zpgammas} we display numerical results 
for the absolute value of the scalar part of the 
1PI effective vertices $Z^{'*}\gamma Z$ and $Z^{'*}ZZ$
in \eqs{ZpgZ}{ZpZZ} for $M_{Z'}= 1~\mathrm{TeV}$ and
$\gzp = \alpha_{em}$. Figs.~(\ref{fig:Zpgammas}a,b) refer
to $Z_{B-L}$ model, Figs.~(\ref{fig:Zpgammas}c,d) to $Z_{\chi}$ models
and, finally, Figs.~(\ref{fig:Zpgammas}e,f) to $Z_{3R}$ models.
For the values of $M_{Z'}$ and $\gzp$ chosen, 
fits to electroweak observables and direct searches are satisfied. We also present results
when adding a sequential 4th generation 
of fermions with the same masses (and the reasoning) as we did for the SM case of 
section~\ref{sec:sm}. We observe that there is an enhancement 
of the vertices by a factor of  2 for $Z_{B-L}$, and a factor of 10-15
for $Z_{\chi}$. Numerically, we can define analogous quantities 
$h_{3}^{Z'}$ and $f_{5}^{Z'}$ by simply replacing $Z$ with $Z'$
in the definition given by footnote~5. As an example, for the $B-L$ model
we obtain,
\begin{align}
h_{3}^{Z'}(\sqrt{s}= 200~{\rm GeV}) &= -2.7 \times 10^{-5}\;, \nonumber \\
h_{3}^{Z'}(\sqrt{s}= 500~{\rm GeV}) &= (-2.7+5.3 i) \times 10^{-4}\;, \nonumber \\
f_{5}^{Z'}(\sqrt{s}= 200~{\rm GeV}) &= -7.2 \times 10^{-6}\;,
\nonumber \\
f_{5}^{Z'}(\sqrt{s}= 500~{\rm GeV}) &= (-7.7+18 i) \times 10^{-5}\;.
\end{align}
Numerical results for the vertices presented above
 and in Fig.~\ref{fig:Zpgammas} are based on
 various analytical approximations for form factors
 described in Appendix~\ref{app:int}. 

Now that $Z'$ can be heavy it is interesting to study
its decay width into $Z\gamma$ and $ZZ$ modes.
Based on (\ref{lint}) and on \eqs{ZpgZ}{ZpZZ} the decay 
widths of the $Z'$ can be read from 
\begin{widetext}
\begin{align}
 \Gamma(Z'\rightarrow \gamma Z) 
 &=\frac{1}{48 \pi}\left |  \sum_{f=u,d,e,\nu} \left [ m_{Z}^{2} (A_{3 f} 
+ A_{5 f }) 
- \frac{m_{f}^{2} \beta_{f}^{Z}}{\pi^{2}} \, I_{2f} 
 +   \frac{(z+1)}{4\pi^{2}}\,(\alpha_{f}^{Z'}\beta_{f}^{Z}+
 \alpha_{f}^{Z}\beta_{f}^{Z'})\,\alpha_{f}^{\gamma}  \right ] \right |^{2}
\nonumber \\
&\times \frac{m_{Z'}^{3}}{m_{Z}^{2}}\,
(1-\frac{m_{Z}^{2}}{m_{Z'}^{2}})^{3}\, (1+\frac{m_{Z}^{2}}{m_{Z'}^{2}})\;,
\\[3mm]
 \Gamma(Z'\rightarrow Z Z) &=\frac{1}{96 \pi}\left |
\sum_{f=u,d,e,\nu} \left [ m_{Z}^{2} (A_{3 f} 
- A_{4 f }) 
+ \frac{m_{f}^{2} \beta_{f}^{Z}}{\pi^{2}} \, I_{1f} 
 -  \frac{(w-1)}{4\pi^{2}} \,[
  (\alpha_{f}^{Z})^{2}\beta_{f}^{Z'}+
 (\beta_{f}^{Z})^{2}\beta_{f}^{Z'}+ 2\, \alpha_{f}^{Z'}\alpha_{f}^{Z}
 \beta_{f}^{Z}]  \right ]
\right |^{2}
\nonumber \\
&\times \frac{m_{Z'}^{3}}{m_{Z}^{2}} \, (1-\frac{4
\,m_{Z}^{2}}{m_{Z'}^{2}})^{5/2} \;,
\\[3mm]
\Gamma(Z'\rightarrow W^{+} W^{-} ) &=\frac{\alpha_{em}\,  m_{Z'}\,
\sin^{2}{\theta'}}{48 \tan^{2}{\theta_{w}}}\, \left (1- 
4\, \frac{m_{W}^{2}}{m_{Z'}^{2}} \right )^{3/2}\, 
\left [1 + 20\, \frac{m_{W}^{2}}{m_{Z'}^{2}} + 
12 \, \frac{m_{W}^{4}}{m_{Z'}^{4}} \right ]  \; \left ( 
\frac{m_{W}^{2}}{m_{Z'}^{2}} \right )^{-2}\;,\label{wZpWW}
\\[3mm]
\Gamma(Z'\rightarrow\bar{f}f) &=\frac{N_{c}\, m_{Z'}}{12\,
\pi}\bigg[(\alpha_{f}^{Z'\,2}+\beta_{f}^{Z'\,2})-\frac{3
m_{f}^{2}}{m_{Z'}^{2}}(\alpha_{f}^{Z'\,2}-\beta_{f}^{Z'\,2})\bigg] \; \sqrt{1-\frac{4
m_{f}^{2}}{m_{Z'}^{2}}}\;,
\end{align}
\end{widetext}
where $N_{c}$ is the color factor (3 for
quarks and 1 for leptons)
and the tree level decay width for $Z' \to WW$ has been taken 
from~\Ref{Deshpande:1988py} and is dominant over the loop-induced 
ones. For $\gzp= \alpha_{em}$, $M_{Z'}= 1~\mathrm{TeV}$
and SM spectrum with three generations
we obtain for the $B-L$ ($\chi$) [$3R$] models:
\begin{align}
\mathrm{Br}(Z'\to \nu\nu) &=37.7\: (42.3) \: [12.5]\; \% \;, \nonumber \\
\mathrm{Br}(Z'\to \ell\ell) &=37.7 \: (12.5)\: [12.6]\; \% \;, \nonumber \\ 
\mathrm{Br}(Z'\to qq) &=24.5 \: (45.1)\: [74.8] \; \% \;, \nonumber \\
\mathrm{Br}(Z' \to W W) &= 0.03\: (3.2)\:[8.1] \times 10^{-5}  \;, \label{ZpBrs}\\
\mathrm{Br}(Z' \to Z \gamma) &= 5.8\: 
(\sim10^{-3})\: [8.7]\times 10^{-6}  \;, \nonumber \\
\mathrm{Br}(Z' \to Z Z) &= 3.0\: (2.5)\: [0.9] \times 10^{-7} \;. \nonumber
\end{align}
These results are pretty much the same for bigger  $M_{Z'}$ values. 
As we see, the  branching fraction for $Z'\to \gamma Z$ is in the 
region of $10^{-5}-10^{-6}$ while for $Z'\to Z Z$ in the region 
$\sim 10^{-7}$. These are very challenging 
numbers even for LHC@14 TeV.

In  coordinate space representation,
the vertices (\ref{ZpgZ}) and (\ref{ZpZZ}) arise on-shell from  the following operators
\begin{align}
\mathcal{O}_{Z'\gamma Z} &\sim \varepsilon^{\mu\nu\rho\sigma} Z'_{\mu} Z_{\nu} \: F_{\rho\sigma}\;, \\
\mathcal{O}_{Z' ZZ} &\sim \varepsilon^{\mu\nu\rho\sigma} Z'_{\mu}  Z_{\nu} \: \partial_{\rho} Z_{\sigma}\;,\label{xZpZZ}
\end{align}
which are both P-odd but CP-invariant. 
Although not present in the SM and in the $Z'$-models under 
consideration there may be P-even but 
CP-violating  operators of the form  
$\mathcal{O}_{Z'ZZ} \sim Z^{'\mu}(\partial^{\nu}Z_{\mu})Z_{\nu}$
induced by a triple scalar loop instead. The latter would interfere
with (\ref{xZpZZ}) and there is a proposal in \Ref{Keung:2008ve} on
how their effects can be separated at the LHC. However, within minimal $Z'$-models
considered here this looks very difficult due to tiny $\mathrm{Br}(Z' \to VV)$ of
\eq{ZpBrs}.







\section{Conclusions}
\label{conc}

We construct an effective 1PI vertex for triple gauge bosons for 
every renormalized theory making explicit mentioning to the 
chiral anomalies and their synergy with heavy fermion
decoupling phenomena. Our method for calculating the vertex is based
on \Ref{Rosenberg}. 
It is quite general and can be divided in four steps:
\begin{enumerate}
\item Write down the most general, Lorentz (and/or possibly other symmetry)
invariant  effective vertex $\Gamma^{\mu\nu\rho}$ [like \eq{rose2}] 
with unknown form factors.
\item Isolate the -potentially- infinite form factors and calculate only the finite
 parts.
\item Derive  Ward Identities arising from the underlying spontaneously broken 
gauge symmetries  at the quantum level. Apply them to  $\Gamma^{\mu\nu\rho}$
and calculate the ambiguous form factors, thus forcing them to be finite. 
\item If the vertex is still undetermined \ie if arbitrary parameters still remain, try 
to fix them by physical requirements.
If nevertheless  arbitrariness persists, then the model needs completion, perhaps
with new particles or new dynamics.   
\end{enumerate} 
This method, explained in detail in Appendix~\ref{sec:app} and in section~\ref{sec:anal},
 does not require dimensional regularisation or other integral 
regularisation technics. It may require, however,  ``shifting momenta'' technics like \eq{eq:A11}. 
The above steps can be augmented with additional relations. 
Instead of WIs, one could use other identities like for example those arising from
perturbative unitarity sum rules or the 
Goldstone boson equivalence theorem e.g., \eq{eq:GBET}. 

All the above steps are realized when calculating triple gauge boson
vertices in spontaneously broken gauge theories, like for example the SM or
its extensions like minimal $Z'$-models. 
The anomalous terms are arbitrary and can only be fixed by physics. Only then
can we discuss non-decoupling effects in the broken limit. We observe that for 
$V^{*}VV, V=\gamma, Z$ and for $\gamma^{*}WW$ vertices, there are two
arbitrary parameters that are
completely determined by two physical symmetries:
$U(1)_{em}$ and Bose symmetry or CP-invariance. 
We find that at the limit of heavy fermion masses,
non-decoupled terms  cancel exactly those that arise from anomalies.
For example, in the SM, decoupling of the top quark
will leave behind anomalous-terms of  light quarks and leptons plus finite
parts. 
On the other hand vertices like  $Z^{*}WW, Z^{'*}VV$ are in general undetermined
because there are no enough symmetries to fix the arbitrary parameters. 
Of course for anomaly-free models this arbitrariness is removed when 
adding up all fermion contributions.

We made a numerical analysis for 
SM and minimal $Z'$-model vertices. To this end, we made an effort 
to calculate finite integrals in terms of standard functions that are
easy to handle. For example in Appendix~\ref{app:int}, we solved  
analytically the integrals
for $V^{*}\gamma V$-vertices. We then proceeded to SM predictions
for the triple gauge boson vertices.
Unfortunately,  it turns out that  within the
SM these are rather small to be discovered 
even at the LHC with $\sqrt{s}=14$ TeV.
Similar results are obtained in the SM extended by a sequential fourth fermion
generation. The difference w.r.t the SM, is that $|\Gamma_{V^{*}VV}(s)|$ is ``delayed''
to vanish for large $\sqrt{s}$ due to the heavy, 4th generation thresholds 
(see Figs.~\ref{fig:gammas}).
In the best case, the SM + 4th generation predicts a maximum of a few$\times 10^{-3}$ 
for $|\Gamma_{\gamma^{*}\gamma Z}|$ [see dashed lines in Figs.~\ref{fig:gammas}].

We have performed a numerical analysis, shown in 
 Fig.~\ref{fig:Zpgammas},  for minimal $Z'$-models 
with $U(1)_{B-L}$ symmetry, SO(10)-like and $U(1)_{3R}$ also extended
with a 4th fermion generation. For a conservative choice of $M_{Z'}= 1$ TeV
and $\gz = \alpha_{em}$, we find $|\Gamma_{Z'ZZ}|$ and 
$|\Gamma_{Z'\gamma Z}|$ in the regime below a few$\times 10^{-5}$.
We also briefly discussed $Z'$-decays to $Z\gamma$ and $ZZ$.  
Adopting the parameters space above, their branching ratio come out
to be in the neighborhood of  $\sim 10^{-5}$ and $\sim 10^{-7}$,
respectively.

In section~\ref{sub-ano} and Appendix~\ref{app:cate},
we calculated non-decoupling effects that arise 
instantaneously with vanishing anomalies. 
We constructed several toy models with two or three external 
gauge bosons and a number of fermions where this situation could 
take place.
In principle, these models can be used as a basis towards realistic
extensions of the SM.

Our main result, the effective triple gauge boson vertex obtained in
section~\ref{sec:anal}  can  be used in various ways: {\emph {i)}} in models with 
anomalous spectrum, {\emph {ii)}} in realistic anomaly driven 
models of section~\ref{sub-ano}, {\emph {iii)}} 
in MSSM and its extensions, {\emph {iv)}} in dark matter or neutrino - nucleon
scattering processes with a photon in the final state. We will pursue some
of these issues in a forthcoming article.

\medskip

\begin{acknowledgments}
We would like to thank I.~Antoniadis, H.~Haber, P.~Kanti, S.~Martin, A.~Pilaftsis,   
and K.~Tamvakis for useful discussions.
This research Project  is co-financed by the European Union -  
European Social Fund (ESF) and National Sources, in the framework of the  
program ``THALIS" of the ``Operational Program Education and Lifelong  
Learning" of the National Strategic Reference Framework (NSRF)  
2007-2013.
K.S. acknowledges full financial support from Greek State
Scholarships Foundation (I.K.Y).
\end{acknowledgments}

\begin{widetext}
\appendix


\section{A Set-Up Toy Model for Calculations }
\label{sec:toy}

Consider a gauge theory of a complex scalar field
$\Phi$ charged under a local $U(1)$ with charge $Y_{\Phi}$ (in units of e),
 a vector spin-1 abelian gauge boson $A_{\mu}$ and a pair of Dirac fermions $E_{L}$ and $e_{R}$ with $U(1)$-charges $Y_{L}$ and $Y_{R}$
 respectively. This gauge theory is described by
 the Lagrangian~\footnote{Throughout we follow the notation and conventions of
 \Ref{Peskin}.},
\begin{eqnarray}
 \mathscr{L}  =  \mathscr{L}_{g}(\Phi,A_{\mu}) +
  \mathscr{L}_{f}(E_{L},e_{R},A_{\mu})+ \mathscr{L}_{Y}(E_{L},e_{R},\Phi)\;, 
  \label{lag}
\end{eqnarray}
 where the gauge boson-scalar interactions are
\begin{eqnarray}
 \mathscr{L}_{g}(\Phi,A_{\mu})=-\frac{1}{4}F_{\mu\nu}F^{\mu\nu} - \frac{1}{2} \: (G)^{2} +
 |D_{\mu}\Phi|^{2}-V(\Phi)\;,  \label{eq:Lg}
 \end{eqnarray}
 while the chiral
fermion  and the Yukawa interaction parts of the Lagrangian
in \eq{lag} are stored in
\begin{eqnarray}
 \mathscr{L}_{f}(E_{L},e_{R},A_{\mu})
&=&\bar{E}_L\: (i\slashchar{D}) \: E_{L}+\bar{e}_R\: (i\slashchar{D}) \:
e_{R} \;,
\\[3mm]
\mathscr{L}_{Y}(E_{L},e_{R},\Phi)
&=&-\lambda_{e} \: (\bar{E}_{L}\, \Phi \, e_{R} \ + \
\bar{e}_{R}\, \Phi^{*} \,  E_{L} ) \;,
\end{eqnarray}
and $D_{\mu}\Phi = \partial_{\mu}\Phi + i e Y_{\Phi} A_{\mu} \Phi$,
 $D_{\mu}E_{L}=\partial_{\mu}E_{L}+i e Y_{L} A_{\mu}E_{L}$,
 and $D_{\mu}e_{R}=\partial_{\mu}e_{R}+i e Y_{R} A_{\mu}e_{R}$.
$\mathscr{L}_{g}$ is invariant under the local, $U(1)$ gauge-transformation
\begin{eqnarray}
&&\Phi(x) \rightarrow e^{i e Y_{\Phi}\Lambda(x)}\Phi(x) \;, \quad A_{\mu}(x) \rightarrow
A_{\mu}(x) - \partial_{\mu} \Lambda(x) \;,  \\[3mm]
&& E_{L}(x) \rightarrow e^{{i e Y_{L}\Lambda(x)}} E_{L}(x)\;, \quad
e_{R}(x) \rightarrow e^{{i e Y_{R}\Lambda(x)}} e_{R}(x)\;, 
 \label{gt1}
\end{eqnarray}
iff $ Y_{\Phi} = Y_{L}-Y_{R}$.
It is convenient to combine the left and right-handed fermions
 into a single Dirac four-component spinor $\Psi=({E_{L}},e_{R})^{T}$.
Then the interaction Lagrangian relevant to our study for triangle graphs
reads:
\begin{eqnarray}
\mathscr{L}_{int} \ = \ -\lambda_{e} \bar{\Psi} \Phi P_{R} \Psi - 
\lambda_{e} \bar{\Psi} \Phi^{*} P_{L} \Psi  
- e A_{\mu} \bar{\Psi} \gamma^{\mu}\, 
(\alpha + \beta\gamma_{5} ) \Psi \;,
\label{frs}
\end{eqnarray}
where 
\begin{eqnarray}
\alpha=\frac{Y_{L}+Y_{R}}{2}\;, \quad \beta=\frac{Y_{R}-Y_{L}}{2}\;.
\label{ab}
  \end{eqnarray}
Under gauge transformations the 4-component field $\Psi$ transforms as
\begin{subequations}
\label{syms}
\begin{align}
\Psi(x) & \rightarrow e^{ie(\alpha+\beta\gamma_{5})
\Lambda(x)} \Psi(x) \;, \\[3mm]
\bar{\Psi}(x) & \rightarrow \bar{\Psi}(x) e^{-ie (\alpha-\beta\gamma_{5})\Lambda(x)} \;,
\end{align}
\end{subequations}
which together with \eq{gt1} leave $\mathscr{L}$ invariant if
$Y_{\Phi}=-2\beta$.

We choose a renormalizable and gauge invariant potential $V(\Phi)$ such
that the field $\Phi$ acquires a non-vanishing vacuum expectation value,
$\langle \Phi \rangle = v/\sqrt{2}$, which breaks
the local $U(1)$ symmetry spontaneously.
We expand \eq{lag} around the minimum, $\Phi
 =\frac{1}{\sqrt{2}}(v+\emph{h}+i\varphi)$ and choose
 a gauge-fixing function in \eq{eq:Lg},
 \begin{eqnarray}
 G = \frac{1}{\sqrt{\xi}} \, (\partial_{\mu} A^{\mu} - \xi e v \varphi )\;,
 \label{gaf}
 \end{eqnarray}
 which eliminates the Goldstone boson - gauge boson mixing term.
 The mass of the vector boson $A_{\mu}$ and of the unphysical
 Goldstone boson $\varphi$ in this $R_{\xi}$-gauge become
 \begin{eqnarray}
 m_{A} = e v Y_{\Phi}\;, \quad m^{2}_{\varphi} = \xi m^{2}_{A}\;.
 \label{mA}
 \end{eqnarray}
%
The ghost part of $\mathscr{L}$ is not relevant to our discussion 
for the one-loop triangle graphs and is not presented.
 In terms of $\Psi$ and $\bar{\Psi}$,
 $\mathscr{L}_{f}+\mathscr{L}_{Y}$ becomes
\begin{eqnarray}
\mathscr{L}_{f}(\Psi,A_{\mu})+\mathscr{L}_{Y}(\Psi,h,\varphi) &=& \bar{\Psi}i\slashchar\partial\Psi-e A_{\mu}\bar{\Psi}\gamma^{\mu}(\alpha+\beta\gamma^{5})\Psi
\nonumber
\\[3mm] &-& m \bar{\Psi}\Psi- \tilde{\beta}\bar{\Psi}\emph{h}\Psi-i
  \tilde{\beta} \bar{\Psi}\gamma^{5}\varphi\Psi \;,
  \end{eqnarray}
where $m =\emph{v}\: \tilde{\beta}$ and 
$\tilde{\beta}=\frac{\lambda_{e}}{\sqrt{2}}$.

This model, albeit very simple, captures the most important
non-decoupling heavy fermion effects in the trilinear gauge boson vertices
in the Standard Model and its 
extensions.  In the context of chiral anomalies it has been exploited
in~\Ref{Gross}. With a light language deform it 
imitates the Standard Model with the difference that its
WI's for the currents corresponding to the  gauge symmetry
in \eq{gt1} are anomalous as we shall see below. 


\section{Calculation of the  Three Point Gauge Boson Vertex}
\label{sec:app}

In this Appendix we explicitly evaluate the three external gauge boson,
fermionic one-loop amplitude of Fig.~\ref{fig:graph}. The loop function is calculated
directly in four dimensions using standard methods studied in
\Refs{Rosenberg,Adler,Jackiw,Weinberg2,DHM}.
Here, we review this calculation  in detail for
the toy model of Appendix~\ref{sec:toy}. 
At the end we generalise our results to the case of three different
external (massive or massless) gauge bosons.

By naive power counting we observe that the two 
diagrams in Fig.~\ref{fig:graph} are linearly divergent.
This means that their quantum amplitudes depend on
the routing of the internal momenta circulating in  the loop.
In each of the two diagrams we shift the internal momenta
with arbitrary four vectors $a^{\mu}$ and $b^{\mu}$, respectively. 
By  reading  Feynman rules from \eq{frs},
the graphs in Fig.~\ref{fig:graph} become
  \begin{eqnarray}
  \Gamma^{\mu\nu\rho}(k_{1},&k_{2}&;a,b) \ =\  (-1) \: e^{3} \times \nonumber \\[3mm]
& &  \mathrm{Tr} \left \{ \int\frac{d^{4}p}{(2 \pi)^4}\frac{\gamma^{\mu}(\alpha+\beta \gamma^{5})(\slashchar{p}-\slashchar{k_{2}}+
  \slashchar{a}+m)\gamma^{\rho}(\alpha+\beta \gamma^{5})(\slashchar{p}+\slashchar{a}+m)\gamma^{\nu}(\alpha+\beta
  \gamma^{5})(\slashchar{p}+\slashchar{k_{1}}+\slashchar{a}+m)}{[(p-k_{2}+a)^{2}-m^{2}][(p+a)^{2}-m^{2}][(p+k_{1}+a)^{2}-m^{2}]} \right.
  \nonumber\\[3mm]
  &+& \left. \int\frac{d^{4}p}{(2 \pi)^4}\frac{\gamma^{\mu}(\alpha+\beta \gamma^{5})(\slashchar{p}-\slashchar{k_{1}}+\slashchar{b}+m)\gamma^{\nu}(\alpha+\beta \gamma^{5})(\slashchar{p}+\slashchar{b}+m)\gamma^{\rho}(\alpha+\beta
  \gamma^{5})(\slashchar{p}+\slashchar{k_{2}}+\slashchar{b}+m)}{[(p-k_{1}+b)^{2}-m^{2}][(p+b)^{2}-m^{2}][(p+k_{2}+b)^{2}-m^{2}]} \right \}\;,
  \nonumber \\[3mm]
  \label{gmnr}
   \end{eqnarray}
 where $m$ is the  fermion mass  and (-1) is a fermionic loop factor.
 The  integral in the second line  is
  the same as the first with only the difference that the upper two
  external legs in Fig.\ref{fig:graph} are interchanged, \ie $\{\nu,\rho \} \leftrightarrow \{\rho,\nu \}$ and $k_{1}\leftrightarrow k_{2}$. Dimensional
  regularization is a scheme not well suited in calculating (\ref{gmnr})
  due to the problems in defining $\gamma_{5}$ and 
  $\epsilon^{\mu\nu\rho\sigma}$ in $d>4$ spacetime
  dimensions. We here follow a method for calculating (\ref{gmnr})
  first presented by Rosenberg in \Ref{Rosenberg} and later used
  by Adler in his classic paper on chiral anomaly~\cite{Adler}. Basically,
  this method relies on the fact that the abiguous part of the integral
  is stored in two form factors in $\Gamma^{\mu\nu\rho}$ expansion,
  $A_{2}$ and $A_{1}$, that multiply the external momenta $k_{1}$ and $k_{2}$, respectively. We then exploit
 physical arguments like for example conservation of charge,  in order to determine the form factors $A_{1},A_{2}$ - all others, $A_{3}...A_{6}$ are finite and can be calculated directly in 4-dimensions.

Our next step is to write down the WIs. This can be done in many ways, probably the
most insightful is the use of functional methods (see for instance Chapter~9.6 in the textbook of \Ref{Peskin}).
One finds the classical WIs of \eq{WIs}, but not the last term on the r.h.s.  
We show below how to calculate this last term.
We need first to calculate the divergence of the 1PI vertex:
 $q_{\mu}\Gamma^{\mu\nu\rho} =
 (k_{1}+k_{2})_{\mu}\Gamma^{\mu\nu\rho}$. It is useful to employ the
 following algebraic identity:
  \begin{eqnarray}
  \slashchar{q}(\alpha+\beta \gamma^{5})=-(\alpha-\beta
  \gamma^{5})(\slashchar{p}-\slashchar{k_{2}}+\slashchar{a}-m)+2
  \beta\gamma^{5}m+(\slashchar{p}+\slashchar{k_{1}}+\slashchar{a}-m)(\alpha+\beta
  \gamma^{5})\;,
 \end{eqnarray}
  in the first integral of (\ref{gmnr}) and a similar identity with $a\to b$ and $k_{1}\to k_{2}$ in the
  second one. These identities split $q_{\mu}\Gamma^{\mu\nu\rho}$ into two parts,
   \begin{eqnarray}
q_{\mu}\Gamma^{\mu\nu\rho}(k_{1},k_{2};a,b) \ =\ -\frac{2 m \beta e
i}{\tilde{\beta}}\: \Gamma^{\nu\rho}(k_{1},k_{2};a,b) + \Pi^{\nu\rho}(k_{1},k_{2};a,b)\;, \label{eq:A3}
 \end{eqnarray}
  a part that is proportional to the fermion mass $m$ and
  a part which contains divergent two-point functions that \emph{would had
  been zero if shifting of the momenta variable was allowed}. The latter
   integrals will be responsible for the failure of the
  axial vector  WI's. Explicitly $\Gamma^{\rho\nu}$ and
  $\Pi^{\rho\nu}$ in \eq{eq:A3} read,
  \begin{eqnarray}
  \Gamma^{\nu\rho}(k_{1},&k_{2}&;a,b) = -i \,e^{2}\, \tilde{\beta}\times \nonumber
  \\[3mm]
  &&\mathrm{Tr}\left \{ \int\frac{d^{4}p}{(2 \pi)^4}\frac{\gamma^{5}(\slashchar{p}-\slashchar{k_{2}}+\slashchar{a}+m)\gamma^{\rho}(\alpha+\beta \gamma^{5})(\slashchar{p}+\slashchar{a}+m)\gamma^{\nu}(\alpha+\beta
  \gamma^{5})(\slashchar{p}+\slashchar{k_{1}}+\slashchar{a}+m)}{[(p-k_{2}+a)^{2}-m^{2}][(p+a)^{2}-m^{2}][(p+k_{1}+a)^{2}-m^{2}]} \right.
  \nonumber\\[3mm] &&\left.
  +\int\frac{d^{4}p}{(2 \pi)^4}\frac{\gamma^{5}(\slashchar{p}-\slashchar{k_{1}}+\slashchar{b}+m)\gamma^{\nu}(\alpha+\beta \gamma^{5})(\slashchar{p}+\slashchar{b}+m)\gamma^{\rho}(\alpha+\beta
  \gamma^{5})(\slashchar{p}+\slashchar{k_{2}}+\slashchar{b}+m)}{[(p-k_{1}+b)^{2}-m^{2}][(p+b)^{2}-m^{2}][(p+k_{2}+b)^{2}-m^{2}]}\right \}\nonumber
  \\[3mm] &=& \frac{-i\, e^{2}\, m\, \tilde{\beta}}{2
\pi^{2}}\varepsilon^{\lambda\nu\rho\sigma}\, k_{1\lambda}\, k_{2\sigma}\, I_{0}(k_{1},k_{2},m)\;,
  \label{gnr}
 \end{eqnarray}
  where
  \begin{eqnarray}
I_{0}(k_{1},k_{2},m)=\int_{0}^{1} dx \int_{0}^{1-x} dy \, \frac{(\alpha^{2}-\beta^{2})+2(x+y)\beta^{2}}{x(x-1)k_{2}^{2}+y(y-1)k_{1}^{2}-2
x y k_{1} \cdot k_{2}+m^{2}} \;. \label{I0} 
 \end{eqnarray}
Obviously, the integral in $\Gamma^{\nu\rho}$ in \eq{gnr}
is obtained from  $\Gamma^{\mu\nu\rho}$ in \eq{gmnr}
with the replacement  $\gamma^{\mu}(\alpha + \beta\gamma^{5}) \rightarrow \gamma^{5}$, that is a replacement of a vector-axial vector
coupling with a pseudoscalar. This validates  the PCAC relation in \eq{eq:A3}.
Note that $\Gamma^{\nu\rho}$ is finite and  independent on the arbitrary vectors  $a^{\mu}$ and $b^{\mu}$ : $\Gamma^{\nu\rho}(k_{1},k_{2};a,b)=
\Gamma^{\nu\rho}(k_{1},k_{2})$.

The divergent part $\Pi^{\nu\rho}$ in the WI of \eq{eq:A3}  contains, among others,
the anomalous term. It is written
explicitly as,
  \begin{eqnarray}
\Pi^{\nu\rho}(k_{1},k_{2};a,b) &=&(- e^{3})\, \mathrm{Tr}
\int\frac{d^{4}p}{(2 \pi)^4}\left
\{-\frac{(\alpha-\beta\gamma^{5})(\alpha-\beta
  \gamma^{5})\gamma^{\rho}(\slashchar{p}+\slashchar{a}+m)\gamma^{\nu}(\alpha+\beta
  \gamma^{5})(\slashchar{p}+\slashchar{k_{1}}+\slashchar{a}+m)}{[(p+a)^{2}-m^{2}][(p+k_{1}+a)^{2}-m^{2}]} \right. \nonumber \\[3mm]
  &+& \left.
  \frac{(\slashchar{p}-\slashchar{k_{2}}+\slashchar{a}+m)\gamma^{\rho}(\alpha+\beta \gamma^{5})(\slashchar{p}+\slashchar{a}+m)\gamma^{\nu}(\alpha+\beta
  \gamma^{5})(\alpha
  +\beta
  \gamma^{5})}{[(p+a)^{2}-m^{2}][(p-k_{2}+a)^{2}-m^{2}]} \right.
  \nonumber \\[3mm] &-&\left. \frac{(\alpha-\beta
  \gamma^{5})(\alpha-\beta
  \gamma^{5})\gamma^{\nu}(\slashchar{p}+\slashchar{b}+m)\gamma^{\rho}(\alpha+\beta
  \gamma^{5})(\slashchar{p}+\slashchar{k_{2}}+\slashchar{b}+m)}{[(p+b)^{2}-m^{2}][(p+k_{2}+b)^{2}-m^{2}]} \right.
  \nonumber \\[3mm] &+& \left.
  \frac{(\slashchar{p}-\slashchar{k_{1}}+\slashchar{b}+m)\gamma^{\nu}(\alpha+\beta \gamma^{5})(\slashchar{p}+\slashchar{b}+m)\gamma^{\rho}(\alpha+\beta
  \gamma^{5})(\alpha
  +\beta
  \gamma^{5})}{[(p+b)^{2}-m^{2}][(p-k_{1}+b)^{2}-m^{2}]} \right \}\;.
  \label{pinr}
   \end{eqnarray}
This is an integral that is divided into two parts : a chiral expression \ie the one that contains $\gamma^{5}$ and a non-chiral expression that does not contain $\gamma^{5}$. Since the 
anomalous term is originated from the chiral part we start from
there. Hence,
   \begin{eqnarray}
  \Pi^{\nu\rho}_\mathrm{chiral}(k_{1},k_{2};a,b)
  &=&(\beta^{3}+3 \alpha^{2} \beta)e^{3}\times \nonumber \\[3mm]
&& \mathrm{Tr}  \int\frac{d^{4}p}{(2
\pi)^4}\left \{\frac{(\slashchar{p}+\slashchar{k_{1}}+\slashchar{a})\gamma^{\rho}(\slashchar{p}+\slashchar{a})\gamma^{\nu}
  \gamma^{5}}{[(p+k_{1}+a)^{2}-m^{2}][(p+a)^{2}-m^{2}]}-\frac{(\slashchar{p}+\slashchar{a})\gamma^{\nu}(\slashchar{p}-\slashchar{k_{2}}+\slashchar{a})\gamma^{\rho}\gamma^{5}}{[(p+a)^{2}-m^{2}][(p-k_{2}+a)^{2}-m^{2}]}\right.\nonumber \\[3mm]
  &+& \left . \frac{(\slashchar{p}+\slashchar{k_{2}}+\slashchar{b})\gamma^{\nu}(\slashchar{p}+\slashchar{b})\gamma^{\rho}
  \gamma^{5}}{[(p+k_{2}+b)^{2}-m^{2}][(p+b)^{2}-m^{2}]}-\frac{(\slashchar{p}+\slashchar{b})\gamma^{\rho}(\slashchar{p}-\slashchar{k_{1}}+\slashchar{b})\gamma^{\nu}\gamma^{5}}{[(p+b)^{2}-m^{2}][(p-k_{1}+b)^{2}-m^{2}]} \right \}. \label{eq:B7}
 \end{eqnarray}
  Grouping together the first and the fourth as well as 
   the third and
  the second terms in the integrand of \eq{eq:B7}, we arrive at,
  \begin{eqnarray}
&&\Pi^{\nu\rho}_\mathrm{chiral}(k_{1},k_{2};a,b) =
(\beta^{3}+3 \alpha^{2}
\beta)e^{3} \times \nonumber \\[3mm] && \int\frac{d^{4}p}{(2
\pi)^4}\left \{ \mathrm{Tr}(\gamma^{\kappa}\gamma^{\rho}\gamma^{\lambda}\gamma^{\nu}\gamma^{5})\left
(\frac{(p+k_{1}+a)_{\kappa}(p+a)_{\lambda}}{[(p+k_{1}+a)^{2}-m^{2}][(p+a)^{2}-m^{2}]}-\frac{(p+b)_{\kappa}(p-k_{1}+b)_\lambda}{[(p+b)^{2}-m^{2}][(p-k_{1}+b)^{2}-m^{2}]} \right )\right. \nonumber\\[3mm]
&+& \left. \mathrm{Tr}(\gamma^{\kappa}\gamma^{\nu}\gamma^{\lambda}\gamma^{\rho}\gamma^{5})\left (\frac{(p+k_{2}+b)_{\kappa}(p+b)_{\lambda}}{[(p+k_{2}+b)^{2}-m^{2}][(p+b)^{2}-m^{2}]}-\frac{(p+a)_{\kappa}(p-k_{2}+a)_{\lambda}}{[(p+a)^{2}-m^{2}][(p-k_{2}+a)^{2}-m^{2}]}\right ) \right \}\;. \label{eq:A8}
 \end{eqnarray}
 Following the steps described in \Ref{Weinberg2},
 we first define a function and an integral,
  \begin{eqnarray}
f_{\kappa\lambda}(p ; c,d)\ = \ \frac{(p+c)_{\kappa}(p+d)_{\lambda}}{[(p+c)^{2}-m^{2}][(p+d)^{2}-m^{2}]} \;,
 \end{eqnarray}
and
\begin{eqnarray}
I_{\kappa\lambda}(k;c,d) \ = \ \int\frac{d^{4}p}{(2
\pi)^{4}} \biggl [ f_{\kappa\lambda}(p+k ; c,d)-
f_{\kappa\lambda}(p ; c,d) \biggr ] \;,
 \end{eqnarray}
where $c,d$ are arbitrary four vectors. 
By exploiting the following ``momentum shift'' 
integral relation (see the lecture by R.~Jackiw in \Ref{Jackiw} 
and \Refs{DHM,Rohrlich,Pugh:1969kn})
  \begin{eqnarray}
\int\frac{d^{4}p}{(2 \pi)^{4}}[f(p+a)-f(p)]=\frac{i}{(2
\pi)^{4}}\biggl [2\pi^{2}a_{\mu}\lim_{p\rightarrow\infty}p^{\mu}p^{2}
f_{o}(p)+\pi^{2}a_{\mu}a_{\nu}\lim_{p\rightarrow\infty}p^{\mu}p^{2}\frac{\partial{f_{e}(p)}}{\partial{p_{\nu}}}\biggr ] \;, \label{eq:A11}
 \end{eqnarray}
where only the first term on the r.h.s is relevant to linearly divergent
diagrams, and,
  \begin{eqnarray}
 f_{o}(p)=\frac{1}{2}[f(p)-f(-p)] \;, \qquad f_{e}(p)=\frac{1}{2}[f(p)+f(-p)]  \;,
 \end{eqnarray}
 are the odd and even parts of $f(p)$ respectively,
 we obtain,\footnote{There is a typographical error
 in the corresponding expression of  a classic textbook 
 written by S. Weinberg in \Ref{Weinberg2}. We thank Steve Martin and 
 Howie Haber for communication related to this point.}
  \begin{eqnarray}
 I_{\kappa\lambda}(k;c,d) \ =\  \frac{i}{96
\pi^{2}}\biggl [2 k_{\lambda}c_{\kappa}+2 k_{\kappa}
d_{\lambda}-k_{\lambda} d_{\kappa}-k_{\kappa}
c_{\lambda}-g_{\kappa\lambda} k\cdot (k+c+d)+k_{\lambda} k_{\kappa}
\biggr ] \;. \label{eq:A13}
 \end{eqnarray}
 Now we have all the necessary machinery 
 to calculate $\Pi^{\nu\rho}$ in \eq{eq:A8} 
 by applying to it \eqs{eq:A11}{eq:A13}.
For the non-chiral part of $\Pi^{\nu\rho}$  the choice $b=-a$ results in
$\Pi^{\nu\rho}_\mathrm{non-chiral}=0$ as we expect,
 since there should be {\emph no} non-chiral anomalies.
With this assignment for vector $b$ we finally obtain for the chiral part:
  \begin{eqnarray}
\Pi^{\nu\rho}_\mathrm{chiral}(k_{1},k_{2};a,-a)=\frac{e^{3}(\beta^{3}+3
\alpha^{2}\beta)}{4\pi^{2}}\varepsilon^{\kappa\nu\lambda\rho}a_{\kappa}(k_{1}+k_{2})_{\lambda}\;. \label{eq:A14}
 \end{eqnarray}
Plugging in \eqs{gnr}{eq:A14}  into \eq{eq:A3},
 the WI associated to the leg $-{\mu}-$ becomes:
  \begin{eqnarray}
q_{\mu}\Gamma^{\mu\nu\rho}(k_{1},k_{2}; a,-a) \ = \ -\frac{2  m  e
\beta i}{\tilde{\beta}}\: \Gamma^{\nu\rho}(k_{1},k_{2})
+ \frac{e^{3}(\beta^{3}+3\alpha^{2}\beta)}{4
\pi^{2}}\: \varepsilon^{\kappa\nu\lambda\rho}\:
a_{\kappa}\: (k_{1}+k_{2})_{\lambda} \;.\label{eq:A15}
 \end{eqnarray}
Along the same lines we can build in the WIs
for the other vertices. For example, the WI referring to the
conservation of current in vertex $-\nu -$ (see Fig.\ref{fig:graph})
reads:
\begin{eqnarray}
-k_{1\nu}\widetilde{\Gamma}^{\nu\rho\mu}(k_{1},k_{2};a,-a)
\ =\ -\frac{2 m \beta e
i}{\tilde{\beta}}\,\widetilde{\Gamma}^{\rho\mu}(k_{1},k_{2})
-\frac{e^{3}(\beta^{3}+3\alpha^{2}\beta)}{4\pi^{2}}\varepsilon
^{\kappa\rho\lambda\mu}\: (a-k_{2})_{\kappa}\: k_{1\lambda}\;.
\label{WI2}
\end{eqnarray}
Vertices $\widetilde{\Gamma}^{\nu\rho\mu}(k_{1},k_{2};a,b)$
and $\widetilde{\Gamma}^{\rho\mu}(k_{1},k_{2})$ are obtained
from ${\Gamma}^{\mu\nu\rho}(k_{1},k_{2};a,b)$
and ${\Gamma}^{\nu\rho}(k_{1},k_{2})$ in \eqs{gmnr}{gnr}, respectively,
after the following replacements
\begin{eqnarray}
\mu\rightarrow\nu,\quad \nu\rightarrow\rho, \quad \rho\rightarrow\mu,
\quad a\rightarrow
a-k_{2}, \quad b\rightarrow b+k_{2}, \quad k_{1}\rightarrow
k_{2}, \nonumber \\ k_{2}\rightarrow -k_{1}-k_{2}, \qquad
q=k_{1}+k_{2}\rightarrow
k_{2}-k_{1}-k_{2}=-k_{1}\Rightarrow q\rightarrow-k_{1}\;. \label{eq:A17}
\end{eqnarray}
It is straightforward to see from \eq{eq:A17} that the non-chiral part of
$-k_{1\nu}\widetilde{\Gamma}^{\nu\rho\mu}(k_{1},k_{2};a,b)$
vanishes again 
for the choice $b=-a$.  Similarly the WI for the current conservation in the $- \rho -$ vertex,
\begin{eqnarray}
-k_{2\rho}\widehat{\Gamma}^{\rho\mu\nu}(k_{1},k_{2};a,-a)\ = \
-\frac{2 m \beta e
i}{\tilde{\beta}}\, \widehat{\Gamma}^{\mu\nu}(k_{1},k_{2})
-\frac{e^{3}(\beta^{3}+3
\alpha^{2}\beta)}{4
\pi^{2}}\: \varepsilon^{\kappa\mu\lambda\nu}\: (a+k_{1})_{\kappa}\:
k_{2\lambda}. \label{WI3}
\end{eqnarray}
As previously, 
$\widehat{\Gamma}^{\rho\mu\nu}(k_{1},k_{2};a,b)$
and   $\widehat{\Gamma}^{\mu\nu}(k_{1},k_{2})$  can be obtained from
\eqs{gmnr}{gnr} by making the following replacements:
\begin{eqnarray}
\mu\rightarrow\rho, \quad \nu\rightarrow\mu, \quad \rho\rightarrow\nu, \quad a\rightarrow
a+k_{1}, \quad b\rightarrow
b-k_{1}, \quad k_{1}\rightarrow-k_{2}-k_{1}, \nonumber \\[3mm] k_{2}\rightarrow
k_{1}, \quad q=k_{1}+k_{2}\rightarrow-k_{2}-k_{1}+k_{1}\Rightarrow
q\rightarrow-k_{2} \;.
\end{eqnarray}
These replacements leave invariant the  choice $b=-a$ so that finally, the
non-chiral part of
$-k_{2\rho}\widehat{\Gamma}^{\rho\mu\nu}(k_{1},k_{2};a,-a)$
vanishes identically everywhere. Furthermore, by direct calculation the
vertices $\widetilde{\Gamma}^{\rho\mu}$ and
$\widehat{\Gamma}^{\mu\nu}$ are found to be,
\begin{eqnarray}
\widetilde{\Gamma}^{\rho\mu}(k_{1},k_{2})\ = \
\frac{i e^{2}m\tilde{\beta}}{2
\pi^{2}}\: \varepsilon^{\lambda\mu\xi\rho}
\: k_{1\lambda}\: k_{2\xi} \: I_{1}(k_{1},k_{2},m) \;, \label{eq:A20}
\end{eqnarray}
and
\begin{eqnarray}
\widehat{\Gamma}^{\mu\nu}(k_{1},k_{2}) \ =\
\frac{i e^{2}m\tilde{\beta}}{2
\pi^{2}}\: \varepsilon^{\lambda\mu\xi\nu}\: k_{1\lambda}\:
k_{2\xi} \: I_{2}(k_{1},k_{2},m) \;, \label{eq:A21}
\end{eqnarray}
respectively, where the corresponding integrals $I_{1,2}$ are written explicitly as,
\begin{eqnarray}
I_{1}(k_{1},k_{2},m)\ = \
\int_{0}^{1}dx\int_{0}^{1-x}dy\frac{-(\alpha^{2}+\beta^{2})+2
x\beta^{2}}{x(x-1)k_{2}^{2}+y(y-1)k_{1}^{2}-2xyk_{1}\cdot k_{2}+m^{2}}
\;, \label{I1}
\end{eqnarray}
and
\begin{eqnarray}
I_{2}(k_{1},k_{2},m) \ = \
\int_{0}^{1}dx\int_{0}^{1-x}dy\frac{(\alpha^{2}+\beta^{2})-2
y\beta^{2}}{x(x-1)k_{2}^{2}+y(y-1)k_{1}^{2}-2xyk_{1}\cdot k_{2}+m^{2}}
\;. \label{I2}
\end{eqnarray}
The  three-point vertex obeys the following equality,
\begin{equation}
\Gamma^{\mu\nu\rho}=
\widetilde{\Gamma}^{\nu\rho\mu}=
\widehat{\Gamma}^{\rho\mu\nu}\;,
\end{equation}
as the property of trace to remain invariant under cyclic
permutations.
It is instructive to write the arbitrary vector $a^{\mu}$,
appearing in the WIs, as a linear combination of the two independent  momenta
$k_{1}$ and $k_{2}$, 
\begin{equation}
 a^{\mu}\ = \ z \, k_{1}^{\mu} + w \, k_{2}^{\mu}\;,
 \label{al}
 \end{equation}
with $z,w$ arbitrary real numbers.
Then the WIs in \eqss{eq:A15}{WI2}{WI3} can be written
explicitly in terms of the three integrals $I_{0},I_{1}$, and $I_{2}$ 
and the real numbers $w$ and $z$ as,
  \begin{eqnarray}
q_{\mu}\Gamma^{\mu\nu\rho}(k_{1},k_{2}; w, z)  &=&
-\frac{e^{3} \beta m^{2}}{\pi^{2}}\:
\varepsilon^{\lambda\nu\rho\sigma}\, k_{1\lambda}\, k_{2\sigma}
\, I_{0}(k_{1},k_{2};m)
+ \frac{e^{3}(\beta^{3}+3\alpha^{2}\beta)}{4
\pi^{2}}\: \varepsilon^{\lambda\nu\rho\sigma}\:
\: k_{1\lambda}\,k_{2\sigma}(w-z)\;.\label{FWI1}
 \\[3mm]
-k_{1\nu}\widetilde{\Gamma}^{\nu\rho\mu}(k_{1},k_{2};w)
 &=& - \frac{ e^{3} \beta m^{2}}{\pi^{2}}\,
 \varepsilon^{\lambda\mu\rho\sigma}\, k_{1\lambda}\, k_{2\sigma}\,
 I_{1}(k_{1},k_{2};m)
+\frac{e^{3}(\beta^{3}+3\alpha^{2}\beta)}{4\pi^{2}}\varepsilon
^{\lambda\mu\rho\sigma}\:  (w-1) \, k_{1\lambda} k_{2\sigma}\;,
\label{FWI2}
 \\[3mm]
-k_{2\rho}\widehat{\Gamma}^{\rho\mu\nu}(k_{1},k_{2};z) &=&
-\frac{e^{3} \beta m^{2}}{\pi^{2}}\,
\varepsilon^{\lambda\mu\nu\sigma} \, k_{1\lambda}\, k_{2\sigma}\, I_{2}(k_{1},k_{2};m)
+\frac{e^{3}(\beta^{3}+3
\alpha^{2}\beta)}{4
\pi^{2}}\: \varepsilon^{\lambda\mu\nu\sigma}\: (z+1)k_{1\lambda} k_{2\sigma}\;. \label{FWI3}
\end{eqnarray}
Obviously, even if we choose $w=1$ and $z=-1$ so that the second and third 
 anomalous terms vanish it cannot be done so for the first one. 
 The second term on the r.h.s of  \eq{FWI1},
 remains.  It is quite interesting to note
that in the limit where $k_{1}^{2}, k_{2}^{2}, k_{1}\cdot k_{2} \ll m\to \infty$,
there is a choice for $w=-z=1/3$
such that the right hand side of \eqss{FWI1}{FWI2}{FWI3} vanishes
identically. For this choice the fermions get decoupled completely.

Our goal is still to calculate the three gauge boson vertex $\Gamma^{\mu\nu\rho}(k_{1},k_{2};a,-a)$. The idea is to first write down the most
general, Lorentz invariant vertex, as\footnote{There are two more terms
allowed in the expansion,
\begin{equation}
A_{7}(k_{1},k_{2})\varepsilon^{\rho\nu\beta\delta}\, k_{2}^{\mu}\, k_{1\beta}\, k_{2\delta} +
A_{8}(k_{1},k_{2})\varepsilon^{\rho\nu\beta\delta}\, k_{1}^{\mu}\, k_{1\beta}\, k_{2\delta}\;.
\end{equation}
However, by exploiting the following, very useful,  identities
\begin{eqnarray}
k_{1}^{\mu} \varepsilon^{\rho\nu\beta\delta} k_{1\beta} k_{2\delta} &=&
-\varepsilon^{\mu\rho\beta\delta} k_{1}^{\nu} k_{1\beta} k_{2\delta} +
\varepsilon^{\mu\nu\beta\delta} k_{1}^{\rho} k_{1\beta} k_{2\delta} \nonumber \\
&+& \varepsilon^{\mu\nu\rho\alpha} [ (k_{1}\cdot k_{2})\: k_{1\alpha} - k_{1}^{2} \: k_{2\alpha}]
\;, \\[3mm]
k_{2}^{\mu} \varepsilon^{\rho\nu\beta\delta} k_{1\beta} k_{2\delta} &=&
-\varepsilon^{\mu\rho\beta\delta} k_{2}^{\nu} k_{1\beta} k_{2\delta} +
\varepsilon^{\mu\nu\beta\delta} k_{2}^{\rho} k_{1\beta} k_{2\delta} \nonumber \\
&-& \varepsilon^{\mu\nu\rho\alpha} [ (k_{1}\cdot k_{2}) \: k_{2\alpha} - k_{2}^{2}\:
 k_{1\alpha}]\;, 
\end{eqnarray}
we arrive at the six form factors given in  \eq{rose}.}
\begin{eqnarray}
\Gamma^{\mu\nu\rho}(k_{1},k_{2};a,-a) &=&  \biggl
[A_{1}(k_{1},k_{2};a,-a)\, \varepsilon^{\mu\nu\rho\sigma}\, k_{2\sigma} \ + \
A_{2}(k_{1},k_{2};a,-a)\,\varepsilon^{\mu\nu\rho\sigma}\, k_{1\sigma}
\ +\
A_{3}(k_{1},k_{2})\,\varepsilon^{\mu\rho\beta\delta}\, k_{2}^{\nu}\, k_{1\beta}\, k_{2\delta}
\nonumber \\[3mm] &+&
 A_{4}(k_{1},k_{2})\,
\varepsilon^{\mu\rho\beta\delta}\, k_{1}^{\nu}\, k_{1\beta}\, k_{2\delta}
\ + \
A_{5}(k_{1},k_{2})\, \varepsilon^{\mu\nu\beta\delta}\, k_{2}^{\rho}\, k_{1\beta}\, k_{2\delta} \ + \ A_{6}(k_{1},k_{2})\, \varepsilon^{\mu\nu\beta\delta}\, k_{1}^{\rho}\,k_{1\beta}\, k_{2\delta}\biggr ] \;. \nonumber \\
\label{rose}
\end{eqnarray}
The form factors $A_{1}$ and $A_{2}$ are dimensionless 
and, by naive power counting, at most linearly
divergent  while
all the rest, $A_{3}...A_{6}$ possess dimension of $m^{-2}$ and are finite.
The latter can be calculated directly in four dimensions from \eq{gmnr}.
We find explicitly:
\begin{eqnarray}
A_{3}(k_{1},k_{2})&=&-A_{6}(k_{1},k_{2})=-\frac{e^3(\beta^{3}+3 \alpha^{2}
\beta)}{\pi^{2}}\int_{0}^{1}dx\int_{0}^{1-x}dy\frac{x
y}{\Delta} \;, \label{eq:A28} \\[3mm]
A_{4}(k_{1},k_{2})&=&\frac{e^3(\beta^{3}+3 \alpha^{2}
\beta)}{\pi^{2}}\int_{0}^{1}dx\int_{0}^{1-x}dy\frac{y(y-1)
}{\Delta}\;, \label{A4} \\[3mm]
A_{5}(k_{1},k_{2})&=&-\frac{e^3(\beta^{3}+3 \alpha^{2}
\beta)}{\pi^{2}}\int_{0}^{1}dx\int_{0}^{1-x}dy\frac{x(x-1)
}{\Delta}, \label{A5}
\end{eqnarray}
where the integrand denominator is common for all $A_{3}...A_{6}$ and 
reads:
\begin{equation}
\Delta \equiv x(x-1)k_{2}^{2}+y(y-1)k_{1}^{2}-2 x y
k_{1}\cdot k_{2}+m^{2}\;. \label{del}
\end{equation}
 To estimate the two divergent integrals,
$A_{1}$ and $A_{2}$,   we apply the Ward Identities  for the vertices ${\nu}$ and ${\rho}$, {\it i.e.,}  \eqs{FWI2}{FWI3} in the expansion (\ref{rose}) and
obtain,
\begin{eqnarray}
A_{1}(k_{1},k_{2};w) \ = \
(k_{1} \cdot k_{2})\, A_{3}(k_{1},k_{2})+ k_{1}^{2}\, A_{4}(k_{1},k_{2}) -\frac{m^{2}
e^3\beta}{\pi^{2}}I_{1}(k_{1},k_{2},m)+\frac{e^3(\beta^{3}+3
\alpha^{2} \beta)}{4\pi^{2}}(w-1)\;, \label{A1}
\end{eqnarray}
and,
\begin{eqnarray}
A_{2}(k_{1},k_{2};z) \ = \
(k_{1} \cdot k_{2}) \, A_{6}(k_{1},k_{2}) + k_{2}^{2}\,
A_{5}(k_{1},k_{2}) -\frac{m^{2}
e^3\beta}{\pi^{2}}I_{2}(k_{1},k_{2},m)+\frac{e^3(\beta^{3}+3
\alpha^{2} \beta)}{4\pi^{2}}(z+1)\;. \label{eq:A33}
\end{eqnarray}
Equations  (\ref{I1}-\ref{I2},\ref{eq:A28}-\ref{eq:A33})  \emph{complete} the evaluation of
the vertex $\Gamma^{\mu\nu\rho}(k_{1},k_{2},w,z)$ in \eq{rose}.
In Appendix~\ref{app:int} we present  analytical expressions
of the integrals $A_{3..6}$ and $I_{0,1,2}$ in various limits. 

Even if  the form factors  $A_{i=1...6}$ had not been calculated 
explicitly there is much to say about their structure by exploiting possible
Bose symmetries. Hence, referring to the notation of Fig.~\ref{fig:graph}, 
Bose symmetry among $j$ and $k$ legs implies, 
\begin{subequations}
\label{Bose}
\begin{align}
A_{1}(k_{1},k_{2}) \ &= \ - A_{2}(k_{2},k_{1})\;, \label{Bose1} \\
A_{3}(k_{1},k_{2}) \ &= \ -A_{6}(k_{2},k_{1})\;, \label{Bose2} \\
A_{4}(k_{1},k_{2}) \ &= \ -A_{5}(k_{2},k_{1})\;, \label{Bose3} 
\end{align}
\end{subequations}
while in $i$ and $j$ legs, 
\begin{subequations}
\label{Bosee}
\begin{eqnarray}
A_{1}(k_{1},k_{2}) \ &=& \  -A_{1}(-q,k_{2}) + A_{2}(-q,k_{2}) - (k_{1}\cdot k_{2}) \left [ ( A_{3}(-q,k_{2})- A_{4}(-q,k_{2}) \right ] + k_{1}^{2}A_{4}(-q,k_{2}) \;, \label{B21} \\[3mm]
A_{2}(k_{1},k_{2}) \ & = & \ A_{2}(-q,k_{2}) + k_{2}^{2} \left [A_{3}(-q,k_{2}) - A_{4}(-q,k_{2}) \right ] - (k_{1}\cdot k_{2}) A_{4}(-q,k_{2}) \;,
 \label{B22} \\[3mm]
A_{3}(k_{1},k_{2})  \ &=& \  A_{4}(-q,k_{2}) - A_{3}(-q,k_{2})\;,
 \label{B23} \\[3mm]
A_{4}(k_{1},k_{2}) \ &=& \ A_{4}(-q,k_{2}) \;,
\label{B24} \\[3mm]
A_{5}(k_{1},k_{2}) \ &=& \ A_{5}(-q,k_{2}) - A_{6}(-q,k_{2}) + A_{3}(-q,k_{2})-A_{4}(-q,k_{2}) \;,
\label{B25} \\[3mm]
A_{6}(k_{1},k_{2})  \ &=& \  -A_{4}(-q,k_{2}) - A_{6}(-q,k_{2})\;,
 \label{B26} 
\end{eqnarray}
\end{subequations}
and, finally,  in $i$ and $k$ legs we find,
\begin{subequations}
\label{Boseee}
\begin{eqnarray}
A_{1}(k_{1},k_{2}) \ &=& \  A_{1}(k_{1,}-q)  - k_{1}^{2} \left [ ( A_{5}(k_{1},-q)- A_{6}(k_{1},-q) \right ]  -(k_{1}\cdot k_{2}) A_{5}(k_{1},-q) \;, \label{B31} \\[3mm]
A_{2}(k_{1},k_{2}) \ & = & \ A_{1}(k_{1},-q) - A_{2}(k_{1},-q) + (k_{1}\cdot k_{2}) \left [A_{5}(k_{1},-q) - A_{6}(k_{1},-q) \right ] + k_{2}^{2} A_{5}(k_{1},-q) \;,
 \label{B32} \\[3mm]
A_{3}(k_{1},k_{2})  \ &=& \  -A_{3}(k_{1},-q) - A_{5}(k_{1},-q)\;,
 \label{B33} \\[3mm]
A_{4}(k_{1},k_{2}) \ &=& \ A_{4}(k_{1},-q) - A_{3}(k_{1},-q) - A_{5}(k_{1},-q) +
A_{6}(k_{1},-q) \;,
\label{B34} \\[3mm]
A_{5}(k_{1},k_{2}) \ &=& \ A_{5}(k_{1},-q) \;,
\label{B35} \\[3mm]
A_{6}(k_{1},k_{2})  \ &=& \  A_{5}(k_{1},-q) - A_{6}(k_{1},-q)\;.
 \label{B36} 
\end{eqnarray}
\end{subequations}
The above relations  have been repeatedly used in section~\ref{TGBV} when determining 
the anomaly  parameters $w$ and $z$.
The reader  should notice that in addition 
to  relations due to Bose symmetry, 
there are few more relations
originated  solely from  fermionic triangle: 
\begin{eqnarray}
A_{3}(k_{1},k_{2}) \ &=& \ A_{3}(k_{2},k_{1})\;, \quad A_{6}(k_{1},k_{2}) \ = \ A_{6}(k_{2},k_{1})\;.
\end{eqnarray}
We can now exploit Bose symmetry to set constraints 
on the arbitrary parameters $w$ and $z$. For example, if the 
gauge bosons associated with
legs $j$ and $k$  in Fig.~\ref{fig:graph} are identical then 
 \eq{Bose}  impose  the following relation, 
 \begin{equation}
 w+z \ = \ 0 \;,
 \end{equation}
among the undefined (momentum route dependent) parameters.
One last remark is that we can rediscover Bose symmetries by using one of
the following equivalent representations 
(\ie they leave the double integral measure invariant) of the integrals $A_{3}...A_{6}$ by noting that
\begin{eqnarray}
\label{inttricks}
\Delta(k_{1},k_{2}) & \xrightarrow[]{x \leftrightarrow y} & \Delta(k_{2}, k_{1}) \;, \\[3mm]
\Delta(k_{1},k_{2}) & \xrightarrow[x \rightarrow x]{y \rightarrow 1-x-y} & \Delta(k_{1},-q) \;, \\[3mm]
\Delta(k_{1},k_{2}) & \xrightarrow[x \rightarrow 1-x-y]{y \rightarrow y} & \Delta(-q,k_{2}) \;,
\end{eqnarray}
where $\Delta(k_{1},k_{2})$ is a function defined in \eq{del}.

As a generalisation of \eqss{eq:A15}{WI2}{WI3}
 we can proceed to the situation
 where there are three, in general different,
external gauge bosons with different couplings to fermions.
As in (\ref{gmnr}), we write the general three point vertex in Fig.~\ref{fig:graph} as:
\begin{eqnarray}
&&\Gamma^{\mu\nu\rho}(k_{1},k_{2};a,b) =
\tilde{\Gamma}^{\nu\rho\mu}(k_{1},k_{2};a,b)=
\hat{\Gamma}^{\rho\mu\nu}(k_{1},k_{2};a,b)=
-e^{3}
\int\frac{d^{4}p}{(2\pi)^{4}}\times \nonumber \\[3mm]
&& \mathrm{Tr} \left \{ \frac{\gamma^{\mu}(\alpha_{i}+
\beta_{i}\gamma^{5})(\slashchar{p}-\slashchar{k_{2}}+\slashchar{a}+m)\gamma^{\rho}(\alpha_{j}+
\beta_{j}\gamma^{5})(\slashchar{p}+\slashchar{a}+m)\gamma^{\nu}(\alpha_{k}
+\beta_{k}\gamma^{5})(\slashchar{p}+\slashchar{k_{1}}+\slashchar{a}+m)}{[(p-k_{2}+a)^{2}-m^{2}][(p+a)^2-m^{2}][(p+k_{1}+a)^2-m^{2}]} \right.
\nonumber \\[3mm] &+& \left. \frac{\gamma^{\mu}(\alpha_{i}+
\beta_{i}\gamma^{5})(\slashchar{p}-\slashchar{k_{1}}+\slashchar{b}+m)\gamma^{\nu}(\alpha_{k}+
\beta_{k}\gamma^{5})(\slashchar{p}+\slashchar{b}+m)\gamma^{\rho}(\alpha_{j}+\beta_{j}\gamma^{5})(\slashchar{p}+\slashchar{k_{2}}
+\slashchar{b}+m)}{[(p-k_{1}+b)^{2}-m^{2}][(p+b)^2-m^{2}][(p+k_{2}+b)^2-m^{2}]}\right \}\;,  \label{GG3}
\end{eqnarray}
and the corresponding two point vertex functions as:
\begin{eqnarray}
\Gamma^{\nu\rho}(k_{1},k_{2}) &=&\frac{-i
e^{2}m\tilde{\beta}}{2\pi^{2}}\varepsilon^{\lambda\nu\rho\sigma}k_{1\lambda}k_{2\sigma}\int_{0}^{1}dx\int_{0}^{1-x}dy\frac{(\alpha_{j}\alpha_{k}-\beta_{j}\beta_{k})+2\beta_{j}\beta_{k}(x+y)}{\Delta} \;,
\nonumber \\[3mm]
\tilde{\Gamma}^{\rho\mu}(k_{1},k_{2}) &=&\frac{i
e^{2}m\tilde{\beta}}{2\pi^{2}}\varepsilon^{\lambda\mu\xi\rho}k_{1\lambda}k_{2\xi}\int_{0}^{1}dx\int_{0}^{1-x}dy\frac{-(\alpha_{i}\alpha_{k}+\beta_{i}\beta_{k})+2x\beta_{i}\beta_{k}}{\Delta}\;, \label{GG2}\\[3mm]
\hat{\Gamma}^{\mu\nu}(k_{1},k_{2})&=&\frac{i
e^{2}m\tilde{\beta}}{2\pi^{2}}\varepsilon^{\lambda\mu\xi\nu}k_{1\lambda}k_{2\xi}\int_{0}^{1}dx\int_{0}^{1-x}dy\frac{(\alpha_{i}\alpha_{j}+\beta_{i}\beta_{j})-2y\beta_{i}\beta_{j}}{\Delta} \;, \nonumber
\end{eqnarray}
where as before $\Delta\equiv \Delta(k_{1},k_{2})$ is given by \eq{del}.
The complete $\Gamma^{\mu\nu\rho}(k_{1},k_{2},w,z)$ in this general case
 is presented 
in section~\ref{sec:anal}.

\section{Charged Gauge boson Vertex}
\label{app:W}

The calculation for $V^{*}W^{-}W^{+}, V=\gamma, Z$ is slightly 
more complicated than the one for neutral triple gauge boson vertices
for two reasons: first, the appearance in the loop of two, in general, 
different fermion masses  and second, the appearance of different 
$Vf\bar{f}$ vertex for each particle contribution  
(see Fig.~\ref{fig:graphWW}). Although the first complication leads to only technical
difficulties the latter one is more serious: it does not allow for an 
obvious exploitation of the 
master 4D ``momentum shift'' equation (\ref{eq:A11}).

Our method for calculating this vertex follows exactly the same steps 
as  described in detail in Appendix~\ref{sec:app} and in section~\ref{sec:anal}. 
The chiral part of the $V^{*}WW$ vertex is still given by \eq{rose2}. The
finite form factors $A_{3}...A_{6}$ for the first  diagram in 
Fig.~\ref{fig:graphWW} are exactly the half of the corresponding ones 
in (\ref{A3456}) but with 
the replacement of $\Delta(k_{1},k_{2})$ into 
\begin{align}
\Delta(k_{1},k_{2};m_{f_{u}}^{2}, m_{f_{d}}^{2}) &\equiv
x(x-1)k_{2}^2+y(y-1)k_{1}^2
- 2xy k_{1}\cdot k_{2} - (x+y)\, \Delta m^{2} +
m_{f_{u}}^{2} \;,
\label{deltaWW}
\end{align}
with the mass squared difference being $\Delta m^{2} \equiv m_{f_{u}}^{2} - m_{f_{d}}^{2}$.
$f_{u}$ and $f_{d}$ here denote each of the fermion 
pair $(u, \nu)$ and $(d,e)$ for leptons and quarks,
respectively.
Obviously, the contribution of  the crossed diagram 
\ie the second diagram in Fig.~\ref{fig:graphWW}, 
requires the  replacement, $f_{u} \leftrightarrow f_{d}$.
Our calculation here is quite
general and is not confined only in to $V^{*}WW$ vertex. For example, it could be used for 
the vertex $VW_{L}W_{R}$ in an $SU(2)_{L}\times SU(2)_{R}\times U(1)$ gauge model.

\begin{figure}[t]
   \centering
   \includegraphics[height=1.6in]{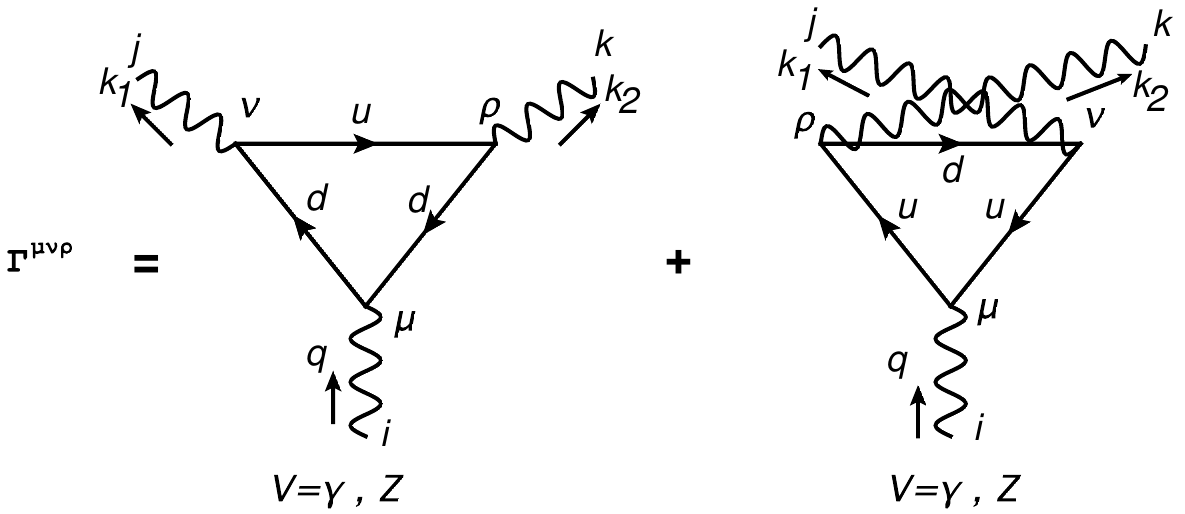} 
   \caption{The one-loop effective triple gauge boson vertex,
   $\Gamma^{\mu\nu\rho}_{VW^{-}W^{+}}, V=\gamma , Z$. As in Fig.~\ref{fig:graph}, 
   indices $\{i,j,k\}$ denote
distinct external gauge bosons in general.}
   \label{fig:graphWW}
\end{figure}
%

As before, the ``infinite'' form factors, $A_{1,2}$ are fixed by the Ward Identities. The
calculation of the first diagram of Fig.~\ref{fig:graphWW} results in, 
\begin{subequations}
\label{A12W}
\begin{align}
A_{1}(k_{1},k_{2}) &= (k_{1}\cdot k_{2}) A_{3} + k_{1}^{2} A_{4} 
- \frac{\alpha_{j} (m_{f_{u}}-m_{f_{d}})}{4\pi^{2}} \, I_{11}(m_{f_{u}}^{2}, m_{f_{d}}^{2})  - 
\frac{\beta_{j} (m_{f_{u}}+m_{f_{d}})}{4\pi^{2}} \, I_{12}(m_{f_{u}}^{2}, m_{f_{d}}^{2}) + \frac{c}{8\pi^{2}}\: (w-1)\;,  \\[2mm]
A_{2}(k_{1},k_{2}) &= (k_{1}\cdot k_{2}) A_{6} + k_{2}^{2} A_{5} 
+ \frac{\alpha_{k} (m_{f_{u}}-m_{f_{d}})}{4\pi^{2}} \, I_{21}(m_{f_{u}}^{2}, m_{f_{d}}^{2})  - 
\frac{\beta_{k} (m_{f_{u}}+m_{f_{d}})}{4\pi^{2}} \, I_{22}(m_{f_{u}}^{2}, m_{f_{d}}^{2})  + \frac{c}{8\pi^{2}}\: (z+1)\;, 
\end{align}
\end{subequations}
where $c \equiv (\alpha_{i}\alpha_{j}+\beta_{i}\beta_{j})\beta_{k}
+(\alpha_{i}\beta_{j}+\alpha_{j}\beta_{i})\alpha_{k}$ is the usual anomaly factor. 
Again, the result depends upon two arbitrary four vectors, $a^{\mu}$ and $b^{\mu}$, that
parameterize the momentum routing in the loop. 
For chiral gauge anomalies to cancel after summing over all fermions,
 the arbitrary vectors $a^{\mu}$ and $b^{\mu}$ 
need to be set at $a^{\mu}=-b^{\mu}$. 
As before, we  write $a^{\mu}$ as a linear combination of independent 
four vectors as $a^{\mu}=z\: k_{1}^{\mu} +w\: k_{2}^{\mu}$, with $z,w$ arbitrary real parameters.
This includes $\gamma,Z,W$-self energy corrections.
The latter depend on their own routing momenta arbitrary vectors that 
can be taken as such in order to eliminate their anomalous contributions.
One then expects that this relation renders the non-chiral part
independent of $a^{\mu}$ as it does for the neutral vertices $VVV$, for $V=\gamma, Z$ 
[see Appendix~\ref{sec:app}].
However, for $VWW$-vertices  there are additional contributions to the non-chiral part 
of $\Gamma^{\mu\nu\rho}$
from $Z,\gamma, W$-self  energy corrections  that depend on routing momentum
arbitrary vectors.  When all these corrections are added one expects the result to
be independent on these arbitrary vectors.

Then the ``non-decoupling'' integrals,   
$I_{ij} \equiv I_{ij}(m_{f_{u}}^{2}, m_{f_{d}}^{2})$ with $i,j=1,2$, 
appearing in \eq{A12W} are
given by
\begin{subequations}
\label{I12W}
\begin{align}
I_{11} &= \int_{0}^{1} dx \int_{0}^{1-x}dy\: \frac{(\alpha_{i}\beta_{k} + \alpha_{k}\beta_{i})\,
 m_{f_{d}}\, y
+ (\alpha_{i}\beta_{k} + \alpha_{k}\beta_{i})\, m_{f_{u}}\, (x+y-1) +
(\alpha_{i}\beta_{k} - \alpha_{k}\beta_{i})\, m_{f_{d}}\, x}{
\Delta(k_{1},k_{2};m^{2}_{f_{u}},m^{2}_{f_{d}})}
\;,  \\[3mm]
I_{12} &= \int_{0}^{1} dx \int_{0}^{1-x}dy\: \frac{-(\alpha_{i}\alpha_{k} + \beta_{i}\beta_{k})\,
 m_{f_{d}}\, y
+ (\alpha_{i}\alpha_{k} + \beta_{i}\beta_{k}) \, m_{f_{u}}\, (x+y-1) -
(\alpha_{i}\alpha_{k} - \beta_{i}\beta_{k}) \, m_{f_{d}}\, x}{
\Delta(k_{1},k_{2};m^{2}_{f_{u}},m^{2}_{f_{d}})}\;, \\[3mm]
%
%
%
I_{21} &= \int_{0}^{1} dx \int_{0}^{1-x}dy\: \frac{(\alpha_{i}\beta_{j} - \alpha_{j}\beta_{i})\,
 m_{f_{d}}\, y
+ (\alpha_{i}\beta_{j} + \alpha_{j}\beta_{i})\, m_{f_{u}}\, (x+y-1) +
(\alpha_{i}\beta_{j} + \alpha_{j}\beta_{i})\, m_{f_{d}}\, x}{
\Delta(k_{1},k_{2};m^{2}_{f_{u}},m^{2}_{f_{d}})}
\;,  \\[3mm]
I_{22} &= \int_{0}^{1} dx \int_{0}^{1-x}dy\: \frac{(\alpha_{i}\alpha_{j} - \beta_{i}\beta_{j})\,
 m_{f_{d}}\, y
- (\alpha_{i}\alpha_{j} + \beta_{i}\beta_{j}) \, m_{f_{u}}\, (x+y-1) +
(\alpha_{i}\alpha_{j} + \beta_{i}\beta_{j}) \, m_{f_{d}}\, x}{
\Delta(k_{1},k_{2};m^{2}_{f_{u}},m^{2}_{f_{d}})}\;,
%
%
%
\end{align}
\end{subequations}
where $\alpha_{i}\equiv \alpha_{f_{d}}, \beta_{i} \equiv \beta_{f_{d}}$,..etc,  
follow  the  first diagram of Fig.~\ref{fig:graphWW}. 
%
The corresponding expressions  for the crossed diagram are easily obtained from
those in \eqs{A12W}{I12W} with the replacement $f_{u}\leftrightarrow f_{d}$. Note that
CP-invariance is maintained since  $A_{1}(k_{1},k_{2}) = -A_{2}(k_{2},k_{1})$.

For reasons we explained at the beginning of this Appendix, finding the anomalous
terms \ie the last terms in  \eq{A12W}, is not a straightforward task. The trick here 
is to add a Lorentz invariant but vanishing integral that generates exactly the anomaly integrals by 
momentum shift. It is then straightforward to use the 4-D expression (\ref{eq:A11}).

To complete our analysis for the chiral fermionic triangle with general external 
charged and neutral gauge bosons,
we append here the relevant WI's analogous to those presented in \eq{WIs}
for neutral external gauge bosons:
\begin{subequations}
\label{WIsW}
\begin{align} 
q_{\mu}\, \Gamma^{\mu\nu\rho}(k_{1},k_{2}) \ &= \ -\frac{\beta_{i}}{2\pi^{2}} \, m_{f_{d}} 
\epsilon^{\nu\rho\lambda\sigma}\, k_{1\lambda} k_{2\sigma}\, I_{01}(m_{f_{u}}^{2}, m_{f_{d}}^{2}) +
\frac{c}{8\pi^{2}}\, \epsilon^{\nu\rho\lambda\sigma}\, k_{1\lambda} k_{2\sigma}\,
(w-z) \;,  \\[3mm]
-k_{1\nu}\, \Gamma^{\mu\nu\rho}(k_{1},k_{2}) \ &= \ -\frac{\alpha_{j}}{4\pi^{2}} \,
(m_{f_{u}} - m_{f_{d}} )\, \epsilon^{\mu\rho\lambda\sigma}\, k_{1\lambda} k_{2\sigma}\,
I_{11}(m_{f_{u}}^{2}, m_{f_{d}}^{2}) -
\frac{\beta_{j}}{4\pi^{2}}\, (m_{f_{u}} + m_{f_{d}} )\,
\epsilon^{\mu\rho\lambda\sigma}\, k_{1\lambda} k_{2\sigma}\,
I_{12}(m_{f_{u}}^{2}, m_{f_{d}}^{2}) \nonumber \\[2mm]
&+ \frac{c}{8\pi^{2}}\, \epsilon^{\mu\rho\lambda\sigma}\, k_{1\lambda} k_{2\sigma}\,
(w-1) \;,  \\[3mm]  
-k_{2\rho}\, \Gamma^{\mu\nu\rho}(k_{1},k_{2}) \ &= \ \frac{\alpha_{k}}{4\pi^{2}} \,
(m_{f_{u}} - m_{f_{d}} )\, \epsilon^{\mu\nu\lambda\sigma}\, k_{1\lambda} k_{2\sigma}\,
I_{21}(m_{f_{u}}^{2}, m_{f_{d}}^{2}) -
\frac{\beta_{k}}{4\pi^{2}}\, (m_{f_{u}} + m_{f_{d}} )\,
\epsilon^{\mu\nu\lambda\sigma}\, k_{1\lambda} k_{2\sigma}\,
I_{22}(m_{f_{u}}^{2}, m_{f_{d}}^{2}) \nonumber \\[2mm]
&+ \frac{c}{8\pi^{2}}\, \epsilon^{\mu\nu\lambda\sigma}\, k_{1\lambda} k_{2\sigma}\,
(z+1) \;.
\end{align}
\end{subequations}
Again, the corresponding expressions for the crossed diagram in Fig.~\ref{fig:graphWW}
are obtained from \eq{WIsW} after the replacement $f_{u} \leftrightarrow f_{d}$.
The integral $I_{01}\equiv I_{01}(m_{f_{u}}^{2}, m_{f_{d}}^{2})$ is given by 
\begin{equation}
I_{01} = \int_{0}^{1} dx \int_{0}^{1-x} dy \,
\frac{(\alpha_{j}\alpha_{k} + \beta_{j}\beta_{k})\, m_{f_{d}}\, y 
-(\alpha_{j}\alpha_{k} -\beta_{j}\beta_{k})\, m_{f_{u}}\, (x+y-1)
+(\alpha_{j}\alpha_{k} +\beta_{j}\beta_{k}) \, m_{f_{d}} \, x}{\Delta(k_{1},k_{2};m^{2}_{f_{u}},m^{2}_{f_{d}})}\;.
\end{equation}
As a check, 
note that in the limit of equal masses $m^{2}_{f_{u}} =m^{2}_{f_{d}}$ all the above integral 
expressions reduce to the corresponding ones in \eqss{A12}{I12}{A3456} for 
the neural gauge boson vertex.

\section{Some useful analytical integral expressions}
\label{app:int}

In this Appendix we present analytical expressions for
integrals related to $A_{3..6}$, and,  $I_{1,2}$ in the limit 
where $k_{1}^{2}, k_{2}^{2}\rightarrow  0$ as well as their
approximate expressions in various limits. 
We make an effort to write the latter 
 in terms of standard
functions \ie {\em not} dilogarithms, which are easy to handle
both symbolically and numerically.  We start out with integrals related to \eq{A3456}, 
\begin{eqnarray}
\widetilde{A_{3}}(\xi)=\int_{0}^{1}dx\int_{0}^{1-x}dy\frac{x y}{x y- \xi/4}=\frac{1}{2}[1+\xi J(\xi)] \;,
 \end{eqnarray}
where $\xi\equiv  \frac{4 m^{2}}{s}$,  $m$ is the loop fermion mass, and $s=(k_{1}+k_{2})^{2}$,
while,
\begin{subequations}
\label{exjxi}
\begin{align}
J(\xi) \ &= \
 -\arctan^{2}\bigg(\frac{1}{\sqrt{\xi-1}}\bigg) \;, \quad \xi\geq1  \;, \\[3mm] 
\ &= \  
\frac{1}{4}\bigg[\ln \bigg(\frac{1-\sqrt{1-\xi}}{1+\sqrt{1-\xi}}\bigg)-i \pi\bigg]^{2}\;, \quad \xi\leq1 \;.
\end{align} 
\end{subequations}
This integral has also been calculated in \Ref{Rudaz:1989ij} and we
find agreement. 
In the same limit  the integral related to $A_{4}$ and $A_{5}$ is:
\begin{eqnarray}
\widetilde{A_{4}}(\xi)= \widetilde{A_{5 }}(\xi) = \int_{0}^{1}dx\int_{0}^{1-x}dy \frac{x(x-1)}{x y -\xi/4}=\int_{0}^{1}dx\int_{0}^{1-x}dy \frac{y(y-1)}{x y -\xi/4} \;,
\end{eqnarray}
%
with its exact answer written like
\begin{eqnarray}
\widetilde{A_{4}}(\xi) &=&
1-\sqrt{\xi-1} \arctan\bigg(\frac{1}{\sqrt{\xi-1}}\bigg)\;, \quad \xi\geq 1 \;,   \\[3mm]
 &=& 
 1+\frac{\sqrt{1-\xi}}{2} \bigg[\ln \bigg(\frac{1-\sqrt{1-\xi}}{1+\sqrt{1-\xi}}\bigg)-i \pi\bigg]\;, \quad \xi\leq1. 
\end{eqnarray}
Integrals that are related  to   $I_{1}$ and $I_{2}$ of \eq{I12} are:
\begin{eqnarray}
\widetilde{I_{1}}(\xi) &=& \int_{0}^{1}dx \int_{0}^{1-x}dy 
\frac{1}{x y -\xi/4} \\[3mm]
&=& -2 \arctan^{2}\bigg(\frac{1}{\sqrt{\xi-1}}\bigg)\;, \quad \xi\geq 1 \\[3mm]
&=&\frac{1}{2}\bigg[\ln\bigg(\frac{1-\sqrt{1-\xi}}{1+\sqrt{1-\xi}}\bigg)-
 i \pi\bigg]^{2} \;, \quad \xi\leq1 \;,
\end{eqnarray}
and 
\begin{eqnarray}
\widetilde{I_{1}'}(\xi)&=& 
\int_{0}^{1}dx\int_{0}^{1-x}dy 
\frac{x}{x y -\xi/4}=\int_{0}^{1}dx\int_{0}^{1-x}dy \frac{y}{x y -\xi/4} \\[3mm]
 &=& 2 \bigg[\sqrt{\xi-1} \: \arctan \bigg(\frac{1}{\sqrt{\xi-1}} \bigg)-1\bigg]\;,\quad \xi\geq1 \\[3mm]
&=&-2-\sqrt{1-\xi}\: \bigg[\ln \bigg(\frac{1-\sqrt{1-\xi}}{1+\sqrt{1-\xi}} \bigg)-
 i \pi\bigg]\;,\quad \xi\leq1 \;.
\end{eqnarray}
These integrals  are related to  standard ones, $A_{3}..A_{6}$, $I_{1,2}$, 
and in the limit where $m_{Z}^{2} \ll s < m^{2}$, become
\begin{subequations}
\label{APR}
\begin{align}
A_{3}(s;m^{2}) &=-A_{6}(s;m^{2})
 = \frac{c}{s} \: \widetilde{A_{3}}(\frac{4 m^{2}}{s}) =
-\frac{c}{m^{2}} \left [\frac{1}{24} +\frac{1}{180}\frac{s}{m^{2}} + O(s^{2}/m^{4}) \right] \;,\label{ApA3} \\[3mm]
A_{4}(s;m^{2}) &= -A_{5}(s;m^{2}) 
= -\frac{c}{s} \widetilde{A_{4}}(\frac{4m^{2}}{s})  
= -\frac{c}{m^{2}}\left [ \frac{1}{12} + \frac{1}{120} \frac{s}{m^{2}}
+ O(s^{2}/m^{4}) \right] \;, \label{ApA4}\\[3mm]
I_{1}(s;m^{2}) &= \frac{\alpha_{i}\alpha_{k} + \beta_{i}\beta_{k}}{s} \: \widetilde{I_{1}}(\frac{4 m^{2}}{s}) - \frac{2\beta_{i}\beta_{k}}{s}\: \widetilde{I_{1}'}(\frac{4 m^{2}}{s}) \nonumber \\
&= -\frac{1}{m^{2}} \left [ \frac{\beta_{i}\beta_{k} + 3 \alpha_{i}\alpha_{k}}{6}+\frac{\beta_{i}\beta_{k} + 5 \alpha_{i}\alpha_{k}}{120} \frac{s}{m^{2}}  + O(s^{2}/m^{4}) \right] \;, \label{ApI1} \\[3mm]
I_{2}(s;m^{2}) &= -\frac{\alpha_{i}\alpha_{j} + \beta_{i}\beta_{j}}{s} \: \widetilde{I_{1}}(\frac{4 m^{2}}{s}) + \frac{2\beta_{i}\beta_{j}}{s}\: \widetilde{I_{1}'}(\frac{4 m^{2}}{s}) \nonumber \\
&= \frac{1}{m^{2}} \left [ \frac{\beta_{i}\beta_{j} + 3 \alpha_{i}\alpha_{j}}{6}+\frac{\beta_{i}\beta_{j} + 5 \alpha_{i}\alpha_{j}}{120} \frac{s}{m^{2}}  + O(s^{2}/m^{4}) \right] \;, \label{ApI2}
\end{align}
\end{subequations}
where $c = \frac{e^3[(\alpha_{i}\alpha_{j}+\beta_{i}\beta_{j})\beta_{k}
+(\alpha_{i}\beta_{j} +\beta_{i}\alpha_{j})
\alpha_{k}]}{\pi^{2}}$ is the anomaly factor. 
These expressions are in 
agreement with the corresponding ones presented in \Ref{Peskin:delayed}.
In the high energy limit
$m^{2} \ll s$, we obtain,
\begin{subequations}
\label{D14}
\begin{align}
A_{3}(s;m^{2}) &= -A_{6}(s; m^{2}) \simeq
c \left \{\frac{1}{2s} + \frac{m^{2}}{2 s^{2}} \left [ \ln^{2}\frac{s}{m^{2}} - \pi^{2} \right ] + i  \pi \frac{m^{2}}{s^{2}} \ln \frac{s}{m^{2}} + O(m^{4}/s^{3}) 
\right \} \;, \label{AsA3}\\[3mm]
A_{4}(s;m^{2}) &= -A_{5}(s; m^{2}) \simeq
c \left \{\frac{1}{s}\left [ -1 +\frac{1}{2}\ln\frac{s}{m^{2}}\right ]
 - \frac{m^{2}}{ s^{2}} \left [ \ln\frac{s}{m^{2}} +1 \right ] + i  \pi 
 \left [ \frac{1}{2s} - \frac{m^{2}}{s^{2}} \right ] + O(m^{4}/s^{3}) 
\right \} \;, \label{AsA4}\\[3mm]
I_{1}(s; m^{2}) &\simeq  \frac{(\alpha_{i}\alpha_{k}+\beta_{i}\beta_{k})}{s} \left[ \frac{1}{2} \left (\ln^{2}\frac{s}{m^{2}}-\pi^{2} \right) 
-2 \frac{m^{2}}{s}\ln\frac{s}{m^{2}} \right ]
-\frac{2\beta_{i}\beta_{k}}{s} \left [ \ln\frac{s}{m^{2}} - 2 - \frac{2 m^{2}}{s}
\left ( \ln\frac{s}{m^{2}}+1 \right ) \right ] \nonumber \\
&+ i \pi\: \left \{ \frac{(\alpha_{i}\alpha_{k}+\beta_{i}\beta_{k})}{s}
\left [\ln\frac{s}{m^{2}} -  \frac{2 m^{2}}{s}\right ]
 - \frac{2\beta_{i}\beta_{k}}{s}
\left [ 1- \frac{2 m^{2}}{s} \right ] \right \} + O(m^{4}/s^{3}) \;, 
\label{AsI1}\\[3mm]
I_{2}(s; m^{2}) &\simeq  -\frac{(\alpha_{i}\alpha_{j}+\beta_{i}\beta_{j})}{s} \left[ \frac{1}{2} \left (\ln^{2}\frac{s}{m^{2}}-\pi^{2} \right) 
-2 \frac{m^{2}}{s}\ln\frac{s}{m^{2}} \right ]
+\frac{2\beta_{i}\beta_{j}}{s} \left [ \ln\frac{s}{m^{2}} - 2 - \frac{2 m^{2}}{s}
\left ( \ln\frac{s}{m^{2}}+1 \right ) \right ] \nonumber \\
&- i \pi\: \left \{ \frac{(\alpha_{i}\alpha_{j}+\beta_{i}\beta_{j})}{s}
\left [\ln\frac{s}{m^{2}} -  \frac{2 m^{2}}{s}\right ]
 - \frac{2\beta_{i}\beta_{j}}{s}
\left [ 1- \frac{2 m^{2}}{s} \right ] \right \} + O(m^{4}/s^{3}) \;. \label{AsI2}
\end{align}
\end{subequations}
Only the real parts of these expressions have been presented in  \Ref{Peskin:delayed} 
and we find agreement\footnote{For notational matter, 
our integrals are related to  those in \Ref{Peskin:delayed} 
like $A_{3} = -c_{6}, \quad A_{4} = \frac{1}{2} (c_{4} -c_{3} -2 c_{6})$, 
where  for example 
$A_{3}\equiv A_{3}(k_{1}^{2}=k_{2}^{2}=m_{W}^{2},s,m_{f_{u}}^{2},m_{f_{d}}^{2})$,...etc.}. 
Other useful identities among $A$'s  that have been used in 
our numerical code for calculating the $V^{*}ZZ$-vertex are,
\begin{eqnarray}
(A_{3}-A_{4}){(k_{1}=m_{Z},k_{2}=m_{Z}, s; m=0)} 
=-\frac{1}{4 m_{Z}^{2}} +
\frac{s}{2 m_{Z}^{2}} A_{3}{(k_{1}=m_{Z},k_{2}=m_{Z}, s; m=0)} \;,
\end{eqnarray}
and  for the $V^{*}\gamma Z$-vertex, 
\begin{eqnarray}
A_{3}(k_{1}=0,k_{2}=m_{Z},s;m=0) &=& \frac{1}{2 (s-m_{Z}^{2})} - 
\frac{m_{Z}^{2}}{2(s-m_{Z}^{2})^{2}} 
\ln \biggl ( \frac{s}{m^{2}_{Z}} \biggr )\;,\label{approx1}
\\[3mm]
A_{5}(k_{1}=0,k_{2}=m_{Z},s;m=0) &=&- \frac{1}{2 (s-m_{Z}^{2})} 
\ln \biggl ( \frac{s}{m^{2}_{Z}} \biggr ) \;.\label{approx2}
\end{eqnarray}
%
Finally, we derive full analytical expressions in the case $k_{1}^{2}=0$, where 
one of the external gauge bosons is massless e.g., the $V^{*}\gamma Z$-vertex. 
To this end it is useful to  define an auxiliary function, 
\begin{eqnarray}
\mathscr{F}(m_{Z},s,m)\equiv \int_{0}^{1}dx\int_{0}^{1-x}dy \,\  \ln[x(x-1) m_{Z}^{2}-x y (s-m_{Z}^{2})+m^{2}]\;,
 \end{eqnarray}
out of which we read $A_{3}..A_{6}, I_{1,2}$ by simply taking appropriate derivatives 
w.r.t $s, k_{2}^{2}=m_{Z}^{2}$ or $m^{2}$. 
Depending on the region of parameters $s, m^{2}, m_{Z}^{2}$ 
we have found the function $\mathscr{F}$ to be,
\begin{eqnarray}
\mathscr{F}(m_{Z},s,m)&=&-\frac{3}{2}+\frac{\ln(m^{2})}{2}-\bigg(\frac{1}{m_{Z}^{2}-s}\bigg) \bigg \{s \sqrt{\frac{4 m^{2}}{s}-1}
\: \arctan\bigg(\frac{1}{\sqrt{\frac{4 m^{2}}{s}-1}}\bigg)- \nonumber \\
&-&m_{Z}^{2} \sqrt{\frac{4 m^{2}}{m_{Z}^{2}}-1}
\: \arctan\bigg(\frac{1}{\sqrt{\frac{4 m^{2}}{m_{Z}^{2}}-1}}\bigg)+\nonumber \\
&+& 2 m^{2}\bigg[\arctan^{2}\bigg(\frac{1}{\sqrt{\frac{4 m^{2}}{s}-1}}\bigg)-
\arctan^{2}\bigg(\frac{1}{\sqrt{\frac{4 m^{2}}{m_{Z}^{2}}-1}}\bigg) \bigg]\bigg\} \;, \quad  \frac{4 m^{2}}{s}> 1,\: 
\frac{4 m^{2}}{m_{Z}^{2}}> 1 \;,
\end{eqnarray}
\begin{eqnarray}
\mathscr{F}(m_{Z},s,m)&=&-\frac{3}{2}+\frac{\ln(m^{2})}{2}-\bigg(\frac{1}{m_{Z}^{2}-s}\bigg) \bigg \{s \sqrt{\frac{4 m^{2}}{s}-1}
\: \arctan\bigg(\frac{1}{\sqrt{\frac{4 m^{2}}{s}-1}}\bigg)+ \nonumber \\
&+&m_{Z}^{2} \bigg[\frac{1}{2}\sqrt{1-\frac{4 m^{2}}{m_{Z}^{2}}}
\:\bigg(\ln(\frac{1-\sqrt{1-\frac{4 m^{2}}{m_{Z}^{2}}}}{1+\sqrt{1-\frac{4 m^{2}}{m_{Z}^{2}}}})-i \pi\bigg)\bigg]+\nonumber \\
&+&  m^{2}\bigg[2 \arctan^{2}\bigg(\frac{1}{\sqrt{\frac{4 m^{2}}{s}-1}}\bigg)+
\frac{1}{2}\bigg(\ln(\frac{1-\sqrt{1-\frac{4 m^{2}}{m_{Z}^{2}}}}{1+\sqrt{1-\frac{4 m^{2}}{m_{Z}^{2}}}})+i \pi\bigg)^{2} \bigg]\bigg\} \;, 
\quad \frac{4 m^{2}}{s}> 1,\: 
\frac{4 m^{2}}{m_{Z}^{2}}< 1 \;,
\end{eqnarray}
\begin{eqnarray}
\mathscr{F}(m_{Z},s,m)&=&-\frac{3}{2}+\frac{\ln(m^{2})}{2}+\bigg(\frac{1}{m_{Z}^{2}-s}\bigg) \bigg \{s 
 \bigg[\frac{1}{2} \sqrt{1-\frac{4 m^{2}}{s}}
\:\bigg(\ln(\frac{1-\sqrt{1-\frac{4 m^{2}}{s}}}{1+\sqrt{1-\frac{4 m^{2}}{s}}})-i \pi\bigg) \bigg]+ \nonumber \\
&+&m_{Z}^{2} \bigg[\sqrt{\frac{4 m^{2}}{m_{Z}^{2}}-1}
\:\ \arctan\bigg(\frac{1}{\sqrt{\frac{4 m^{2}}{m_{Z}^{2}}-1}}\bigg)\bigg]+\nonumber \\
&+&  m^{2}\bigg[2 \arctan^{2}\bigg(\frac{1}{\sqrt{\frac{4 m^{2}}{m_{Z}^{2}}-1}}\bigg)+
\frac{1}{2}\bigg(\ln(\frac{1-\sqrt{1-\frac{4 m^{2}}{s}}}{1+\sqrt{1-\frac{4 m^{2}}{s}}})-i \pi\bigg)^{2} \bigg]\bigg\} \;, 
\quad \frac{4 m^{2}}{s}< 1,\: 
\frac{4 m^{2}}{m_{Z}^{2}}> 1 \;,
\end{eqnarray}
\begin{eqnarray}
\mathscr{F}(m_{Z},s,m)&=&-\frac{3}{2}+\frac{\ln(m^{2})}{2}+\bigg(\frac{1}{m_{Z}^{2}-s}\bigg) \bigg \{s 
 \bigg[\frac{1}{2} \sqrt{1-\frac{4 m^{2}}{s}}
\:\bigg(\ln(\frac{1-\sqrt{1-\frac{4 m^{2}}{s}}}{1+\sqrt{1-\frac{4 m^{2}}{s}}})-i \pi\bigg) \bigg]- \nonumber \\
&-&m_{Z}^{2} \bigg[\frac{1}{2} \sqrt{1-\frac{4 m^{2}}{m_{Z}^{2}}}
\:\bigg(\ln(\frac{1-\sqrt{1-\frac{4 m^{2}}{m_{Z}^{2}}}}{1+\sqrt{1-\frac{4 m^{2}}{m_{Z}^{2}}}})-i \pi\bigg)\bigg]+\nonumber \\
&+&  m^{2}\bigg[
\frac{1}{2}\bigg(\ln(\frac{1-\sqrt{1-\frac{4 m^{2}}{s}}}{1+\sqrt{1-\frac{4 m^{2}}{s}}})\pm i \pi\bigg)^{2}-
\frac{1}{2}\bigg(\ln(\frac{1-\sqrt{1-\frac{4 m^{2}}{m_{Z}^{2}}}}{1+\sqrt{1-\frac{4 m^{2}}{m_{Z}^{2}}}})\pm i \pi\bigg)^{2} \bigg]\bigg\} \;, 
\quad \frac{4 m^{2}}{s}< 1,\:  
\frac{4 m^{2}}{m_{Z}^{2}}< 1\;. \label{eq:C21} \nonumber \\
 \end{eqnarray}
In \eq{eq:C21}, the plus sign corresponds to  $s<m_{Z}^{2}$ while the minus sign
to $s>m_{Z}^{2}$. As an example the full analytical expressions for 
$A_{3}$ and $A_{5}$ can be obtained by taking appropriate derivatives of function 
$\mathscr{F}$ like, 
$A_{3}=c\,\ \frac{\partial \mathscr{F}}{\partial s}$ and $A_{5}=-c\,\ 
(\frac{\partial \mathscr{F}}{\partial s} + 
\frac{\partial \mathscr{F}}{\partial m_{Z}^{2}})$, where, as above, 
$c$ is a factor related to the couplings in 
the corresponding vertex.
As a cross check, taking the limit $m\rightarrow 0$ in \eq{eq:C21} we arrive at,
\begin{eqnarray}
 \mathscr{F}(m_{Z},s,0)=-\frac{3}{2}-\frac{1}{2(m_{Z}^{2}-s)}\bigg [s \ln(s)-m_{Z}^{2} \ln(m_{Z}^{2})\bigg ]+\frac{i \pi}{2}\;,
\end{eqnarray}
and differentiating w.r.t $s$ and $m_{Z}^{2}$  we  reproduce the expressions 
\eqs{approx1}{approx2}.

\section{Conditions for non-decoupling effects in $X,Y,Z$ model}
\label{app:cate} 
In this appendix we present  necessary conditions for anomaly cancellation
and non-decoupling heavy fermion effects in a model 
with three different $U(1)$'s corresponding to three distinct 
massive or massless gauge bosons $X$,$Y$, and $Z$. For this model to
be anomaly-free, the following conditions among couplings [see \eq{lint}]:
\begin{eqnarray}
  \sum_{i=1}^{n}(\beta_{X}^{3}+3 \alpha_{X}^{2}\beta_{X})_{i}=
   \sum_{i=1}^{n}(\beta_{Y}^{3}+3 \alpha_{Y}^{2}\beta_{Y})_{i}= 
   \sum_{i=1}^{n}(\beta_{Z}^{3}+3 \alpha_{Z}^{2}\beta_{Z})_{i} &=&0 \;, 
   \nonumber \\
   \sum_{i=1}^{n}(\beta_{X}^{2}\beta_{Y}+2 \alpha_{X}\alpha_{Y}\beta_{X}+\alpha_{X}^{2}\beta_{Y})_{i}=
    \sum_{i=1}^{n}(\beta_{X}^{2}\beta_{Z}+2 \alpha_{X}\alpha_{Z}\beta_{X}+\alpha_{X}^{2}\beta_{Z})_{i}= 
   \sum_{i=1}^{n}(\beta_{Y}^{2}\beta_{X}+2 \alpha_{X}\alpha_{Y}\beta_{Y}+\alpha_{Y}^{2}\beta_{X})_{i} &=& 0 \;, \nonumber \\
   \sum_{i=1}^{n}(\beta_{Y}^{2}\beta_{Z}+2 \alpha_{Z}\alpha_{Y}\beta_{Y}+\alpha_{Y}^{2}\beta_{Z})_{i}=
   \sum_{i=1}^{n}(\beta_{Z}^{2}\beta_{X}+2 \alpha_{ X}\alpha_{Z}\beta_{Z}+\alpha_{Z}^{2}\beta_{X})_{i}=
    \sum_{i=1}^{n}(\beta_{Z}^{2}\beta_{Y}+2 \alpha_{Z}\alpha_{Y}\beta_{Z}+\alpha_{Z}^{2}\beta_{Y})_{i} &=& 0\;,  \nonumber \\
    \sum_{i=1}^{n}(\beta_{X}\beta_{Y}\beta_{Z} +\alpha_{X}\alpha_{Z}\beta_{Y}+\alpha_{X}\alpha_{Y}\beta_{Z}+\alpha_{Z}\alpha_{Y}\beta_{X})_{i}&=&0\;, \nonumber \\ \label{eq:aXYZ}
\end{eqnarray}
must hold.
Non-decoupling effects in $XYZ$-vertex are activated if, in addition
to the requirements in \eq{eq:aXYZ},
at least one of the following expressions is  non-zero:
\begin{eqnarray}
&  &
 \sum_{i=1}^{n}(\beta_{X}^{2}\beta_{Y}+3\,\ \alpha_{X}\alpha_{Y}\beta_{X})_{i} \;, 
 \qquad \nonumber \\
& &  \sum_{i=1}^{n}(\beta_{X}^{2}\beta_{Y}+3\,\ \alpha_{X}^{2}\beta_{Y})_{i} \;, \qquad
\sum_{i=1}^{n}(\beta_{X}^{2}\beta_{Z}+3\,\ \alpha_{X}\alpha_{Z}\beta_{X})_{i} \;, \qquad
 \sum_{i=1}^{n}(\beta_{X}^{2}\beta_{Z}+3\,\ \alpha_{X}^{2}\beta_{Z})_{i} \;, \qquad
\sum_{i=1}^{n}(\beta_{Y}^{2}\beta_{X}+3\,\ \alpha_{X}\alpha_{Y}\beta_{Y})_{i} \nonumber \\
& &\sum_{i=1}^{n}(\beta_{Y}^{2}\beta_{X}+3\,\ \alpha_{Y}^{2}\beta_{X})_{i} \;, \qquad
\sum_{i=1}^{n}(\beta_{Y}^{2}\beta_{Z}+3\,\ \alpha_{Y}\alpha_{Z}\beta_{Y})_{i} \;, \qquad
\sum_{i=1}^{n}(\beta_{Y}^{2}\beta_{Z}+3\,\ \alpha_{Y}^{2}\beta_{Z})_{i} \;, \qquad
\sum_{i=1}^{n}(\beta_{Z}^{2}\beta_{X}+3\,\ \alpha_{X}\alpha_{Z}\beta_{Z})_{i} \nonumber \\
& &\sum_{i=1}^{n}(\beta_{Z}^{2}\beta_{X}+3\,\ \alpha_{Z}^{2}\beta_{X})_{i} \;, \qquad
\sum_{i=1}^{n}(\beta_{Z}^{2}\beta_{Y}+3\,\ \alpha_{Y}\alpha_{Z}\beta_{Z})_{i} \;, \qquad
\sum_{i=1}^{n}(\beta_{Z}^{2}\beta_{Y}+3\,\ \alpha_{Z}^{2}\beta_{Y})_{i} \;, \qquad
\sum_{i=1}^{n}(\beta_{X}\beta_{Y}\beta_{Z}+3\,\ \alpha_{X}\alpha_{Z}\beta_{Y})_{i} \nonumber \\
& &\sum_{i=1}^{n}(\beta_{X}\beta_{Y}\beta_{Z}+3\,\ \alpha_{X}\alpha_{Y}\beta_{Z})_{i} \;, \qquad
\sum_{i=1}^{n}(\beta_{X}\beta_{Y}\beta_{Z}+3\,\ \alpha_{Y}\alpha_{Z}\beta_{X})_{i} \;.\label{eq:bXYZ}
\end{eqnarray}

\end{widetext}

\bibliographystyle{h-physrev5}  
\bibliography{biblio}       


\end{document}